\documentclass[prb,twocolumn,notitlepage,superscriptaddress]{revtex4-2}
\usepackage{amsmath,amsfonts,amssymb,amsthm,epsfig,array}
\usepackage{bbold}
\usepackage{dsfont}
\usepackage{slashed}
\usepackage{graphics}
\usepackage{float}
\usepackage{verbatim}
\usepackage{color}
\usepackage[dvipsnames]{xcolor}
\usepackage[hidelinks,colorlinks,linkcolor=blue,
citecolor=blue,urlcolor=blue]{hyperref}
\usepackage[titletoc,title]{appendix}
\usepackage{multirow}
\usepackage{bibentry}
\usepackage{tabularx}
\usepackage[mathscr]{euscript}
\usepackage{mathtools}
\usepackage{physics}
\usepackage{siunitx}
\usepackage{cleveref}

\crefname{appendix}{Appendix.}{Appendices.}
\crefname{equation}{Eq.}{Eqs.}
\crefname{figure}{Fig.}{Figs.}
\crefname{table}{Table}{Tables}
\crefname{section}{Section}{Sections}
\creflabelformat{appendix}{[#2#1#3]}

\newcommand{\bsl}[1]{\boldsymbol{#1}}

\let\vec\mathbf
\renewcommand{\mod}{\,\mathrm{mod}\,}

\newcommand{\bea}{\begin{equation} \begin{aligned}}
\newcommand{\eea}{\end{aligned} \end{equation} }
\DeclareRobustCommand{\EqJHA}[1]{Eq.~(\ref{#1})}
\newcommand{\la}{\lambda}
\newcommand{\be}{\beta}
\newcommand{\al}{\alpha}
\newcommand{\bpm}{\begin{pmatrix}}
\newcommand{\epm}{\end{pmatrix}}
\newcommand{\eps}{\epsilon}

\renewcommand{\th}{\theta}

\newcommand{\lp}{\left(}
\newcommand{\rp}{\right)}

\newcommand{\del}{\partial}

\newcommand{\mbf}[1]{\mathbf{#1}}

\renewcommand{\Tr}{\text{Tr}}

\usepackage{braket}

\newcommand{\ii}{\mathrm{i}}

\newcommand{\dsZ}{\mathbb{Z}}

\newcommand{\dsR}{\mathbb{R}}

\renewcommand{\Re}{\mathop{\mathrm{Re}}}
\renewcommand{\Im}{\mathop{\mathrm{Im}}}

\newcommand{\SU}{\mathrm{SU}}
\newcommand{\U}{\mathrm{U}}

\newcommand{\eqnref}[1]{Eq.\,\eqref{#1}}
\newcommand{\figref}[1]{Fig.\,\ref{#1}}

\newcommand{\appref}[1]{Appendix.\,\ref{#1}}
\newcommand{\refcite}[1]{Ref.\,\cite{#1}}

\newcommand{\mat}[1]{\left(\begin{matrix}#1\end{matrix}\right)}

\newcommand{\eq}[1]{\begin{equation} #1 \end{equation}}

\newcommand{\eqa}[1]{\begin{align}\begin{split} #1 \end{split}\end{align}} 

\usepackage{environ}
\NewEnviron{eqs}{%
\begin{equation}\begin{split}
    \BODY
\end{split}\end{equation}
}

\let\oldAA\AA
\renewcommand{\AA}{\text{\normalfont\oldAA}}

\newcommand{\ie}{{\emph{i.e.}}}

\newcommand{\TR}{\mathcal{T}}

\newcommand{\C}{\mathcal{C}}

\newcommand{\X}{\mathcal{X}}

\newcommand{\M}{\mathcal{M}}
\newcommand{\N}{\mathcal{N}}

\newcommand{\Q}{\mathcal{Q}}
\newcommand{\V}{\mathcal{V}}
\newcommand{\K}{\text{K}}

\newcommand{\MBZ}{\text{MBZ}}

\newcommand{\Ch}{\text{Ch}}

\newcommand{\cre}[2]{\hat{#1}^\dagger_{#2}}
\newcommand{\des}[2]{\hat{#1}_{#2}}
\newcommand{\tTBG}{\text{TBG}}
\newcommand{\tD}{\text{D}}
\newcommand{\tI}{\text{I}}

\newcommand{\thc}{\text{h.c.}}
\newcommand{\zero}{\mathbb{0}}

\newcommand{\normord}[1]{:\mathrel{#1}:}
\newcommand{\bSigma}{\boldsymbol{\sigma}}

\newcommand{\brak}[1]{\left[#1\right]}
\newcommand{\vk}{\vec{k}}

\newcommand{\vK}{\vec{K}}
\newcommand{\vQ}{\vec{Q}}
\newcommand{\vq}{\vec{q}}
\newcommand{\vG}{\vec{G}}
\newcommand{\ut}{\tilde{u}}

\newcommand{\ies}[1]{i_{#1} \eta_{#1} s_{#1}}

\newcommand{\iesC}[1]{i_{#1}, \eta_{#1}, \sigma_{#1}}

\newcommand{\expec}[1]{\left\langle #1 \right\rangle}
\allowdisplaybreaks
\begin{document}
\title{Topological Heavy Fermion Principle For Flat (Narrow) Bands With Concentrated Quantum Geometry}

\author{Jonah Herzog-Arbeitman}
\thanks{These authors contributed equally.}
\affiliation{Department of Physics, Princeton University, Princeton, New Jersey 08544, USA}

\author{Jiabin Yu}
\thanks{These authors contributed equally.}
\affiliation{Department of Physics, Princeton University, Princeton, New Jersey 08544, USA}

\author{Dumitru C\u{a}lug\u{a}ru$^*$}
\thanks{These authors contributed equally.}
\affiliation{Department of Physics, Princeton University, Princeton, New Jersey 08544, USA}

\author{Haoyu Hu}
\thanks{These authors contributed equally.}
\affiliation{Donostia International Physics Center, P. Manuel de Lardizabal 4, 20018 Donostia-San Sebastian, Spain}

\author{Nicolas Regnault}
\affiliation{Department of Physics, Princeton University, Princeton, New Jersey 08544, USA}
\affiliation{Laboratoire de Physique de l’Ecole normale sup\'erieure,
ENS, Universit\'e PSL, CNRS, Sorbonne Universit\'e,
Universit\'e Paris-Diderot, Sorbonne Paris Cit\'e, 75005 Paris, France}

\author{Chaoxing Liu}
\affiliation{Department of Physics, the Pennsylvania State University, University Park, Pennsylvania 16802, USA}

\author{Oskar Vafek}
\affiliation{National High Magnetic Field Laboratory, Tallahassee, Florida, 32310, USA}
\affiliation{Department of Physics, Florida State University, Tallahassee, Florida 32306, USA}

\author{Piers Coleman}
\affiliation{Center for Materials Theory, Department of Physics and Astronomy, Rutgers
University, Piscataway, NJ 08854 USA}
\affiliation{Department of Physics, Royal Holloway University of London, Egham, Surrey TW20
0EX, UK}

\author{Alexei Tsvelik}
\affiliation{Division of Condensed Matter Physics and Materials Science,
Brookhaven National Laboratory, Upton, NY 11973-5000, USA}

\author{Zhi-da Song}
\affiliation{International Center for Quantum Materials, School of Physics, Peking University, Beijing 100871, China}
\affiliation{Hefei National Laboratory, Hefei 230088, China}
\affiliation{Collaborative Innovation Center of Quantum Matter, Beijing 100871, China}

\author{B. Andrei Bernevig}
\email{bernevig@princeton.edu}
\affiliation{Department of Physics, Princeton University, Princeton, New Jersey 08544, USA}
\affiliation{Donostia International Physics Center, P. Manuel de Lardizabal 4, 20018 Donostia-San Sebastian, Spain}
\affiliation{IKERBASQUE, Basque Foundation for Science, Bilbao, Spain}

\begin{abstract}

We propose a general principle for the low-energy theory of narrow bands with concentrated Berry curvature and Fubini-Study metric in the form of a map to Anderson-``+" models composed of heavy fermions hybridizing and interacting with semi-metallic modes. This map resolves the obstruction preventing topological bands from being realized in a local Hamiltonian acting on the low-energy degrees of freedom. The concentrated quantum geometry is reproduced through band inversion with a dispersive semi-metal, leaving a nearly flat, trivial band which becomes the heavy fermion. This representation is natural when the narrow band is not energetically isolated on the scale of the interaction and an enlarged Hilbert space is inescapable, but also provides analytical insight into the projected-interaction limit. First exemplified in twisted bilayer graphene (TBG), we extend it to (1) the twisted checkerboard, which we find has a chiral symmetric stable anomaly that forbids a lattice realization at all energies, and (2) the Lieb lattice with gapless flat bands, where we show the heavy fermions can be obtained by minimizing a Euclidean instanton action to saturate its BPS bound. The heavy fermion approach is widely applicable and physically transparent: heavy electrons carry the strong correlations and dispersive electrons carry the topology. This simple picture unifies the dichotomous phenomena observed in TBG and points to connections between moir\'e and stoichiometric materials. 

\end{abstract}

\maketitle

\section{Introduction}


Flat or narrow bands resulting from quantum interference can appear in any dimension, in both stoichiometric and engineered (moire) materials, and engender strongly correlated phases of matter \cite{2023arXiv231210659C,Balents2020TBGSCReview,2022Natur.603..824R}. In stoichiometric compounds, exactly flat bands arise in local tight binding models on bipartite lattices and their descendants\cite{Lieb1989LiebLattice, Dumitru2022GeneralConstructionFlatBand,Chiu2020FragileFlat,PhysRevB.99.045107,PhysRevB.104.085144}, providing a principle for the existence of flat or narrow band materials \cite{yin2022topological,kang2020topological,li2018realization,kang2020dirac,2023arXiv230409066H,wakefield2023three}. Moreover, flat bands arising from strong hoppings are likely to be (fragile) topological if they are gapped from dispersive bands \cite{Dumitru2022GeneralConstructionFlatBand}. Topology, and more generally quantum geometry, is a crucial feature for flat bands to host exotic correlated states, since trivial flat bands can be formed from spatially decoupled atomic limits. In moir\'e systems, flat bands typically arise in continuum models away from the tight-binding limit \cite{Bistritzer2011BMModel,PhysRevLett.122.086402}, and exhibit emergent symmetries in the small twist angle limit which frequently render them topological. In these cases, no exponentially localized Wannier description which faithfully represents the symmetries of the flat bands is possible. Since continuum Hamiltonians are infinite, analytic expressions for the wavefunctions are difficult to find, a notable exception being the idealized ``chiral" limit \cite{PhysRevLett.122.106405,PhysRevB.108.075126,ledwith2021tb,PhysRevResearch.3.023155,KaiSun2023THF} of twisted bilayer graphene (TBG) which remains separated from experimentally relevant parameters.

\begin{figure}
    \centering
    \includegraphics[width=\columnwidth]{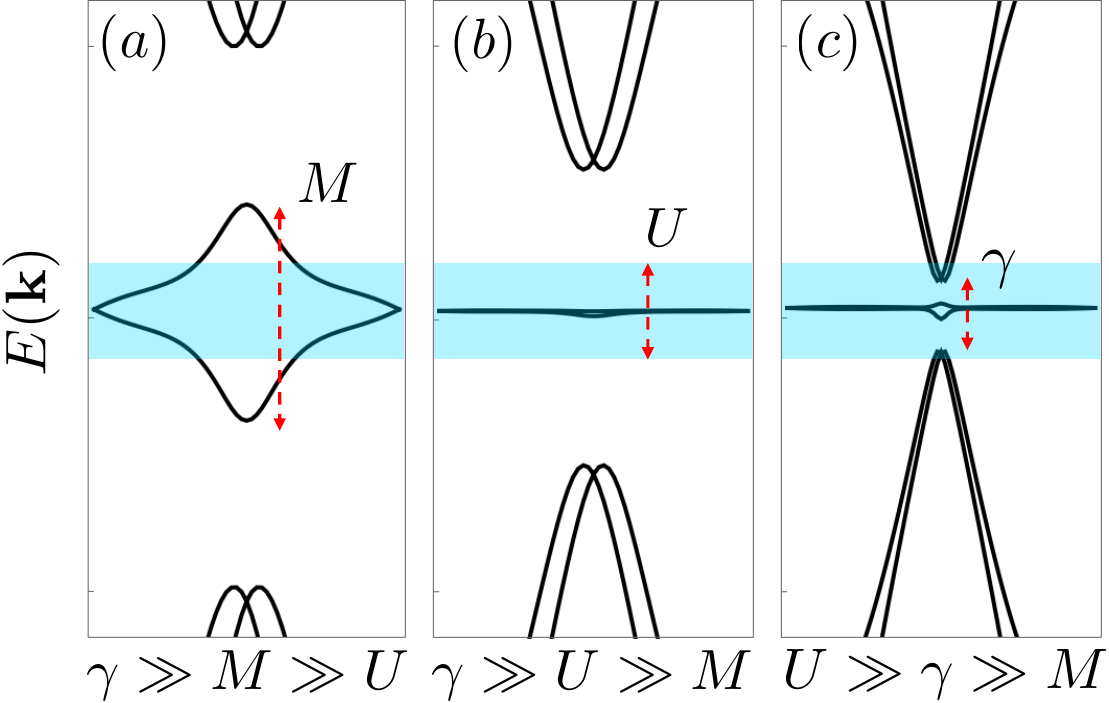}
    \caption{Characteristic energy scales of a Fermi liquid $(a)$, a projected/strong coupling ferromagnet $(b)$, and a heavy fermion model $(c)$.}
    \label{fig:main}
\end{figure}

A common characteristic of these flat bands (outside the chiral limit) is the presence of concentrated, generically non-abelian, quantum geometry in isolated regions of the Brillouin zone. The question then arises: what is the interacting physics particular to this type of flat band system? There are several limits, schematized in \figref{fig:main}.  If the interaction scale $U$ is weaker than the narrow band bandwidth $M$, then a Fermi liquid (with its own instabilities at Van Hove singularities) is expected and weak coupling Hartree-Fock methods can be used, though analytic progress is dependent upon expressions for the eigenstates which may not be available. Second, if the interaction is much larger than the bandwidth $M$ but smaller than the gap to dispersive bands $\gamma$, then one can project the interaction to the flat bands and solve a strong coupling interacting problem with exact Slater groundstates \cite{KAN19,LIA21,BER21b,Zaletel2019Nov5MATBGIntegerFilling,2024arXiv240207171K} in the repulsive case at integer filling, or exact eta-pairing groundstates in the attractive case \cite{PhysRevB.94.245149,PhysRevB.106.184517,2022arXiv220900007H,2023arXiv230815963S,2024arXiv240104163H} in the spin-0 sector. These exact results are also in agreement with Hartree-Fock, and analytical insight again requires knowledge of the flat eigenstates to obtain the form factors. Third, if the interaction $U$ is larger than both the bandwidth $M$ and the gap to the dispersive bands $\gamma$ one cannot perform any meaningful restriction and the problem is both strongly coupled and not prone to projection. The $\bf{k}$-space formalism obfuscates any local description. Realistic models of TBG, as well as gapless flat bands including the Lieb lattice \cite{Lieb1989LiebLattice} and twisted trilayer graphene \cite{Vishwanath20190129MATMG}, and many stoichiometric materials fall under this latter category.

In twisted bilayer graphene, detailed microscopic calculations\cite{Song20211110MATBGHF,Dai20220920HFTBGTMG,CAL23,2023arXiv230508171S,2023arXiv230302670L} have shown that the topological flat bands can be mapped into a set of trivial and highly localized heavy fermions ($f$-electrons) hybridized with untwisted bilayer graphene $c$-electrons of mass $\pm M \approx 4$ meV via a matrix element $\gamma \approx 25$meV. Their band structure matches that of the Bistritzer-MacDonald (BM) model \cite{Bistritzer2011BMModel} almost perfectly within the range of the Coulomb interaction $U\approx 40-50$ meV, obtained assuming a dielectric constant $\eps = 6$. $2M$ becomes the bandwidth of the narrow (almost flat) bands and $\gamma$ is the gap between the flat bands and a set of dispersive bands in the BM model. This is a faithful, symmetric, unitary transformation of the low-energy bands of the model, where strain, relaxation, and other effects can be included \cite{Koshino2018TBGFragile,PhysRevB.96.075311,Bi2019TBG,PhysRevB.107.075123,2023PhRvB.107g5408K,PhysRevB.102.155136}. The immediate benefit of this transformation is that it identifies the heavy carriers of the interaction $U$ ---the $f$-electrons--- and the itinerant carriers of the topology --- the $c$-electrons. The interaction can be rewritten in this basis, and one obtains a generalized Anderson model consisting of a large Hubbard $U \approx 50$ meV on the $f$-electrons, as well as other interactions.  This model is suitable is suitable for DMFT implementation \cite{Hu2023KondoMATBG,HU23,Chou20221128MATBGKondo,chou2023scaling,2023NatCo..14.5036D,ZHO24,RAI23a,2024arXiv240214057C}, and can match in a natural way a large amount of the experimental features observed in a wide array of experiments. Furthermore, it provides for a Kondo mechanism \cite{2024arXiv240200869W,HH_TBG_SC} for screening of the repulsive $U$ as a potential pathway to understand superconductivity --- the core mystery of TBG.   

Twisted bilayer and trilayer graphene\cite{2023PhRvB.108c5129Y} are not the only systems whose flat bands can be mapped into a topological heavy fermion model. Recently, a quadratic touching model that is related to the twisted checkerboard model \cite{PhysRevResearch.4.043151} has also been mapped to a heavy fermion problem \cite{KaiSun2023THF}. Here we present a wider perspective, arguing that a broad class of narrow/flat band models with concentrated Berry curvature or quantum geometry admit such a heavy fermion description. This entails a single particle Hamiltonian that separates the bands into atomic $f$-electrons coupled to topological semi-metals, as well as periodic Anderson-``+" interacting terms. This broad class includes gapped and gapless bands in continuum or tight-binding models as we will exemplify. 

We first review the emergence of heavy fermions in TBG, and give a simple two-band example on the lattice of a Chern band hybridizing with a metal. We then consider the twisted checkerboard lattice in the chiral limit \cite{KaiSun2023THF}, completing the heavy fermion theory by (i) showing the stable anomaly that obstructs a chiral-symmetric tight-binding model at all energies and (ii) specifying the forms of the Anderson-``+" interacting terms. We also consider a family of gapless flat bands on bipartite lattices whose most famous representative is the Lieb lattice. There we examine the heavy fermion Wannier state wavefunction, showing that it can be solved analytically by mapping the Wannier spread localization problem to a Euclidean instanton action. 

Finally, we present arguments that heavy fermion physics arises in interacting narrow bands with concentrated (even divergent, in the gapless case) quantum geometry when the interaction strength is larger than the gap and the narrow band width. To this end, we provide simple estimates for when this regime is reached, and also characterize flat bands that cannot be mapped to heavy fermion models.

\section{Gapped Topological Bands}

\subsection{$C_{2z} T$ protected obstruction: Heavy Fermions in Twisted Bilayer Graphene}

\begin{figure}
    \centering
    \includegraphics[width=\columnwidth]{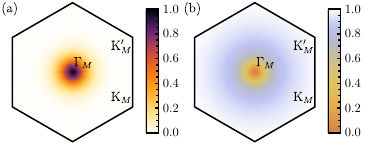}
    \caption{ 
    (a) The non-abelian Berry curvature, expressed in arbitrary units, of the flat Chern band in TBG at the magic angle. 
    (b) The $f$-electron orbital weight within the active TBG bands at the magic angle.}
    \label{fig:berry_curvature_THF}
\end{figure}

Twisted bilayer graphene (TBG) is formed by stacking together two layers of graphene with a slight twist~\cite{Bistritzer2011BMModel} and exhibits unique electronic properties, such as correlated insulators~\cite{Cao2018TBGMott} and superconducting phases~\cite{Cao2018TBGSC}. Its non-interacting band structure, which can be described by the Bistritzer-MacDonald (BM) model ~\cite{Bistritzer2011BMModel}, exhibits eight flat bands with stable/fragile topology (in the presence/absence of eigenstate particle-hole symmetry)~\cite{Song2019TBGFragile,Po2019TBGFragile,Song2020TBGII} around charge neutrality. Importantly, the nonabelian Berry curvature associated with these flat bands is predominantly localized near the $\Gamma$ point~\cite{Song20211110MATBGHF}, as shown in~\cref{fig:berry_curvature_THF} (a). This observation enables one to model the TBG as a topological heavy fermion (THF) system, which consists of strongly correlated, localized $f$-orbitals and topological, dispersive $c$-electrons~\cite{Song20211110MATBGHF}. The concentrated Berry curvature allows over 95\% of the states within the flat bands to be accurately described by the $f$ orbitals as shown in~\cref{fig:berry_curvature_THF} (b). The non-trivial topology of the flat bands arises from the hybridization between $f$ and topological $c$ electrons. Remarkably, this mapping also enables one to analytically obtain the wavefunctions of the active TBG flat bands in the zero-bandwidth limit. These flat bands exhibit a chiral symmetry which differs from the conventional chiral symmetry of the BM model in the unrealistic $w_0/w_1 = 0$ limit~\cite{Tarnopolsky2019MagicAngleChiralLimit}. Hence the resulting wavefunctions (see~\cref{app:sec:anal:wavf})~\cite{Song20211110MATBGHF}
\begin{align}
     \cre{f}{\vk,1,\eta,s} -  \frac{\left( \gamma e^{-\frac{|\vk|^2\lambda^2}{2}}/v_\star \right) \cre{c}{\vk,3,\eta,s} }{\eta k_x + i k_y }, \label{creationopsBM}
\end{align}
are a good approximation of the realistic TBG bands. Here $\cre{d}{\vk, \zeta=+, \eta, s}$ is the electron operator in the Chern band basis with Chern number $\zeta \cdot\eta$, valley $\eta \in \{+,-\}$, spin $s \in \{\uparrow,\downarrow\}$, while $\cre{f}{\vk,\alpha,\eta,s}$ and $\cre{c}{\vk+\vG,a,\eta,s}$ are the corresponding electron operators of the $f$- and $c$-fermions, respectively. $\alpha \in \{1,2\},a\in \{1,2,3,4\} $ denote the orbital indices of $f$- and $c$-fermions respectively. $\gamma$ characterizes the $f$-$c$ hybridization strength with a decay factor $\lambda$. $v_{\star}$ is the Fermi velocity of non-interacting $c$ electrons. 
$\cre{d}{\vk, \zeta=+, \eta, s}$ electron operators admit analytical expressions in the plane-wave BM basis~\cite{Song20211110MATBGHF,CAL23} using the form of the $f$-electron Wannier states and of the $c$-electron basis. This allows for computing form factors of the $\cre{d}{\vk, +, \eta, s}$ operators. Approximate analytic formulas can be obtained
as shown in~\cref{app:sec:anal:wavf:ff} and were used in Ref.~\cite{liu2023electronkphonon} to compute the electron-phonon interaction. 

There are two interesting limits of this heavy fermion formalism. First, if the Hubbard interactions $U_1$ of $f$-fermions is larger than the $f$-$c$ hybridization strength $|\gamma|$ (which is also the gap between flat active bands and remote bands), we should include both flat and dispersive bands and treat the system as an effective heavy fermion system with both $f$ and $c$ fermions. In the opposite limit, where $|\gamma|>U_1$, a flat-band-projected approach~\cite{Zaletel2019Nov5MATBGIntegerFilling,LIA21} is more suitable for the problem. However, even in this limit, the THF model can still be employed to derive the analytical expressions of the flat-band wavefunction and to calculate the excitation spectrum as we discuss in~\cref{app:sec:anal}.

\subsection{Chern bands with localized Berry curvature}
A similar heavy fermion construction can also be obtained for gapped Chern bands with concentrated Berry curvature. To see this, we consider a simple model with $s$ and $p_z$ orbitals located at the $1a$ position of a square lattice (see~\cref{app:sec:gap_chern_bands})~\cite{bhz_model}. We use $\cre{f}{\vk,\sigma}$ and $\cre{c}{\vk,\sigma}$ to denote the creation operator for the $s$ and $p_z$ orbitals, respectively, with $\vk$ being the momentum and $\sigma $, the spin index. 
The $p_z$ orbitals host a doubly-degenerate quadratic band near the $\Gamma$ point with dispersion $\epsilon_{\vk,p_z} \approx -\Delta \epsilon + \frac{t_p \abs{\vk}^2}{2}$, where $t_p$ denotes the nearest-neighbor hopping of the $p_z$ orbitals, and $\Delta\epsilon$ controls the energy level of the $p_z$ orbitals. The $s$ orbitals have vanishing kinetic energy, but hybridize with the $p_z$ orbitals.

The single-particle Hamiltonian can be written as 
\begin{align} 
   &H_0 = \sum_{|\vk|<\Lambda_c,\sigma } (\Delta \epsilon -\frac{t_p}{2}|k|^2) \cre{c}{\vk,\sigma}\des{c}{\vk,\sigma} \nonumber\\ 
   &  +\frac{t_{sp}}{\sqrt{N}} \sum_{|\vk|<\Lambda_c,\vec{R},\sigma}\left[e^{-i\vk\cdot\vec{R}}(\sigma k_y+ik_x)\cre{c}{\vk,\sigma} \des{f}{\vec{R},\sigma}+\text{h.c.}\right] \, .
\end{align} 
where $N$ is the number of unit cells, $t_{sp}$ characterizes the $f$-$c$ hybridization strength and $\Lambda_c$ is the momentum cutoff of $c$ electrons.  
We consider the limit where $|\Delta\epsilon|, |t_{sp}| \ll |t_{p}|$ and $\Delta\epsilon <0$. Then the system consists of a two-fold degenerate narrow Chern band with bandwidth $\sim |\Delta\epsilon|$ and Chern numbers $\pm 1$, and dispersive bands that are mostly formed by the dispersive $p_z$ orbitals. The non-trivial topology of the narrow band comes from the band inversion that happens near the $\Gamma$ point, where we can observe that the Chern bands are mostly formed by $p_z$ orbitals near $\Gamma$ and $s$ orbitals away from it. The Berry curvature is also concentrated near the $\Gamma$ point (see~\cref{app:sec:gap_chern_bands}). 
Therefore, our simple model indeed realizes a system with both narrow Chern bands and dispersive bands. 
We can already observe that, this system is equivalent to a heavy fermion system where the $s$ orbitals correspond to the localized $f$ electrons, while the $p_z$ orbitals represent dispersive $c$ electrons. The flatness of the bands comes from the localized nature of $s$ orbitals ($f$ fermions) and the non-trivial topology emerges from the hybridization between $s$ and $p_z$ orbitals. 
After further introducing the on-site Hubbard interaction of the $s$ orbital ($U$) and $p$ orbital ($V$), along with the Coulomb repulsion between $s$ and $p$ orbitals ($W$), the full Hamiltonian adapts a formula similar to the THF Hamiltonian of the TBG (~\cref{app:sec:gap_chern_bands}).
Therefore, we have established the duality between a topological heavy fermion systems and a system that contains narrow Chern bands with concentrated Berry curvatures. This duality has also been demonstrated in a moir\'e system featuring Chern flat bands~\cite{CX_Chernband}.

\subsection{Chiral-Symmetry-Protected Wannier Obstruction: Heavy Fermions in  the Twisted Checkerboard}

The stable anomaly giving rise to the topology of  the TBG narrow bands is protected by particle-hole symmetry for small twist angles~\cite{tbg2}.
We now discuss a stable anomaly protected by another symmetry, chiral symmetry, which is present in various models, including the twisted checkerboard model which we now discuss.
Given a generic 2D single-particle matrix Hamiltonian $h(\bsl{k})$ with chiral symmetry $\mathcal{C}$ obeying  $\{\mathcal{C},h(\bsl{k})\}=0 $ and $\mathcal{C}^2 = \C \C^\dagger= 1$, there may still be an obstruction to finding exponentially localized Wannier functions for an isolated set of bands, even if the set of bands have zero total Chern number.
To see this is so, let us consider a generic isolated set of $2N$ chirally symmetric bands: their energies are symmetric with respect to the zero energy and their projection matrix $P(\bsl{k})$ commutes with $\mathcal{C}$. An odd number of chirally symmetric bands would necessarily have an exactly-flat zero-energy band, and is not our focus here.
At a fixed $\bsl{k}$, we can separate the Hilbert space for $P(\bsl{k})$ into the chiral-even and chiral-odd subspaces, each of which has the dimension $N$; in other words, $P(\bsl{k})=P_{+}(\bsl{k}) + P_{-}(\bsl{k})$ with $\C P_{\pm}(\bsl{k}) = \pm P_{\pm}(\bsl{k})$.
We then define the Berry curvature and Chern numbers for $P_{\pm}(\bsl{k})$ as $F_{\pm}(\bsl{k})$ and $\Ch_{\pm}$, respectively, and the total Chern number of the entire isolated set of $2N$ bands reads $\Ch = \Ch_+ + \Ch_-$.
Even if the total Chern number is zero, we can still have nonzero chiral Chern number $\Ch_+ - \Ch_-$, which forbids the construction of the exponentially-localized Wannier functions with strictly local   $\mathcal{C}$  in the real space. See details in \appref{app:chiral_symm_anomaly}.  The nonzero chiral Chern number $\Ch_+ - \Ch_-$ can occur even if there is $C_{2z}\TR$ symmetry, but a stable anomaly cannot be protected by $C_{2z}\TR$ alone.  

In practice, the nonzero value of $\Ch_+ - \Ch_-$ can be indicated by the total winding number of the chiral protected band-touching points at zero energy.
Chiral symmetry can protect stable band touching points at zero energy in 2D~\cite{Chiu2016RMPTopoClas}, labeled  $\bsl{k}_l$. The robustness of the band touching points is indicated by their nonzero chiral-protected winding numbers, labelled by $W_l$. 
As elaborated in \appref{app:chiral_symm_anomaly}, we prove that 
\eq{
\label{eq_main:winding_ch_chiral}
\sum_l W_l = \Ch_+ - \Ch_-\ .
}
Therefore, if $\sum_l W_l \neq 0$, a Wannier obstruction exists for any isolated chiral-symmetric set of $2N$ bands, indicating a stable anomaly protected by chiral symmetry.
Continuously deforming $h(\bsl{k})$ cannot change $\sum_l W_l$, unless the continuous deformation involves making $h(\bsl{k})$ into a vanishing matrix.
Using \eqnref{eq_main:winding_ch_chiral}, we can conclude that the single-valley spinless single-particle Hamiltonian of twisted bilayer graphene in the chiral limit~\cite{tbg2} has chiral-protected anomaly, since it has two Dirac cones at zero energy whose total winding number is 2.

\begin{figure}
    \centering
    \includegraphics[width=\columnwidth]{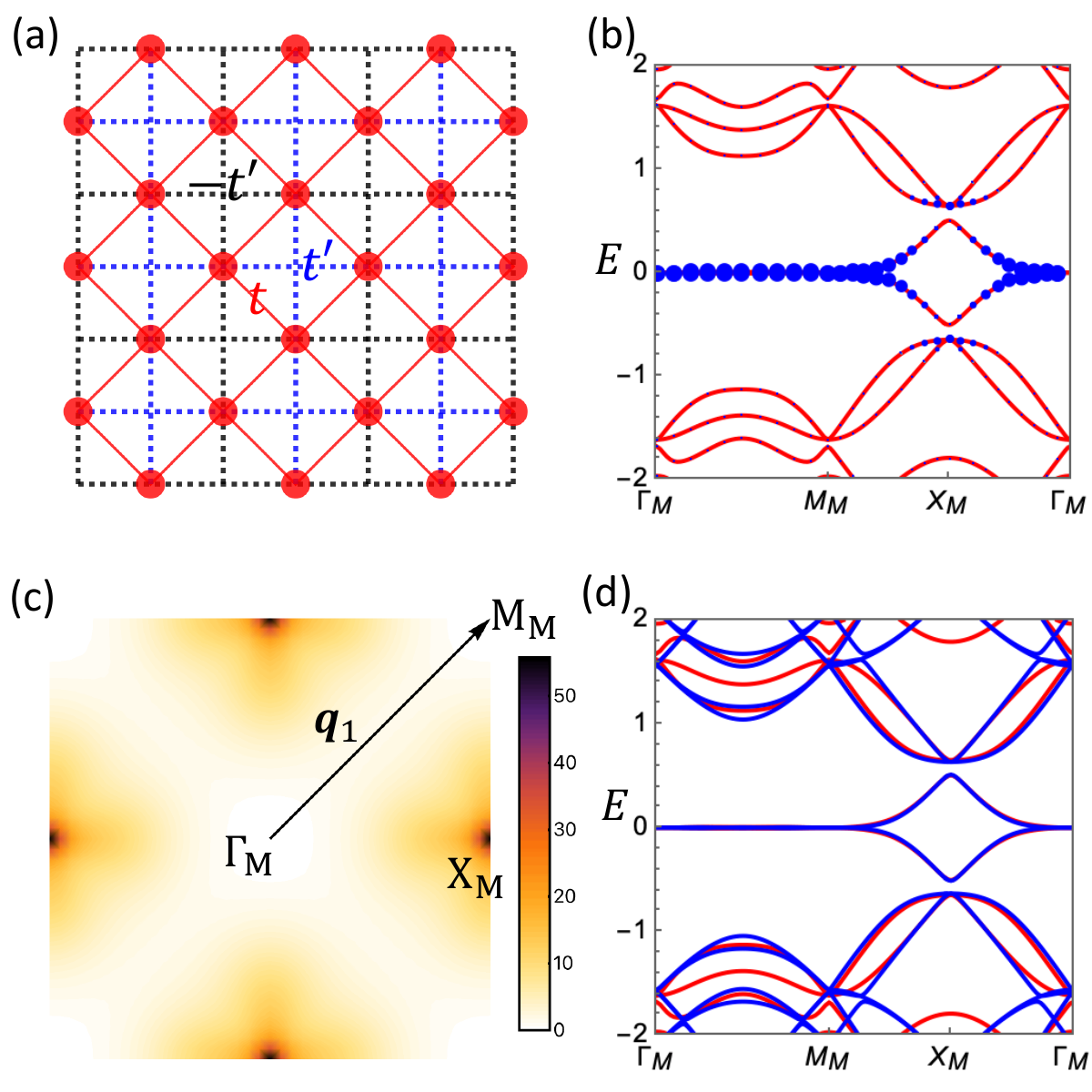}
    \caption{(a) The monolayer checkerboard model parametrized by two hoppings proposed in \refcite{Yao2022TwistedCheckorboard}. (b) The band structure (red) of the twisted checkerboard model with $t = 4$, $t' = 1.26$, $w_1 = 0.66$ and $w_1' = -0.4$. The blue dots mark the probability of the $f$ modes. (c) The chiral Berry curvature $F_+(\bsl{k})-F_{-}(\bsl{k})$ for the isolated set of two bands around zero energy in (b). (d) The comparison between the continuum-model band structure (red) and the heavy fermion-model band structure (blue).}
    \label{fig:CherkerBoard}
\end{figure}

The twisted checkerboard model in the chiral limit proposed in \refcite{Yao2022TwistedCheckorboard} has a stable anomaly protected by chiral symmetry. 
The model consists of two layers of checkerboard lattice with lattice constant $1$; each layer of the checkerboard system has two atoms (A and B) in one unit cell.
For one individual layer, the model is parameterized by two hoppings $t$ and $t'$ (\figref{fig:CherkerBoard}(a)).  Its low-energy physics at half filling happens around $M$ point with an effective Hamiltonian $h_0(\bsl{p})=t p_x p_y\sigma_x + t' (p_x^2 + p_y^2) \sigma_y $ in the basis of $(\widetilde{c}^\dagger_{\bsl{p},+},\widetilde{c}^\dagger_{\bsl{p},-})=(c^\dagger_{\bsl{p},A}-\ii c^\dagger_{\bsl{p},B},-\ii c^\dagger_{\bsl{p},A}+ c^\dagger_{\bsl{p},B})/\sqrt{2}$.
The effective Hamiltonian carries a chiral winding number 2 with chiral operator $\sigma_z$, and the quadratic touching is pinned by $C_{4z}$ symmetry of the system---the chiral winding number 2 only guarantees two linear touching points without pinning their positions (see \cref{app:twisted_chekerboard}.)
The moir\'e model is built from low-energy models around $M$ from both layers, which reads
\eq{
\label{eq_main:twisted_checkerboard}
h_{\bsl{r}} = \mat { h_0(-\ii\nabla_{\bsl{r}})  & W(\bsl{r}) \\
W^\dagger(\bsl{r}) & h_0(-\ii\nabla_{\bsl{r}})} \otimes s_0  \ ,
}
where the basis is $\mat{ \psi^\dagger_{\bsl{r},t} , \psi^\dagger_{\bsl{r},b} }$ with
\eq{
\psi^\dagger_{\bsl{r},l} = (\psi^\dagger_{\bsl{r},l,+,\uparrow}, \psi^\dagger_{\bsl{r},l,+,\downarrow}, \psi^\dagger_{\bsl{r},l,-,\uparrow}, \psi^\dagger_{\bsl{r},l,-,\downarrow})\ ,
}
$l=t,b$ labels the layer, 
\eq{
 W(\bsl{r})=  2 \sum_{i=1}^2 T_i \cos(\bsl{r}\cdot\bsl{q}_i) + 2 \sum_{i=1}^4 T_i'\cos(\bsl{r}\cdot\bsl{g}_i) \ ,
 }
$\bsl{q}_n = C_{4z}^{n-1} \frac{k_\theta}{\sqrt{2}} (1,1)^T$ with $n=1,2,3,4$ (see $\bsl{q}_1$ in \cref{fig:CherkerBoard}(c)), $\bsl{g}_i=\bsl{b}_{M,i}+\bsl{q}_1$ with $i=1,2$, $\bsl{g}_i=C_{4z} \bsl{g}_{i-2}$ with $i=3,4$,  $k_\theta = 2\sqrt{2} \pi \sin(\frac{\theta}{2})$, $\bsl{b}_{M,1}=\bsl{q}_1+\bsl{q}_4$, $\bsl{b}_{M,2}=\bsl{q}_1+\bsl{q}_2$.
Just like the original BM model~\cite{Bistritzer2011BMModel}, the intralayer potential in \eqnref{eq_main:twisted_checkerboard} is neglected for simplicity.
In practice, we choose $T$ and $T'$ to have the following forms which preserve chiral symmetry: $T_1 = \sigma_z T_2 \sigma_z  =  w_1 \sigma_x$ and $T_1'=T_2'= w_1' \sigma_x$ with $T_3'=T_4'= \sigma_z T_1' \sigma_z$.
Besides chiral symmetry, the moir\'e model has spin $\SU(2)$, spinless $C_{4z}$, spinless $C_{2x}$, $\TR$, moir\'e-lattice translations, and an effective $m_z$ symmetry that exchange two layers. (See \cref{app:twisted_chekerboard} for detailed expressions.)
We do not have valleys here since the moir\'e model is constructed from the monolayer M points which are time-reversal symmetric.
If we set $w_1=w_1'=0$, we have two quadratic touching points (from two layers) at the zero energies at $\Gamma_M$ and $M_M$ in the moir\'e Brillouin zone, and each of it has chiral winding number 2, adding uo to a total chiral winding number of $\sum_{l}W_l = 4$.
The nonzero total chiral winding number indicates a stable anomaly, which holds for nonzero $w_1$ and $w_1'$ due to the presence of chiral symmetry.

The stable topology of the nearly flat bands with concentrated curvature present in this model are particularly suited for the construction of topological heavy fermion model \cite{KaiSun2023THF}.
To show this, we choose $t = 4/k_\theta^2$, $t' = 1.26/k_\theta^2$, $w_1 = 0.66$ and $w_1' = -0.4$, resulting in nearly flat bands with nontrivial topology and concentrated chiral Berry curvature $F_{+}(\bsl{k})-F_{-}(\bsl{k})$ as shown in \figref{fig:CherkerBoard}(b-c).
(The topology can also be indicated by the symmetry eigenvalues as discussed in \cref{app:twisted_chekerboard}.)
The concentrated chiral Berry curvature allows us to construct the exponentially-localized Wannier modes  $f_{\bsl{k},\alpha}$, $\alpha=1,2$ by combining states of nearly flat bands near zero energy and the states near $X_M$ and $Y_M$ from the remote bands (see \figref{fig:CherkerBoard}(b)).
The resulting $f$ modes have the symmetry representations of $p_x-p_y$ orbitals at $1a$ position in the moir\'e unit cell.
There are 6 low-energy states at $X_M$ (or $Y_M$) per spin, and the construction of the $f$ modes uses 2 of them; the remaining 4 states at $X_M$ (or $Y_M$) per spin give rise to the itinerant $c$ modes. The resulting  heavy fermion model consists of three parts: the $f$-mode part with zero Hamiltonian, the $c$-mode part, and the $f-c$ coupling.
The $c$-mode part reads
\eq{
H_c = \sum_{\bsl{p}}^{\Lambda} c^\dagger_{X_M+\bsl{p}} h_{cc}^{X_M}(\bsl{p})\otimes s_0 c_{X_M+\bsl{p}} + \text{$C_{4z}$ partner}\ ,
}
where $c^\dagger_{X_M+\bsl{p}}$ $=(c^\dagger_{X_M+\bsl{p},1}$, $c^\dagger_{X_M+\bsl{p},2}$, $c^\dagger_{X_M+\bsl{p},3}$, $c^\dagger_{X_M+\bsl{p},4})$ $\otimes (\uparrow,\downarrow)$, 
\eq{
h_{cc}^{X_M}(\bsl{p}) =
\mat{ 
m \sigma_x & \ii v_2 \left[ (p_x + p_y) \sigma_x + (p_x - p_y) \sigma_y \right]\\
h.c. & 0_{2\times 2} \ .
}
}
to the linear order of $\bsl{p}$, and the $C_{4z}$ partner term stands for the $Y_M$ Hamiltonian:
\eq{
h_{cc}^{Y_M}(\bsl{p}) = 
h_{cc}^{X_M}(C_{4z}^{-1}\bsl{p})\ .
}
For the spinless part of the basis $c^\dagger_{X_M+\bsl{p},\beta}$ with $\beta=1,...,4$, $\beta=1,2$ have opposite spinless-$C_{2z}$ eigenvalues than $\beta=3,4$, explaining the linear-in-$\bsl{p}$ terms that couple them.
Combined with all other symmetries that leave $X_M$ unchanged, $\beta=1,2$ furnish a reducible representation of them while $\beta=3,4$ furnish an irreducible representation, explaining the appearance of the mass term for $\beta=1,2$ but not for $\beta=3,4$ (see \cref{app:twisted_chekerboard}).
The $f-c$ coupling reads
\eqa{
H_{fc} & = \sum_{\bsl{p}}^{\Lambda} f^\dagger_{\bsl{p}} h_{fc}^{X_M}(\bsl{p}) c_{X_M+\bsl{p}} e^{-\frac{|\bsl{p}|^2 \lambda^2 }{2}}  + (C_{4z}\text{ partner}) \\
& \quad+ h.c.\ ,
}
where $\lambda=0.43 a_M$ is square root of the Wannier spread of the $f$ modes with $a_M$ the moir\'e lattice constant, and 
\eq{
h_{fc}^{X_M}(\bsl{p}) = \mat{ \ii v_1 (p_x \sigma_x + p_y \sigma_y) & \gamma (\sigma_x + \sigma_y)}\ .
}
From the tight-binding parameters and the Wannier states, we compute $v_1 = -0.52/k_\theta$, $\gamma=0.45$, $m = 0.51$ and $v_2 =1.53/k_\theta$. The heavy fermion bands in \figref{fig:CherkerBoard}(d) match the continuum model remarkably well.
At the interacting level, we project the Coulomb interaction to obtain an Anderson-``+"  model with nine interaction terms. The larges term is the density-density Hubbard $U$ for $f$ modes which can be treated with Hartree Fock and DMFT methods. Other important terms, which can all be well-approximated in mean field, are the density-density interaction $V/W$ among $c$/ $f-c$ modes, and the flavor-dependent interactions $J$ and $K$ between $f$ and $c$. They are listed in \appref{app:twisted_chekerboard}.

We specify that \refcite{KaiSun2023THF} studied the heavy fermion framework for a quadratic band touching model with a perfectly flat band protected by chiral symmetry. We have here complemented their analysis by uncovering the stable chiral-symmetric anomaly and obtaining the interaction Hamiltonian.

\section{Gapless Flat Band Models}

\begin{figure}
  \includegraphics[width=\columnwidth]{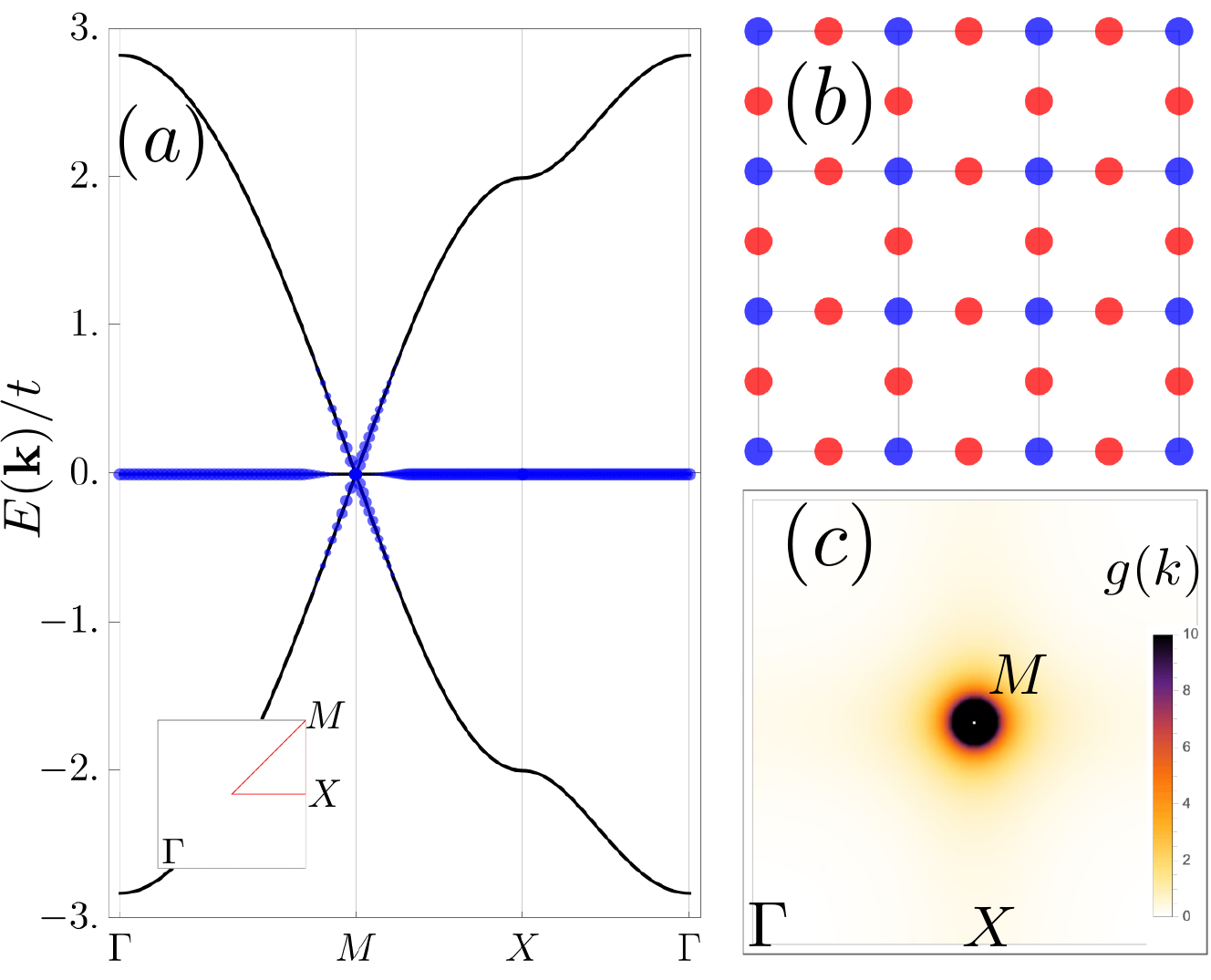}
  \caption{Divergent quantum geometry of the Lieb lattice. The band structure $(a)$ exhibits an exactly flat band protected by the bipartite lattice structure shown in $(b)$ where the $L/L'$ sublattice is red/blue respectively. A singular band crossing at the $M = (\pi,\pi)$ point results in divergent Fubini-Study metric $(c)$ in the flat band, going as $g(M+\mbf{k}) \to 1/|\mbf{k}|^2$. Blue shading in $(a)$ shows the weight of the heavy fermion wavefunction on the dispersive bands.}
  \label{fig:liebmain}
\end{figure}

 We now turn to a family of lattice models known as bipartite crystalline flat bands. 
These models support both gapped topological flat bands as well as flat bands which are degenerate with dispersive bands at isolated momenta, which will be our focus here. 
To introduce this family of models, we study the Lieb lattice (see \figref{fig:liebmain}a).  In momentum space, the Hamiltonian can be written in bipartite form
\bea
\label{eq:Liebh}
h(\mbf{k}) &= 2t \bpm
0 & S^\dag(\mbf{k}) \\
S(\mbf{k}) & 0 
\epm, \quad S(\mbf{k}) = \bpm \cos k_y/2 \\ \cos k_x/2 \epm \ .
\eea
There exist  $N_L - N_{L'}$ flat bands in a general bipartite models if $S(\mbf{k})$ is a rectangular matrix of dimensions $N_L \times N_{L'}$, where $L$ and $L'$ are the two sublattices of the crystal with $N_L$ and $N_{L'}$ orbitals per unit cell respectively. The flat band wavefunction  $U_0(\mbf{k}) = (\psi(\mbf{k}) , 0 )^T$ is supported only on the $L$ sublattice and has exactly zero energy since there always exists a vector $\psi(\mbf{k})$ such that $S^\dag(\mbf{k}) \psi(\mbf{k}) = 0$. In the Lieb lattice,
\bea
\label{eq:lieneigenstate}
\psi(\mbf{k}) = (-\cos \frac{k_x}{2},  \cos \frac{k_y}{2})^T/\sqrt{\cos^2 \frac{k_x}{2} + \cos^2 \frac{k_y}{2}} \ .
\eea
This construction can be generalized to encompass realistic materials \cite{Dumitru2022GeneralConstructionFlatBand}, identifying minimal models for flat and nearly bands away from zero energy. 

At generic $\mbf{k}$, the dimension of the null space of $S^\dag(\mbf{k})$ is  $N_L - N_{L'} = 2 -1 = 1$, but at high-symmetry points the nullity may be larger, indicating band touching points. Plainly, \EqJHA{eq:Liebh} shows a band touching occurs at $\mbf{k} = (\pi,\pi)$, the $M$ point at which $S^\dag(\mbf{k})=0$, as seen in the band structure in \figref{fig:liebmain}b. Its existence fundamentally obstructs a local, low-energy model of the flat band alone, even though most of the flat band is clearly separated from the dispersive bands. From \EqJHA{eq:lieneigenstate}, we see that
\bea
\psi(\mbf{k}) \to (\cos \phi, - \sin \phi)
\eea
has a singular limit at the gap closing point, where $\mbf{k} - M = |k|(\cos \phi,\sin  \phi)$ and $|k| \to 0$. While the Berry curvature in this model is strictly zero due to $C_2\mathcal{T}$ symmetry rendering all eigenstates real, the quantum geometry of the model, measured by the Fubini-Study metric, is small away from $M$ and is peaked (indeed, diverges) at the gap closing, as shown in \figref{fig:liebmain}c. This suggests that a topological heavy fermion model can overcome this obstruction. 

The band structure in \figref{fig:liebmain}a shows a flat band with pierced by a dispersive bands. What is the low-energy theory at partial filling of the flat bands? Naively, one may expect a theory of Dirac-like electrons and coupled to the a localized $s$ orbital in the atomic basis. Instead, we will show analytically that the low-energy degrees of freedom are a quadratic semi-metal with $2\pi$ Berry phase (unlike the linear, $\pi$ Berry phase Dirac) and an obstructed atomic $d$-orbital (unlike the atomic basis). The (unitary) transformation of the microscopic degrees of freedom into this basis is accomplished through the heavy fermion framework. 

Symmetries play a key role in identifying the low-energy degrees of freedom of the heavy fermion model. We will need the rotation symmetry $D[C_{4z}] = 1 \oplus \sigma_1$ and time-reversal $D[\mathcal{T}] = K$, obeying $D[g] h(\mbf{k}) D^\dag[g] = h(g \mbf{k})$, as well as the anti-commuting chiral symmetry $D[\Sigma] = 1 \oplus - \sigma_0$ obeying $D[\Sigma] h(\mbf{k}) D^\dag[\Sigma] = -h(\mbf{k})$. The combination of $C_2$ and $\mathcal{T}$ ensures the Berry curvature is identically zero everywhere in the Brillouin zone when the bands are gapped. The irrep tables in this space group, $p41'$, are \cite{Aroyo2006BilbaoIR}
\bea
\begin{array}{c|ccc}
41' &1& C_4 \\
\hline
\Gamma_1,M_1 &1&1\\
\Gamma_2,M_2 &1 & -1 \\
\Gamma_3\Gamma_4,M_3M_4 &2 & 0 \\
\end{array}, \quad \ \begin{array}{c|cc}
21' &1& C_{2} \\
\hline
X_1 &1&1\\
X_2 &1 & -1 \\
\end{array} \ .
\eea

The band touching at the $M$ point is protected by these symmetries. Ref. \cite{Dumitru2022GeneralConstructionFlatBand} has shown that the irreps of the flat band are given by the formal subtraction of the representations of the two sublattices
\bea
\mathcal{B}_{0} &= \mathcal{B}_L \ominus \mathcal{B}_{L'} = \Gamma_2 + X_1 + (M_3 M_4 \ominus M_2)
\eea
where $\mathcal{B}_L = \Gamma_1 \oplus \Gamma_2 + X_1 \oplus X_2 + M_3 M_4$ and $\mathcal{B}_{L'} = \Gamma_1 + X_2 + M_2$ can be found on the Bilbao crystallographic server\cite{Aroyo2006BilbaoIR}. The incompatibility of the $M$ irrep subtraction is interpreted as a gapless point. In fact, these representations identify a unique atomic (localized) representation available at low energy. At the $\Gamma, X$ points, the dispersive bands are far away in energy from the flat band so the only accessible irreps are $\Gamma_2, X_1$. There is only a single one-band atomic representation consistent with these irreps:
\bea
\Gamma_2 + X_1 + M_2 = B_{1a} \uparrow G
\eea
where $B_{1a}$ is a $d_{x^2-y^2}$-orbital at the origin --- importantly off the sites of the $L$ and $L'$ sublattices. This is the representation of the low-energy heavy fermion. The $M_2$ irrep is induced from the $L'$ sublattice, meaning the wavefunction of the heavy fermion obeys $U_f(\mbf{k}=M) = (1,0,0)$. This requires mixing into the dispersive band Hilbert space since the flat band wavefunction (carrying the $\Gamma_2, X_1$ irreps) is entirely supported on the $L$ sublattice. Thus the heavy fermion wavefunction takes the simple form
\bea \label{liebansatz1}
U_{f}(\mbf{k}) &= U_0(\mbf{k}) \cos \th(\mbf{k}) + (1,0,0)^T \sin \th(\mbf{k})
\eea
where the band mixing is parameterized by $\th(\mbf{k})$. $\th(M)=\pi/2$  and decays to zero away from $M$. We can choose $\th(\mbf{k})$ such that the heavy fermion is maximally localized\cite{marzari2011} in position space in order to obtain an Anderson-``+" model where Hubbard-like interactions are predominantly carried by the heavy fermion, and correlations are local. Usually one first uses numerical routines such as Wannier90 to accomplish this task, or relies on gaussian approximations for the Wannier states \cite{Song20211110MATBGHF}. Remarkably, we now show it is possible to solve for $\th(\mbf{k})$ analytically. This is likely to generalize to other bipartite models with band touchings, as the $L$ and $L'$ support of the flat band and touching point as well as 
ansatz in Eq. \ref{liebansatz1} generalize. 

\begin{figure}
  \includegraphics[width=\columnwidth]{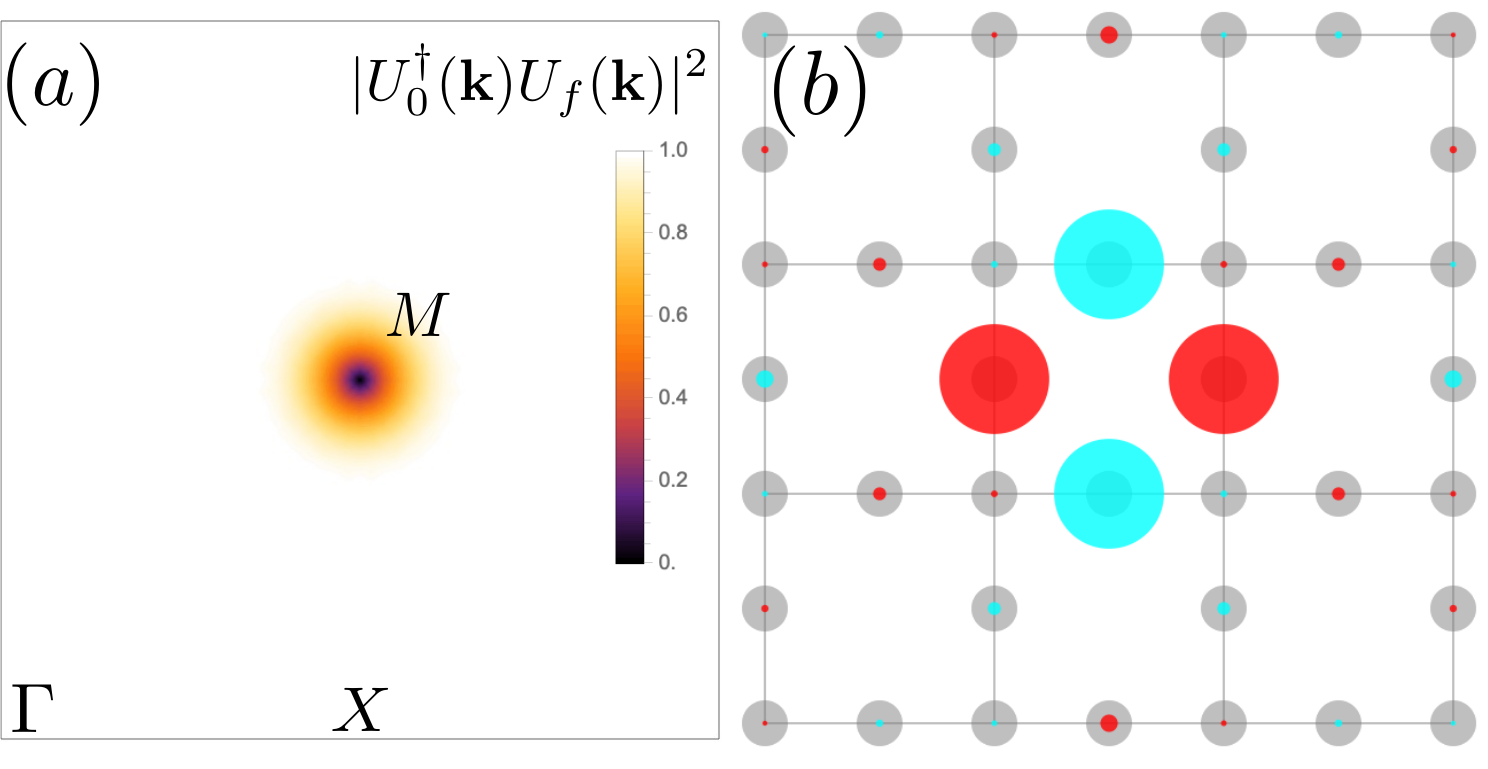}
  \includegraphics[width=\columnwidth]{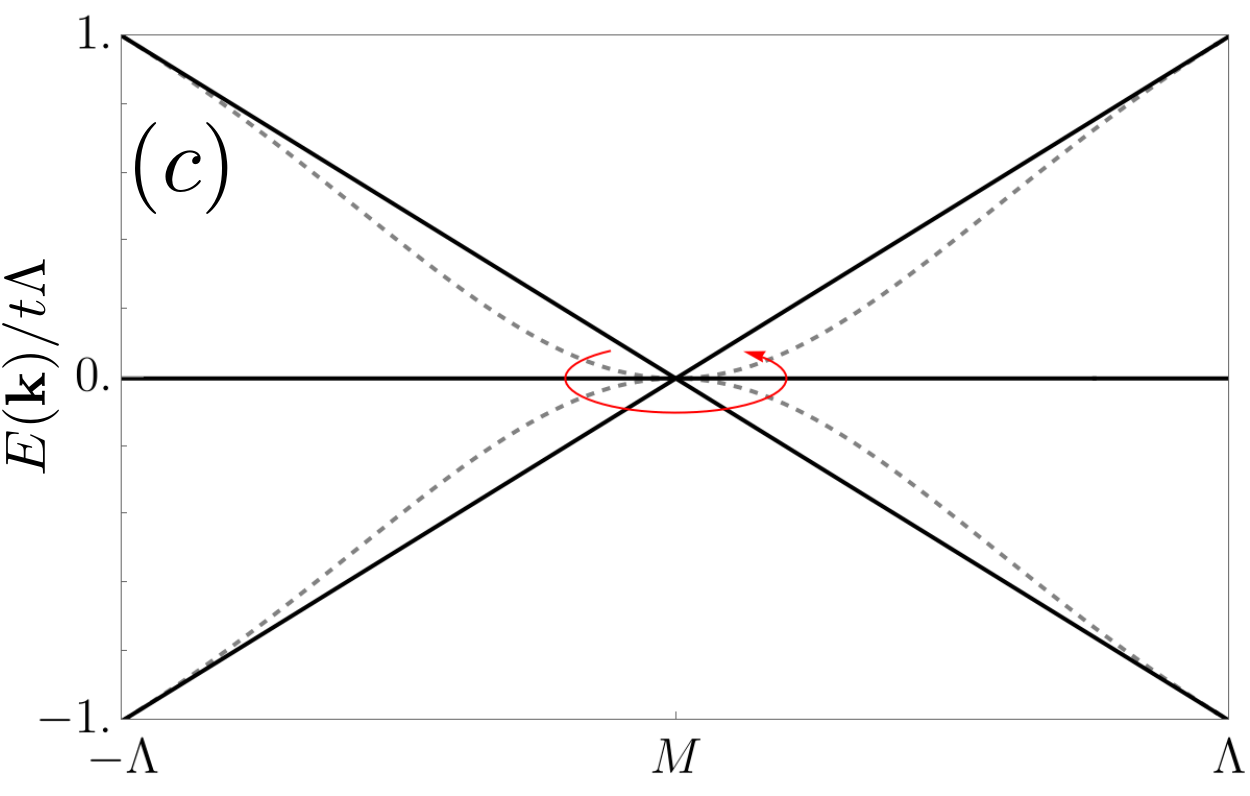}
  \caption{Lieb lattice Heavy Fermion. (a) $c$-electron instanton-tunneling probability in the flat band. (b) Real-space Wannier function, whose $d$-orbital character is apparent from the sign structure (blue/red corresponds to $\pm$ amplitude). The total probability on the $L'$ sublattice is set by the cutoff $\Lambda$. $(c)$ We show the decoupled HF and conduction electron bands (dashed) within $\Lambda$ around the $M$ point. The conduction electrons carry $2\pi$ Berry phase and are a topological semi-metal. Linear $f-c$ coupling restores the Lieb dispersion (black).}
  \label{fig:liebHF}
\end{figure}

We form the Wannier state
\bea
f^\dag_\mbf{R} &= \sum_\al \int \frac{d^2k}{(2\pi)^2} e^{i \mbf{k} \cdot \mbf{R}} c^\dag_{\mbf{k},\al} U_{f,\al}(\mbf{k})
\eea
and compute its spread to be (see \appref{app:liebSP})
\bea
\label{eq:wannierspread}
\frac{1}{2}\braket{r^2} 
&= \int \frac{d^2k}{(2\pi)^2} \lp \frac{1}{2}(\nabla \theta)^2 + \frac{1}{2} g(\mbf{k}) \cos^2 \th \rp
\eea
using the fact that a smooth $C_2\mathcal{T}$-symmetric band can be chosen real with vanishing Berry connection. Here 
\bea
g(\mbf{k}) &= \frac{1}{2} \Tr \, (\nabla P)^2 = \frac{1-\cos k_x \cos k_y}{2 (2 + \cos k_x +\cos k_y)^2}
\eea
is the Fubini-Study (quantum) metric of the flat band with projector $P = U_0U_0^\dag$,  which diverges as $1/|k|^2$ for $\mbf{k} = M + |k|(\cos \phi, \sin \phi)$. Only through mixing into the dispersive bands parameterized by $\sin^2 \th$, which reaches 1 at $M$, can the Wannier spread of the heavy fermion be made finite. We now impose a cutoff $\Lambda \ll \pi $ around the $M$ point, beyond which $\th(\mbf{k}) = 0$ and the heavy fermion wavefunction is supported strictly in the flat bands. This cutoff is analogous to the ``energy window" employed in disentanglement algorithms, and physically corresponds to allowing band mixing only within the a range of energy given by the interaction. The cutoff $\Lambda \sim U/t$ restricts the heavy fermion to be supported on low-energy states only. Minimizing \EqJHA{eq:wannierspread} now corresponds to an Euler-Lagrange problem with Dirichlet boundary conditions given by the action
\bea
S &= \int_\Lambda \frac{d^2k}{(2\pi)^2} \lp \frac{1}{2}(\nabla \theta)^2 + \frac{1}{2k^2} \cos^2 \th \rp \\
&= \int_\Lambda \frac{dt}{(2\pi)^2}  \lp \frac{1}{2}\dot \th^2 + \frac{1}{2} \cos^2 \th \rp
\eea
where in the last line we have taken $\th(\mbf{k}) = \th(|\mbf{k}|)$ since higher angular momentum dependence strictly increases the spread, and written $k = e^{t}$ to bring the action into the famous instanton form. Writing $\dot \th^2 + \cos^2 \th = (\dot \th \pm \cos \th)^2 \mp 2 \frac{d}{dt} \sin \th$ allows us to saturate the Bogomol`nyi–Prasad–Sommerfield (BPS) bound \cite{Bogomolny:1975de}, and find the absolute minimum with solution (see \appref{app:liebSP})
\bea
\cos \th(\mbf{k}) = \frac{2|\mbf{k}|/\Lambda}{(|\mbf{k}|/\Lambda)^2 + 1}
\eea
for $|\mbf{k}|\leq \Lambda$ and $\th(\mbf{k}) = 0$ otherwise. Note that $\th(\mbf{k})$ and its first derivative are continuous at $\Lambda$, as can be seen in \figref{fig:liebHF}a. The real space Wannier wavefunction and its $d$-orbital character are shown in \figref{fig:liebHF}b. The theory of instantons is well developed and solutions are known for higher dimensional, non-abelian actions (see Ref. \cite{2005hep.th....9216T} and references therein), which may find other analogues in heavy-fermion problems. 

The heavy fermion wavefunction carries the $M_2$ irrep, but the $M_3M_4$ irrep must be carried by another particle, the conduction electrons. Their wavefunctions are easily obtained by looking at the two-dimensional orthogonal complement of $U_f(\mbf{k})$, and fixing the gauge freedom with time-reversal symmetry (see \appref{app:liebSP}). To leading order, we find
\bea
\gamma^\dag_{M+\mbf{k},c_\pm} &= \sum_\al c^\dag_{M+\mbf{k},\al} \big(\mp i \sqrt{2} \, \frac{k_x\pm i k_y}{\Lambda}, \mp \frac{i}{\sqrt{2}}, - \frac{1}{\sqrt{2}} \big)_\al \\
\eea
which are explicitly smooth, and have highest probability on the $L$ sublattice (recall at the heavy fermion wavefunction is supported primarily on the $L'$ sublattice within the cutoff). The Hamiltonian in the heavy fermion basis can now be written in a $k.p$ expansion as ($k,\bar{k} = k_x \pm i k_y$)
\bea
\label{eq:HliegSP}
h_{HF}(\mbf{k}) = t \bpm
 & - 2i \bar{k}^2 / \Lambda & \bar{k} /\sqrt{2} \\
2i k^2 / \Lambda & & k /\sqrt{2} \\
k /\sqrt{2} & \bar{k} /\sqrt{2} & 0 \\
\epm
\eea
in the basis $c_{\mbf{k},+}, c_{\mbf{k},-}, f_{\mbf{k}}$. The heavy fermion description reveals the emergent degrees of freedom at low energies to be obstructed $d$-orbitals coupled to a double vortex topological semi-metal (see \figref{fig:liebHF}c). 

We now consider the addition of Hubbard interactions 
\bea
H_{int} &= U \sum_{\mbf{R} \al} n_{\mbf{R} \al,\uparrow}n_{\mbf{R} \al,\downarrow}
\eea
acting on the spin-degenerate bands. Since the HF basis is unitarily related to the band basis, it is simple to rotate between them and compute the periodic Anderson model, which we show in \appref{app:liebSP}. Let us recall that for a highly localized heavy fermion like in TBG, the largest interaction is an onsite Hubbard term, with nearest- and farther-neighbor interactions controlled by the Wannier spread. A surprising feature that distinguishes the Lieb lattice from its continuum heavy fermion counterparts is that there is an obstruction to taking $\lambda \to 0$.  This obstruction is easily explained through topological quantum chemistry\cite{Bradlyn2017TQC}: the representation of the Lieb lattice $f$-mode is a $d$-orbital at a lattice vacancy, so its wavefunction must always be supported on the neighboring orbitals (see \cref{fig:liebHF}) which causes strong overlaps between the densities of neighboring $f$-modes. Hence the $f$-mode density-density interaction always has strong nearest-neighbor components, while farther-neighbor interactions will be suppressed due to the localized Wannier wavefunction. Approximating the Wannier state by keeping only its largest 4 amplitudes (the central 4 $d$-wave like lobes), we find the $f$-$f$ interaction takes the form 
\bea
\label{eq:Huhub}
H_U &= U \sum_{\mbf{R} \al} \bar{n}^f_{\mbf{R},\al,\uparrow}\bar{n}^f_{\mbf{R},\al,\downarrow} \approx \frac{U}{4} \sum_{\braket{\mbf{R} \mbf{R}'}} n_{\braket{\mbf{R} \mbf{R}'},\uparrow}n_{\braket{\mbf{R} \mbf{R}'},\downarrow}
\eea
where $\bar{n}^f_{\mbf{R},\al,\sigma}$ is the projection of $n_{\mbf{R},\al,\sigma}$ to the $f$-modes, $f^\dag_{\braket{\mbf{R} \mbf{R}'},\sigma} = \frac{f^\dag_{\mbf{R},\sigma}+f^\dag_{\mbf{R}',\sigma}}{\sqrt{2}}$ is a bond-centered heavy fermion, $n_{\braket{\mbf{R} \mbf{R}'},\sigma} = f^\dag_{\braket{\mbf{R} \mbf{R}'},\sigma}f_{\braket{\mbf{R} \mbf{R}'},\sigma}$ is its density, and $\braket{\mbf{R}\mbf{R}'}$ denotes all $f$-mode nearest neighbors on the square lattice. The physical meaning of the symmetry obstruction to an onsite Hubbard $f$-mode term is the following: while the tight-binding Hubbard interaction is onsite, it is smeared out in the low-energy degrees of freedom due to the quantum geometry of the heavy fermion. This shows that the effective model at low energies is dramatically altered from the original tight-binding description. 

We remark that $f$-mode sector $\bar{H}^f = H_U$ is an analytically tractable strong coupling problem. This is because projected Hubbard Hamiltonians with flat bands have exactly solvable groundstates \cite{PhysRevB.94.245149} and low-energy excitations \cite{2022arXiv220900007H}, where the minimal Fubini-Study metric \cite{PhysRevB.106.014518} is known to generate the mass of the collective bosonic modes (the spin wave for repulsive interactions, and the Cooper pair for attractive interactions) \cite{PhysRevB.98.220511,2023PhRvB.107v4505I,2023arXiv230805686H,PhysRevLett.132.026002,2024PhRvL.132b6002C}. Projecting the Lieb lattice onto the $f$-mode basis yields such a projected Hubbard Hamiltonian since the $f$-modes have zero kinetic energy in \EqJHA{eq:HliegSP}. However, the hybridization with the conduction electrons cannot be neglected, and the full interacting Hamiltonian 
\bea
\label{eq:Hliebanderson}
H_{int} &= H_U + H_{W} + H_{J} + \dots
\eea
contains additional interactions between the $f$ and $c$ modes, including the $f$-density/$c$-density interaction $H_{W} \sim f^\dag f c^\dag c$ the exchange interaction $H_{J} \sim f^\dag c c^\dag f$, which are usually decoupled in mean-field. (The dots in Eq. \ref{eq:Hliebanderson} refer to odd-parity $f$-terms and the conduction electron Coulomb repulsion which are typically neglected see \appref{app:liebSP}). Together, these terms form an Anderson-``+" model in the low-energy Hilbert space, which can be approached from various directions. The importance of quantum geometry in the Lieb lattice has been studied in seminal work on flat band superconductivity \cite{PhysRevLett.117.045303}, and reveals strongly correlated behavior consistent with the phenomenology of heavy fermions \cite{PhysRevB.103.L220502}. Although our focus is on the partially filled flat band for $U < t$, the heavy fermion representation is simply a unitary rewriting of the model, and can potentially shed light on other densities and parameter regimes of the model \cite{2023PhRvL.130r6404H}. 

\section{General Argument For Heavy Fermion Representations of Flat Bands With Concentrated Quantum Geometry}

We have shown that flat Bands with concentrated non-abelian Berry curvature in moir\'e systems, or concentrated Fubini-Study metric in the Lieb lattice,  are accompanied by dispersive bands whose gap is smallest where the curvature or metric is largest. These systems admit a heavy fermion representation and serve to illustrate the broader proposal that narrow band systems with strongly peaked quantum geometry are candidates for heavy fermion systems. 

Let us recall the two obstacles to an exponentially localized Wannier basis with physical symmetry representations in a given set of bands. The first obstacle arises from symmetry considerations: if the momentum space irreps of the band cannot be induced from atomic limits \cite{Bradlyn2017TQC}, then no such Wannier basis with physical symmetry representations exists. The second obstacle is quantum geometric: the Wannier spread is lower bounded by the integrated Fubini-Study metric \cite{marzari2011}, becoming infinite if the bands carry a nonzero Chern number \cite{2018CMaPh.359...61M} (see Ref. \cite{2022arXiv221011573Z,PhysRevB.104.075143}) or are gapless, as in e.g. the Lieb lattice. These obstacles are intrinsic properties of the Hilbert space of the bands. They cannot be removed without mixing into nearby bands, necessitating a heavy fermion description.  

Symmetries are crucial for restricting the possible $f$-modes. In all the cases studied so far, the symmetries are powerful enough to uniquely identify candidate $f$-modes: the $p_x-p_y$ orbitals in TBG and twisted checkerboard, and the $d_{x^2-y^2}$ orbital in the Lieb lattice. It is worth noting that a class of line graph lattices \cite{mielke1991ferromagnetic,Dumitru2022GeneralConstructionFlatBand,yin2022topological,2023arXiv231109290J} such as the kagome are ruled out as heavy fermion candidates by symmetry. 

The existence of irrep-compatible $f$-modes is a necessary but not sufficient condition, since the quantum geometric obstruction must also be overcome. We will now present a criterion to estimate the $f$-mode spread and determine whether a heavy fermion description is appropriate. 

\begin{figure}
    \centering
    \includegraphics[width=\columnwidth]{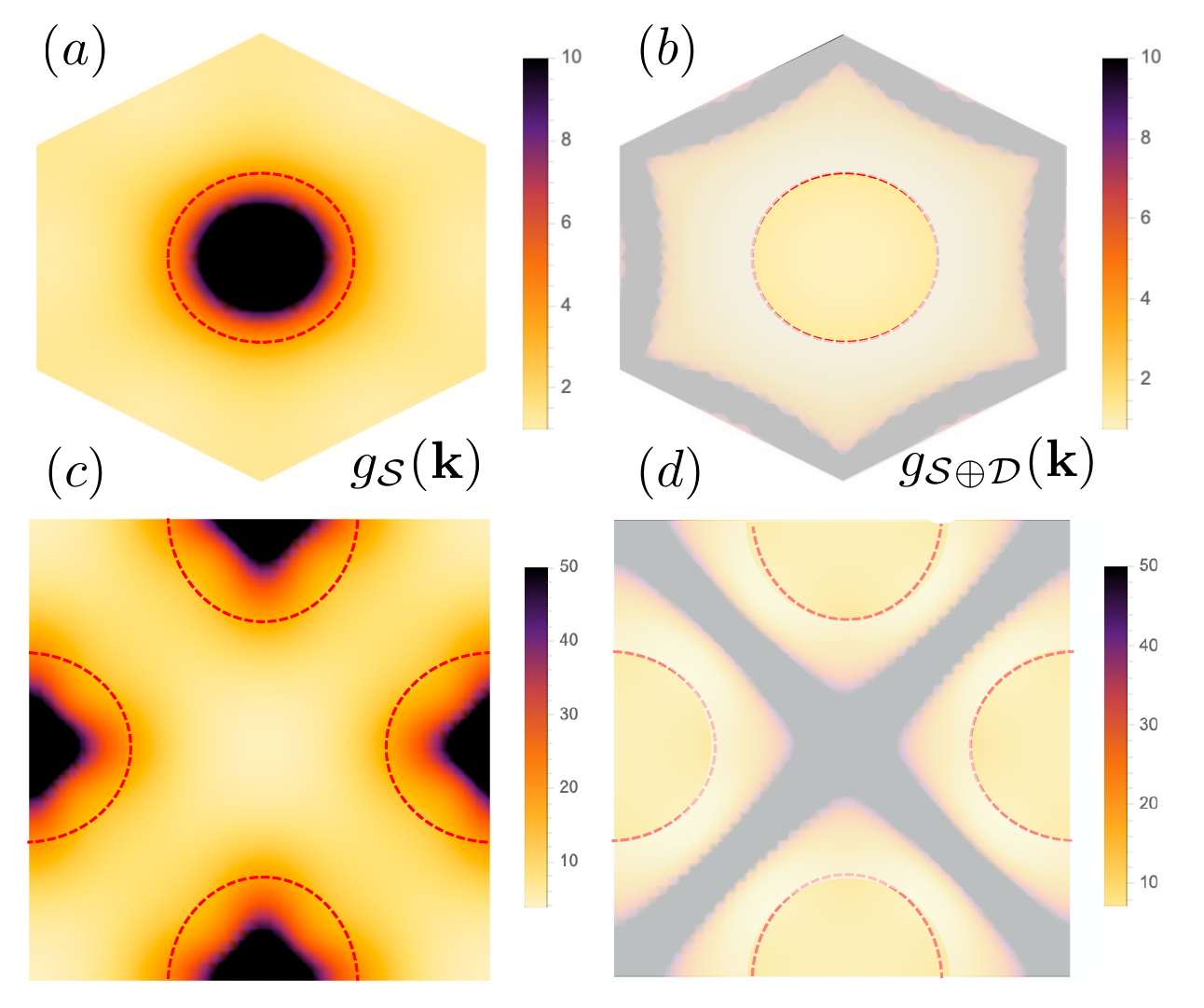}
    \caption{Resolution of the quantum geometric obstruction to localized Wannier functions. The 2-band Fubini-Study metric for TBG $(a)$ and twisted checkerboard $(c)$ are peaked in the regions (dashed red) where the nearest 4 dispersive bands are closest. The 6-band Fubini-Study metric for TBG $(b)$ and twisted checkerboard $(d)$ shows no peak in these regions, demonstrating that mixing into the dispersive bands (within the low-energy, unshaded regions) can yield highly localized Wannier functions. 
    }
    \label{fig:wannierres}
\end{figure}

Given a projector $P_\mathcal{S}(\mbf{k}) = \sum_{n \in \mathcal{S}} U_n(\mbf{k}) U_n^\dag(\mbf{k})$ onto a set of bands indexed by $\mathcal{S}$, their abelian Fubini-Study metric is \cite{Resta2011QuantumGeometry}
\bea
g_\mathcal{S}(\mbf{k}) &= \frac{1}{2}\Tr \, \del_i P_\mathcal{S}(\mbf{k}) \del_i P_\mathcal{S}(\mbf{k}) \geq 0
\eea
where the momentum index $i$ is implicitly summed over. Since $P_\mathcal{S}(\mbf{k})$ is gauge-invariant, $g_\mathcal{S}(\mbf{k})$ can be computed without choosing a gauge. Picking $\mathcal{S}$ to be the set of narrow bands, a large value of $g_\mathcal{S}(\mbf{k})$ at its peak prevents a highly localized $f$-mode by lower-bounding the spread of any possible Wannier functions formed in $\mathcal{S}$. However, by including a set of low-energy dispersive bands $\mathcal{D}$, it may be possible to find a more localized Wannier function if $g_{\mathcal{S}\oplus \mathcal{D}}(\mbf{k}) < g_\mathcal{S}(\mbf{k})$, noting that $P_{\mathcal{S}\oplus\mathcal{D}} = P_{\mathcal{S}} +P_{\mathcal{D}}$, but $g_{\mathcal{S}\oplus\mathcal{D}} \neq  g_{\mathcal{S}} + g_{\mathcal{D}}$. We propose a simple estimate for the total Wannier spread $\la^2_f$ of the $f$-mode
\bea
\label{eq:estimate}
\la^2_f \approx \Omega \int  \frac{d^2k}{(2\pi)^2} \min \{ \, g_{\mathcal{S}}(\mbf{k}) ,  g_{\mathcal{S}\oplus \mathcal{D}}(\mbf{k}) \} 
\eea
where $\Omega$ is the unit cell area. $\la^2_f$ reflects the possibility of reducing the Wannier obstruction by mixing into the dispersive bands if the enlarged Hilbert space $\mathcal{S}\oplus \mathcal{D}$ reduces the Fubini-Study metric. Fig. \ref{fig:wannierres} shows $g_{\mathcal{S}}, g_{\mathcal{S}\oplus\mathcal{D}}$ for TBG and the twisted checkerboard model, demonstrating how the inclusion of the nearby dispersive bands can remove the strong peak in $g_{\mathcal{S}}(\mbf{k})$. For TBG, this estimate gives $0.29 a_M$ compared to the Wannier90 value $0.3 a_M$, and for twisted checkerboard, this estimate gives $0.45 a_M$ compared to $0.42 a_M$. The good numerical agreement between these estimates and the Wannier90 results is evidence that the localization properties of different Hilbert spaces, captured by \EqJHA{eq:estimate}, control the existence of a heavy fermion (assuming there is no symmetry obstruction). Further understanding can be gleaned from the Lieb lattice, where $g_{\mathcal{S}}$ diverges at the gap closing point whereas $g_{\mathcal{S}\oplus\mathcal{D}}$ vanishes, since $\mathcal{S}\oplus\mathcal{D}$ is the entire single-particle Hilbert space whose projector is the identity. One can verify from the value of the instanton action at its BPS bound that \EqJHA{eq:estimate} is exact up to $\frac{\Lambda^2}{2\pi} \ll 1$. Although \EqJHA{eq:estimate} is written in terms of the Fubini-Study metric, the inequality $g(\mbf{k}) \geq |f(\mbf{k})|$ \cite{2012arXiv1208.2055R,PhysRevB.106.165133} ensures that small Fubini-Study metric also constrains the Berry curvature, $f(\mbf{k})$. 

Armed with an approximate value for the Wannier spread $\lambda^2_f$, it is also possible to estimate the onsite Hubbard interaction strength via\cite{CAL23}
\bea
U \approx \frac{e^2}{\eps \la_f}
\eea
assuming a Coulomb interaction (with gate-screening length much larger than $\la_f$). Given the gap $\gamma$ to the dispersive bands, we can determine the regime of the resulting Anderson model based on whether
\bea
\frac{\gamma}{U} &= \frac{\gamma}{e^2/\eps a} \frac{\la_f}{a}
\eea
is less than 1 (heavy fermion regime) or greater than 1 (projected regime). Here $a$ is the lattice constant, and $e^2/(\eps a)$ is the Coulomb energy scale. In TBG, using a value of $\eps = 6$ for the dielectric constant gives $\frac{\gamma}{e^2/\eps a} \sim 1$. Due to the strong localization of the $f$-modes, $\la_f / a \sim .3$ ensures $\frac{\gamma}{U}$ is firmly less than 1, putting TBG into the heavy fermion regime where local moment behavior is expected.

\section{Conclusions}

The low-energy description of a system can be very different from its microscopic model. Untangling these relevant degrees of freedom from the microscopic model is often the key to understanding their emergent properties, both intuitively and computationally. The formation of heavy fermions from the moir\'e coupling of two massless Dirac particles in TBG is a prototypical example of this phenomenon in interacting flat bands. In this work, we have studied a variety of other examples, from moir\'e systems to tight-binding lattices, and observed a simple pattern. While band theoretic considerations would forbid a local description of topological bands, we find that --- if their quantum geometry is concentrated --- a heavy fermion representation is possible by introducing metallic degrees of freedom to carry the obstruction. This single-particle description is extended to a periodic Anderson-``+" model incorporating interactions within this basis. The abundance of  heavy fermions thus far revealed in topological, correlated quantum materials underscores the value in studying this class of models, and sets a direction for future research.

\section{Acknowledgements}

The authors are grateful for important conversations and collaborations with Antoine Georges, Xi Dai, Andy Millis, Francisco Guinea, Roser Valent\'i, Giorgio Sangiovanni, Tim Wehling, Qimiao Si, Jian Kang, Dmitri K. Efetov, Keshav Singh, Liam Lau, Daniel Kaplan, Ryan Lee, Gautam Rai, Lorenzo Crippa, and Michael Scheer. This work was supported by Office of Basic Energy Sciences, Material Sciences and Engineering Division, U.S. Department of Energy (DOE) under Contracts No. DE-SC0016239 (B. A. B.), DE-FG02-99ER45790 (P. C.) and  DE-SC0012704 (A. M. T.). J. H.-A. is supported by a Hertz Fellowship, with additional support from DOE Grant No. DE-SC0016239. J. Y. is supported by the Gordon and Betty Moore Foundation through Grant No. GBMF8685 towards the Princeton theory program and through the Gordon and Betty Moore Foundation’s EPiQS Initiative (Grant No. GBMF11070). D.C. acknowledges the hospitality of the Donostia International Physics Center, at which this work was carried out, and acknowledges support by the Simons Investigator Grant No. 404513, the Gordon and Betty Moore Foundation through Grant No. GBMF8685 towards the Princeton theory program, the Gordon and Betty Moore Foundation’s EPiQS Initiative (Grant No. GBMF11070), Office of Naval Research (ONR Grant No. N00014-20-1-2303), Global Collaborative Network Grant at Princeton University, BSF Israel US foundation No. 2018226, NSF-MERSEC (Grant No. MERSEC DMR 2011750). H.H. was supported by the European Research Council (ERC) under the European Union’s Horizon 2020 research and innovation program (Grant Agreement No. 101020833).
\newpage
\appendix
\onecolumngrid
\tableofcontents
\newpage
\section{Twisted bilayer graphene}\label{app:sec:tgh}
Twisted bilayer graphene (TBG) consist of two graphene layers rotated by a small angle $\theta$ relative to each other. Near the magic angle $\theta = \SI{1.05}{\degree}$, flat bands emerge near charge neutrality, which gives rise to various exotic quantum phases, including correlated insulators and superconductivity. 
Extensive experimental efforts, including both transport~\cite{GHO24,MER24,DIE23,TIA23,DI22a,LIA21c,YU23c,HUB22,GRO22,JAO22,PAU22,GHA22,WU21a,DAS21,PAR21c,LU21,SAI21a,CAO21,CHE20b,SAI21,POL19,CAO20,SER20,Stepanov2019SCCorrTBG,SAI20,YAN19,LU19,CAO18} and spectroscopic measurements~\cite{ZHO23a,BEN21,CAL22d,NUC23,HES21,LIS21,CHO21,TSC21,CHO20,WON20,JIA19,KER19,CHO19,NUC20,XIE19}, have been taken to investigate the nature of these quantum phases. Concurrently, significant theoretical efforts have also been made towards constructing the microscopic models~\cite{VAF23,CAL20,FAN19,CAR19a,KAN23b,CAR19,NAM17,KWA20,LED21,RAD19,SHE21,DAV22,CAO21a,HEJ21,WAN21a,REN21,BER21,CAR20a,WIL20,HUA20a,LIU19a,KOS18,KAN18,EFI18,FU20,JAI16,DAI16,WIJ15,UCH14,ZOU18,SUA10,LOP07,Po2019TBGFragile,BIS11,TAR19,davis2023kinetic,PhysRevB.107.045426,PhysRevB.107.235155}, investigating experimental response~\cite{OCH23,CAL24,HON22,WAN23b,MOO13,CAL22d,LIU20b,KRU23,GAR20,PAD20}, understanding the topology~\cite{LIU19a,HEJ19a,Lian2020LLFragileTBG,HEJ19,AHN19,XIE20,ZOU18,SON19,Po2019TBGFragile} and excitations~\cite{KHA20,KWA22,KAN21,KUM21,XIE21,BER21b,WU20,VAF20}, elucidating the mechanism behind the correlated insulating states~\cite{MAC23,HON22,THO21,BRI22,KWA23,KWA21b,XIE23a,ANG19,KWA21a,BLA22,CAL22d,LED21,RAD18,CHA21,SON22,PAD18,VAF21,ZHA21,HOF22,WAG22,KWA21,KAN21,XIE21a,POT21,CHI20b,PAR21a,CHE21,Zaletel2020SkyrmionTBG,DA21,XIE21,LIA21,REP20,HUA19,EUG20,KEN18,CLA19,DA19,Liu2019May17NematicTSMMATBG,ZHA20,CEA20,XIE20b,CHR20,SOE20,KAN20a,YUA18,DOD18,THO18,WU19,BUL20b,WU20,VEN18,XU18b,OCH18,KOS18,LIU19,VAF20,KAN19,BUL20a,SEO19,PO18a}, and superconductivity~\cite{CHR23,YU22,CHO21d,CHO21c,HU19a,KWA22,WAG23a,WAG23,CHA22,FER21,Chichinadze2019Oct16NemSCTBG,LEW21,KHA21,WAN21,ROY19,WU19a,Guo2018SCMATBG,HUA19,KEN18,CLA19,CHR20,KON20,JUL20,YUA18,GON19,DOD18,WU19,YOU19,VEN18,GUI18,XIE20,LIA19,PEL18,LIU18a,WU18,ISO18,PO18a,XU18}. In this appendix, we briefly review both the Bistritzer-MacDonald (BM) model~\cite{Bistritzer2011BMModel} and the topological heavy fermion (THF) model~\cite{Song20211110MATBGHF} of TBG. 

\subsection{Review of the Bistritzer-MacDonald model}\label{appLsec:tgh:BM}
In this section, we briefly review the Bistritzer-MacDonald (BM) model~\cite{Bistritzer2011BMModel} of twisted bilayer graphene (TBG). Throughout this work, we use the notations introduced in Refs.~\cite{BER21,Song2020TBGII,BER21a,LIA21,BER21b,Xie2021TBGVI,BAB20210211TSTGI,BAB20210628TSTGII}.  
We use $\cre{a}{\vec{p},\alpha,s,l}$ to denote the creation operator of an electron with momentum $\vec{p}$, spin $s=\uparrow,\downarrow$, sublattice index $\alpha = A,B$, and within the graphene layer $l = \pm $. We use $\vec{K}_l$ to label the $K$ point of the $l$-layer graphene Brillouin Zone (BZ), and use $\eta =\pm $ to denote the two valleys of graphene which corresponds to the momenta $\eta \vec{K}_{\pm}$. In addition, we introduce the following auxiliary vectors
\begin{equation}
	\vq_{1}=k_\theta \left( 0,-1 \right)^T, \qquad 
	\vq_{2}=k_\theta \left(\frac{\sqrt{3}}{2},\frac{1}{2} \right)^T,\qquad
	\vq_{3}=k_\theta \left(-\frac{\sqrt{3}}{2},\frac{1}{2} \right)^T,
\end{equation}
where $k_{\theta} = \abs{\vK_+ - \vK_-} = 2 \abs{\vK_+}\sin \frac{\theta}{2}$. 
The reciprocal lattice of the moir\'e system is then given by
\begin{align}
&\mathcal{Q}_0 = \mathbb{Z} \vec{b}_{M1} + \mathbb{Z} \vec{b}_{M2} \nonumber \\ 
&	\vec{b}_{M1}=\vq_3-\vq_1\ ,\qquad  \vec{b}_{M2}=\vq_3-\vq_2,
\end{align}
For future convenience, we also introduce $\mathcal{Q}_{\pm} = \mathcal{Q} \pm \vq_1$. 
The corresponding moir\'e lattice vectors in the real space are defined as $\vec{a}_{M1} = \frac{2\pi}{3 k_\theta} \left( \sqrt{3}, 1 \right)$ and $\vec{a}_{M2} = \frac{2\pi}{3 k_\theta} \left( - \sqrt{3}, 1 \right)$. 
\subsubsection{Single-particle Hamiltonian}\label{app:sec:BM_review:TBG:sp}
To define the Hamiltonian of TBG, we consider the following electron operators
\begin{equation}
	\label{app:eqn:BM_low_energy_ops_TBG}
	\cre{c}{\vk,\vQ,\alpha,\eta,s} \equiv \cre{a}{\eta \vK_l + \vk - \vQ,\alpha,s,l}, \qq{for} \vQ \in \mathcal{Q}_{\eta l }.
\end{equation}
The single-particle Hamiltonian can then be written as~\cite{BER21,Song2020TBGII,BER21a} 
\begin{equation}
	\label{app:eqn:spHamiltonian_BM_TBG}
	\hat{H}^{\tTBG}_{0} = \sum_{\vk} \sum_{\eta, \alpha, \beta, s} \sum_{\vQ,\vQ' \in \mathcal{Q}_{\pm}} \left[h^{\tTBG,\eta}_{\vQ,\vQ'} \left( \vk \right) \right]_{\alpha \beta} \cre{c}{\vk,\vQ,\eta,\alpha,s} \des{c}{\vk,\vQ',\eta,\beta,s},
\end{equation}
We can decompose matrix $h^{\tTBG,\eta}_{\vQ,\vQ'} \left( \vk \right)$ into two parts, which correspond, respectively, to the original Dirac cone graphene Hamiltonian $h^{\tD,\eta}_{\vQ} \left( \vk \right)$ and the inter-layer coupling term $h^{\tI,\eta}_{\vQ,\vQ'}$
\begin{equation}
	\label{app:eqn:BM_TBG_ham}
	h^{\tTBG,\eta}_{\vQ,\vQ'} \left( \vk \right) = h^{\tD,\eta}_{\vQ} \left( \vk \right) \delta_{\vQ,\vQ'} + h^{\tI,\eta}_{\vQ,\vQ'}.
\end{equation}
For valley $+$, we have
\begin{align}
	h^{\tD,+}_{\vQ} \left( \vk \right) &= v_F \left( \vk - \vQ \right) \cdot \bSigma, \label{app:eqn:BM_dirac_ham}\\
	h^{\tI,+}_{\vQ,\vQ'} &= \sum_{j=1}^{3} T_{j} \delta_{\vQ,\vQ' \pm \vq_j} \label{app:eqn:BM_interlayer_ham}.
\end{align}
where $v_F$ is the Fermi velocity of graphene,  $\bSigma = \left( \sigma_x,\sigma_y \right)$ is the Pauli vector, and $T_j$ characterizes the inter-layer tunneling with
\begin{equation}
\label{app:eqn:BM_interlayerT}
	T_j = w_0 \sigma_0 + w_1 \left[ \sigma_x \cos \frac{2 \pi \left( j-1 \right)}{3} + \sigma_y \sin \frac{2 \pi \left( j-1 \right)}{3} \right], \quad \text{for} \quad j=1,2,3.
\end{equation}
In this work, we take $w_1 = \SI{110}{\milli\electronvolt}$, $v_F = \SI{5.944}{\electronvolt \angstrom}$, $\abs{\vec{K}_+} = \SI{1.703}{\angstrom^{-1}}$, and vary $0 \leq w_0/w_1 \leq 1$. The Hamiltonian of valley $\eta = -$ can be obtained from the time-reversal transformation
\begin{equation}
	h^{\tTBG,-}_{\vQ,\vQ'} \left( \vk \right) = \sigma_x h^{\tTBG,+}_{-\vQ,-\vQ'} \left( -\vk \right) \sigma_x .
\end{equation}

\subsubsection{Many-body Hamiltonian}\label{app:sec:BM_review:TBG:many_body}
 For a double-gated experimental setup, the electron-electron interaction potential is given by~\cite{BER21a,LIA21}
\begin{equation}
	\label{app:eqn:double_gate_interaction}
	V \left( \vec{r} \right) = U_{\xi} \sum_{n=-\infty}^{\infty} \frac{(-1)^n}{\sqrt{\left( \abs{\vec{r}}/\xi \right)^2 + n^2}} \qq{and}
	V \left( \vq \right) = \left( \pi U_{\xi} \xi^2 \right) \frac{\tanh \left( \abs{\vq} \xi/2 \right)}{\abs{\vq}\xi/2},
\end{equation}
where $U_{\xi} = \frac{e^2}{4 \pi \epsilon_0 \epsilon \xi}=\SI{24}{\milli\electronvolt}$, $\epsilon = 6$ is the dielectric constant, and $\xi=\SI{10}{\nano\meter}$ is the distance between the two screening gates.
The interaction term takes the form of~\cite{KAN19,Zaletel2019Nov5MATBGIntegerFilling,BER21a} 
\begin{equation}
	\label{app:eqn:BM_interaction_ham_TBG}
	\hat{H}^{\tTBG}_I = \frac{1}{2} \frac{1}{N_0 \Omega_0} \sum_{\vq} \sum_{\vG \in \mathcal{Q}_0} V \left( \vq + \vG \right) \delta \rho \left( - \vq - \vG \right) \delta \rho \left( \vq + \vG \right),
\end{equation}
where the normal-ordered density operator $\delta \rho \left( \vq + \vG \right)$ is defined as
\begin{equation}
	\label{app:eqn:BM_offset_density_op}
	\delta \rho \left( \vq + \vG \right) = \rho \left( \vq + \vG \right) - \frac{1}{2} \delta_{\vq,\vec{0}} \delta_{\vG,\vec{0}} = \sum_{\eta,\alpha,s} \sum_{\vk} \sum_{\vQ \in \mathcal{Q}_\pm} \left( \cre{c}{\vk + \vq, \vQ - \vG, \alpha, \eta, s} \des{c}{\vk, \vQ, \alpha, \eta, s} - \frac{1}{2} \delta_{\vq,\vec{0}} \delta_{\vG,\vec{0}} \right),
\end{equation}
$N_0$ is the number of moir\'e unit cells, and $\Omega_0$ is the area of a single moir\'e unit cell. The full TBG Hamiltonian is written as $\hat{H}^{\tTBG} = \hat{H}^{\tTBG}_0 + \hat{H}^{\tTBG}_I$.

\subsection{Review of the topological heavy fermion model for TBG}\label{app:sec:tgh:THF}

We now review the  topological heavy fermion (THF) description~\cite{Song20211110MATBGHF,LAU23,CAL23,CHO23,HU23i,YU23a,LI23a,RAI23a,HU23,ZHO24,WAN24,CAL24} of the BM low-energy bands. The THF model resolves the stable topological obstruction of the entire BM model~\cite{Song2020TBGII,Song20211110MATBGHF,Song2019TBGFragile,Zaletel2019Nov5MATBGIntegerFilling,Ahn2019TBGFragile,Po2018FragileTopo} by introducing two fermionic species: ``heavy'' ($f$) and ``conduction'' ($c$) electrons. In contrast to the $f$-fermions, which form effective $p_x \pm i p_y$ orbitals localized at the moir\'e AA sites, the $c$-fermions are semi-metallic, dispersive, and form anomalous electronic bands. 

\subsubsection{The THF model fermions}\label{app:sec:tgh:THF:fermions}
In terms of the low-energy operators of the BM model introduced in \cref{app:eqn:BM_low_energy_ops_TBG}, the $f$-fermions are given in momentum space by 
\begin{equation}
\label{app:eqn:f_fermions_mom_def}
	\cre{f}{\vk,\alpha,\eta,s} = \sum_{\vQ,\beta} v^{\eta}_{\vQ\beta;\alpha} \left( \vk \right) \cre{c}{\vk,\vQ,\beta,\eta,s},
\end{equation} 
where $v^{\eta}_{\vQ\beta;\alpha} \left( \vk \right)$ is the $f$-fermion momentum-space wave function in valley $\eta$, with $\alpha = 1,2$ denotes the orbital quantum number. Although usually obtained numerically through a Wannierization and disentanglement procedure, the $f$-fermion wave function has an analytical approximation in terms of Gaussian profiles~\cite{Song20211110MATBGHF,CAL23}. Defining the Fourier transformation for the $f$-fermion wave functions,
\begin{align}
	v^{\eta}_{\vQ\beta;\alpha} \left( \vk \right) &= \frac{1}{\sqrt{\Omega_0}} \int \text{d}^2{r} w^{\eta}_{l\beta;\alpha} \left(\vec{r} \right) e^{-i\left( \vk - \vQ \right) \cdot \vec{r}}, \qq{for} \vQ \in \mathcal{Q}_{\eta l}, \label{app:eqn:fourier_transform_f_fermions_to_k}\\
	w^{\eta}_{l\beta;\alpha} \left(\vec{r}\right) &= \frac{1}{N_0 \sqrt{\Omega_0}} \sum_{\vk} \sum_{\vQ \in \mathcal{Q}_{\eta l},\beta} v^{\eta}_{\vQ\beta;\alpha} \left(\vk \right) e^{i \left( \vk - \vQ \right) \cdot \vec{r}},  \label{app:eqn:fourier_transform_f_fermions_to_r}
\end{align}
their real-space Wannier wave functions can be approximated by~\cite{Song20211110MATBGHF},
\begin{alignat}{4}
	& w_{l1;1}^{\eta}(\vec{r}) & =& \frac{\alpha_1}{\sqrt{2\pi\lambda_1^2}}e^{i\frac{\pi}{4}l\eta - \frac{r^2}{2\lambda_1^2}}, \qquad &&
	w_{l2;1}^{\eta}(\vec{r}) &= &-l\frac{\alpha_2}{\sqrt{2}}\frac{r_x + i\eta r_y}{\lambda_2^2\sqrt{\pi}}e^{i\frac{\pi}{4}l\eta - \frac{r^2}{2\lambda_2^2}},
	\label{app:eqn:wannier_states_I} \\
	& w_{l1;2}^{\eta}(\vec{r}) & =& l\frac{\alpha_2}{\sqrt{2}}\frac{r_x - i\eta r_y}{\lambda_2^2\sqrt{\pi}}e^{-i\frac{\pi}{4}l\eta - \frac{r^2}{2\lambda_2^2}},
	\qquad && 
	w_{l2;2}^{\eta}(\vec{r}) & =& \frac{\alpha_1}{\sqrt{2\pi\lambda_1^2}}e^{-i\frac{\pi}{4}l\eta - \frac{r^2}{2\lambda_1^2}},
	\label{app:eqn:wannier_states_II}
\end{alignat} 
where the amplitudes $\alpha_1$ and $\alpha_2$, as well as the spreads $\lambda_1$ and $\lambda_2$ have been obtained both numerically and analytically across a large parameter regime~\cite{Song20211110MATBGHF,CAL23}. In momentum-spaceacross a large parameter regime [4, 12]. In momentum-space the approximations from Eqs. (A16) and (A17) can be, the approximations from \cref{app:eqn:wannier_states_I,app:eqn:wannier_states_II} can be computed using \cref{app:eqn:fourier_transform_f_fermions_to_r}
\begin{align} 
    v_{\vQ 1; 1}^{(\eta)}(\vk) =& \alpha_{1} \sqrt{ \frac{2\pi \lambda_{1}^2}{\Omega_0 \mathcal{N}_{f,\vk}} } e^{ i\frac{\pi}4 \zeta_\vQ - \frac12 (\vk-\vQ)^2 \lambda_{1}^2}, \nonumber\\
    v_{\vQ 2; 1}^{(\eta)}(\vk) =& \alpha_{2}   \sqrt{ \frac{2\pi \lambda_{2}^4}{\Omega_0 \mathcal{N}_{f,\vk}} }  \zeta_\vQ \brak{ i\eta (k_x-Q_x) -  (k_y-Q_y) } e^{ i\frac{\pi}4 \zeta_\vQ - \frac12 (\vk-\vQ)^2 \lambda_{2}^2} \label{eq:vQ-1}, \\
    v_{\vQ 1; 2}^{(\eta)}(\vk) =& \alpha_{2}  \sqrt{ \frac{2\pi \lambda_{2}^4}{\Omega_0 \mathcal{N}_{f,\vk} } } \zeta_\vQ  \brak{-i\eta (k_x-Q_x) - (k_y-Q_y) } e^{-i\frac{\pi}4 \zeta_\vQ - \frac12 (\vk-\vQ)^2 \lambda_{2}^2} ,\nonumber\\
    v_{\vQ 2; 2}^{(\eta)}(\vk) =&  \alpha_{1} \sqrt{ \frac{2\pi \lambda_{1}^2}{\Omega_0 \mathcal{N}_{f,\vk}} } e^{-i\frac{\pi}4 \zeta_\vQ - \frac12 (\vk-\vQ)^2 \lambda_{1}^2} \label{eq:vQ-2}. 
\end{align}
In \cref{eq:vQ-1,eq:vQ-2}, we have introduced the normalization factor $\mathcal{N}_{f,\vk}$, which can be determined to be 
\begin{equation}
\label{app:eq:def_Nfk}
\mathcal{N}_{f,\vk} = \alpha_1^2 \frac{2\pi \lambda_1^2}{\Omega_0} 
    \sum_{\vQ} e^{-(\vk-\vQ)^2\lambda_1^2} 
+ \alpha_2^2 \frac{2\pi \lambda_2^4}{\Omega_0} \sum_{\vQ} 
    (\vk-\vQ)^2 e^{-(\vk-\vQ)^2 \lambda_2^2} \ . 
\end{equation}

In order to correctly capture the strong topology of the system, Ref.~\cite{Song20211110MATBGHF} also introduces four anomalous $c$-electrons around the $\Gamma_M$ point. The $c$-fermion operators are defined according to 
\begin{equation}
	\label{app:eqn:def:c_electrons}
	\cre{c}{\vk,a,\eta,s} = \sum_{\vQ,\beta} \ut^{\eta}_{\vQ\beta;a} \left( \vk \right) \cre{c}{\vk,\vQ,\beta,\eta,s}, \qq{for} 1 \leq a \leq 4.
\end{equation}
In \cref{app:eqn:def:c_electrons}, $a$ indexes the four $c$-electron states (for each spin and valley), while $\ut^{\eta}_{\vQ\beta;a} \left( \vk \right)$ denotes their wave function.  The $\cre{c}{\vk,a,\eta,s}$ fermions have a large kinetic energy away from the $\Gamma_M$ point, and as a result, they are only defined for momenta below a certain cutoff $\abs{\vk} < \Lambda_c$, and their wave function in the proximity of the $\Gamma_M$ point can be approximated by~\cite{Song20211110MATBGHF}
\begin{equation}
	\label{app:eqn:approx_of_c_wavf}
	\ut^{\eta}_{\vQ\beta;a} \left( \vk \right) \approx \ut^{\eta}_{\vQ\beta;a} \left( \vec{0} \right).
\end{equation}  

Analytical expressions for $c$-electron wave functions were found in Ref.~\cite{CAL23}. These were not directly derived from the BM model approximations~\cite{BER21,CAL23}. The approximations for the wave functions at $\vk = \vec{0}$ are given by~\cite{CAL23}
{\footnotesize\begin{alignat}{4}
	&\ut_{\vQ 1, 1}^{(\eta)}(\vec{0})
	&&= - \alpha_{c1}\lambda_{c1} \sqrt{ \frac{2\pi }{ \Omega_0 \mathcal{N}_{c1} } } e^{- i \frac{\pi}4 \zeta_\vQ - \frac12 \vQ^2 \lambda_{c1}^2 } ,\qquad &&
	\ut_{\vQ 2, 1}^{(\eta)}(\vec{0})
	&&= \alpha_{c2}\lambda_{c2}^3 \sqrt{ \frac{\pi }{ \Omega_0 \mathcal{N}_{c1}} } (i\eta Q_x + Q_y)^2 e^{- i \frac{\pi}4 \zeta_\vQ - \frac12 \vQ^2 \lambda_{c2}^2 }, \label{app:eqn:anal_func_c_1} \\
	&\ut_{\vQ 1, 2}^{(\eta)}(\vec{0})
	&&= \alpha_{c2} \lambda_{c2}^3 \sqrt{ \frac{\pi }{ \Omega_0 \mathcal{N}_{c2}} } (-i\eta Q_x + Q_y)^2 e^{i \frac{\pi}4 \zeta_\vQ - \frac12 \vQ^2 \lambda_{c2}^2 } ,\qquad &&
	\ut_{\vQ 2, 2}^{(\eta)}(\vec{0})
	&&= - \alpha_{c1} \lambda_{c2}\sqrt{ \frac{2\pi }{ \Omega_0 \mathcal{N}_{c2}} } e^{i \frac{\pi}4 \zeta_\vQ - \frac12 \vQ^2 \lambda_{c1}^2 }, \label{app:eqn:anal_func_c_2} \\
	&\ut_{\vQ 1, 3}^{(\eta)}(\vec{0})
	&&= \alpha_{c3}\lambda_{c3}^2 \sqrt{ \frac{2\pi  }{ \Omega_0 \mathcal{N}_{c3}} } \zeta_\vQ (-i \eta Q_x + Q_y) e^{ - i \frac{\pi}4 \zeta_\vQ - \frac12 \vQ^2 \lambda_{c3}^2 },\qquad &&
	\ut_{\vQ 2, 3}^{(\eta)}(\vec{0})
	&&= \alpha_{c4}\lambda_{c4}^3 \sqrt{ \frac{\pi  }{ \Omega_0 \mathcal{N}_{c3}} } (-i\eta Q_x +  Q_y)^2 e^{ - i \frac{\pi}4 \zeta_\vQ - \frac12 \vQ^2 \lambda_{c4}^2 }, \label{app:eqn:anal_func_c_3}\\
	&\ut_{\vQ 1, 4}^{(\eta)}(\vec{0}) &&= \alpha_{c4} \lambda_{c4}^3\sqrt{ \frac{\pi  }{ \Omega_0 \mathcal{N}_{c4}} } (i\eta Q_x +  Q_y)^2 e^{ i \frac{\pi}4 \zeta_\vQ - \frac12 \vQ^2 \lambda_{c4}^2 },\qquad &&
	\ut_{\vQ 2, 4}^{(\eta)}(\vec{0})
	&&= \alpha_{c3} \lambda_{c3}^2\sqrt{ \frac{2\pi  }{ \Omega_0 \mathcal{N}_{c4}} } \zeta_\vQ (i \eta Q_x + Q_y) e^{ i \frac{\pi}4 \zeta_\vQ - \frac12 \vQ^2 \lambda_{c3}^2 }, \label{app:eqn:anal_func_c_4}
\end{alignat}}
where we have used the sublattice factor $\zeta_\vQ = + 1$ ($\zeta_\vQ = - 1$) for $\vQ \in \mathcal{Q}_{+}$ ($\vQ \in \mathcal{Q}_{+}$). 
These expressions are also extended for small momenta around the $\Gamma_M$ point using \cref{app:eqn:approx_of_c_wavf}~\cite{Song20211110MATBGHF}. For $\theta = \SI{1.05}{\degree}$ and $w_0 / w_1 = 0.8$, the parameters appearing in \cref{app:eqn:anal_func_c_1,app:eqn:anal_func_c_2,app:eqn:anal_func_c_3,app:eqn:anal_func_c_4} have been fitted to the numerical solution and are given by~\cite{CAL23}
\begin{alignat}{5}
	& \alpha_{c1} = 0.3958,\qquad 
	&& \alpha_{c2} = 0.9183, \qquad 
	&&\mathcal{N}_{c1} = \mathcal{N}_{c2} = 1.2905,\qquad
 && \lambda_{c1} = 0.2194 \abs{\vec{a}_{M1}},\qquad 
	&& \lambda_{c2} = 0.3299 \abs{\vec{a}_{M1}},\qquad 
 \\
	& \alpha_{c3} = 0.9257,\qquad 
	&& \alpha_{c4} = 0.3783, \qquad 
	&&\mathcal{N}_{c3} = \mathcal{N}_{c4} = 1.1102,\qquad 
 &&\lambda_{c3} = 0.2430 \abs{\vec{a}_{M1}},\qquad 
	&& \lambda_{c4} = 0.2241 \abs{\vec{a}_{M1}}.
\end{alignat}

\subsubsection{The single-particle THF model}\label{app:sec:tgh:THF:sp_ham}

The single-particle THF model Hamiltonian for TBG is obtained by projecting the low-energy fermions from \cref{app:eqn:BM_low_energy_ops_TBG} into the THF model basis via the approximation 
\begin{equation}
	\label{app:eqn:proj_to_THF_basis}
	\cre{c}{\vk,\vQ,\beta,\eta,s} \approx \begin{cases}
		 \sum_{\alpha} v^{*\eta}_{\vQ\beta;\alpha} \left( \vk \right) \cre{f}{\vk,\alpha,\eta,s} + \sum_{a} \ut^{*\eta}_{\vQ\beta;a}\left( \vk \right) \cre{c}{\vk,a,\eta,s} & \qq{for} \abs{\vk} \leq \Lambda_c, \\
		 \sum_{\alpha} v^{*\eta}_{\vQ\beta;\alpha} \left( \vk \right) \cre{f}{\vk,\alpha,\eta,s} & \qq{for} \abs{\vk} > \Lambda_c.
	\end{cases}
\end{equation}
and reads as
\begin{align}
	H^{\tTBG}_{0} =& \sum_{\substack{\abs{\vk} \leq \Lambda_c\\ \eta, s}} \left[ \sum_{a,a'} h^{cc,\eta}_{a a'} \left( \vk \right) \cre{c}{\vk,a,\eta,s} \des{c}{\vk,a,\eta,s} + \left( \sum_{a,\alpha} h^{cf,\eta}_{a \alpha} \left( \vk \right) \cre{c}{\vk,a,\eta,s} \des{f}{\vk,\alpha,\eta,s}  + \thc \right) \right] \nonumber \\
	+&  \sum_{\substack{\vk,\alpha,\alpha' \\ \eta,s}} h^{ff,\eta}_{\alpha \alpha'} \left( \vk \right) \cre{f}{\vk,\alpha,\eta,s} \des{f}{\vk,\alpha',\eta,s}.
	\label{app:eqn:single_part_THF_TBG}
\end{align}
In \cref{app:eqn:single_part_THF_TBG}, ``$+\thc$'' denotes the addition of the Hermitian conjugate. In order to differentiate the BM and THF model Hamiltonians, the former are always denoted with a hat. The matrix blocks appearing in \cref{app:eqn:single_part_THF_TBG} are given by~\cite{Song20211110MATBGHF}
\begin{align}
	h^{cc,\eta} \left( \vk \right) =& \begin{pmatrix}
		\zero & v_{\star} \left(\eta k_x \sigma_0 + i k_y \sigma_z \right) \\ 
		v_{*} \left(\eta k_x \sigma_0 - i k_y \sigma_z \right) & M \sigma_z
	\end{pmatrix} \label{app:eqn:cc_thf_block},\\
	h^{cf,\eta} \left( \vk \right) =& \begin{pmatrix}
		\gamma \sigma_0 + v_{*}' \left( \eta k_x \sigma_x + k_y \sigma_y \right) \\ 
		v_{*}'' \left( \eta k_x \sigma_x - k_y \sigma_y \right)
	\end{pmatrix} e^{-\frac{\abs{\vk}^2 \lambda^2}{2}} \label{app:eqn:cf_thf_block}, \\
	h^{ff,\eta} \left( \vk \right) =& \zero \label{app:eqn:f_thf_block},
\end{align}
with $\zero$ denoting the zero $2 \times 2$ (per spin per valley) matrix. The single-particle parameters appearing in \cref{app:eqn:cc_thf_block,app:eqn:cf_thf_block} depend on the parameters of the BM model of TBG and have been obtained both numerically and analytically across a vast range of twist angles $\theta$ and tunneling ratios $w_0/w_1$~\cite{Song20211110MATBGHF,CAL23}. For $\theta = \SI{1.05}{\degree}$, $w_0 / w_1 = 0.8$, and the BM model parameters chosen around \cref{app:eqn:BM_TBG_ham}, $\lambda = 0.3375 \abs{\vec{a}_{M1}}$, $\gamma = \SI{-24.75}{\milli\electronvolt}$, $M = \SI{3.697}{\milli\electronvolt}$, $v_{\star} = \SI{-4.303}{\electronvolt\angstrom}$, $v'_{\star} = \SI{1.623}{\electronvolt\angstrom}$, and $v''_{\star} = \SI{-0.0332}{\electronvolt\angstrom}$.

\subsubsection{The interaction THF Hamiltonian}\label{app:sec:tgh:THF:int_ham}

The THF interaction Hamiltonian was derived by directly projecting density operator from \cref{app:eqn:BM_offset_density_op} with the aid of \cref{app:eqn:proj_to_THF_basis}. Ref.~\cite{Song20211110MATBGHF} has shown that the resulting expression can be simplified to a sum of just seven terms which depend on only six parameters and the Coulomb interaction potential from \cref{app:eqn:double_gate_interaction}
\begin{equation}
	\label{app:eqn:THF_interaction_TBG}
	H^{\tTBG}_I = H_{U_1} + H_W + H_V + H_J + H_{U_2} + H_{\tilde{J}} + H_K.
\end{equation}
The seven terms of \cref{app:eqn:THF_interaction_TBG} are given by
\begin{align}
    H_{U_1} &= \frac{U_1}{2} \sum_{\vec{R}} \sum_{\substack{\alpha,\eta,s \\ \alpha',\eta',s'}} \normord{\cre{f}{\vec{R},\alpha',\eta',s'} \des{f}{\vec{R},\alpha',\eta',s'}} \normord{\cre{f}{\vec{R},\alpha,\eta,s} \des{f}{\vec{R},\alpha,\eta,s}}, \label{app:eqn:THF_int:U1} \\
	H_W &= \frac{1}{N_0} \sum_{\substack{\vk_1 \\ \abs{\vk_2} \leq \Lambda_c}} \sum_{\substack{\vq \\ \abs{\vk_2-\vq} \leq \Lambda_c}} \sum_{\substack{\alpha,\eta,s \\ a',\eta',s'}} W_{a'} \normord{\cre{f}{\vk_1 + \vq,\alpha,\eta,s} \des{f}{\vk_1,\alpha,\eta,s}} \normord{\cre{c}{\vk_2 - \vq,a',\eta',s'} \des{c}{\vk_2,a',\eta',s'}}, \label{app:eqn:THF_int:W} \\
	H_V &= \frac{1}{2 \Omega_0 N_0} \sum_{\abs{\vk_1},\abs{\vk_1} \leq \Lambda_c} \sum_{\substack{\vq \\ \abs{\vk_1+\vq},\abs{\vk_2+\vq} \leq \Lambda_c}} \sum_{\substack{a,\eta,s \\ a',\eta',s'}} V \left( \vq \right)\normord{\cre{c}{\vk_1 + \vq ,a',\eta',s'} \des{c}{\vk_1,a',\eta',s'}} \normord{\cre{c}{\vk_2 - \vq,a,\eta,s} \des{c}{\vk_2,a,\eta,s}}, \label{app:eqn:THF_int:V} \\
	H_J &= -\frac{J}{2N_0} \sum_{\substack{\vk_1 \\ \abs{\vk_2} \leq \Lambda_c}} \sum_{\substack{\vq \\ \abs{\vk_2+\vq} \leq \Lambda_c}} \sum_{\substack{\alpha,\eta,s \\ a',\eta',s'}} \left[ \eta \eta' + (-1)^{\alpha + \alpha'} \right] \normord{\cre{f}{\vk_1 + \vq,\alpha',\eta',s'} \des{f}{\vk_1,\alpha,\eta,s}} \normord{\cre{c}{\vk_2 - \vq,\alpha+2,\eta,s} \des{c}{\vk_2,\alpha'+2,\eta',s'}}, \label{app:eqn:THF_int:J} \\
	H_{U_2} &= \frac{U_2}{2} \sum_{\left\langle \vec{R}, \vec{R}' \right\rangle} \sum_{\substack{\alpha,\eta,s \\ \alpha',\eta',s'}} \normord{\cre{f}{\vec{R}',\alpha',\eta',s'} \des{f}{\vec{R}',\alpha',\eta',s'}} \normord{\cre{f}{\vec{R},\alpha,\eta,s} \des{f}{\vec{R},\alpha,\eta,s}}, \label{app:eqn:THF_int:U2} \\
	H_{\tilde{J}} &= -\frac{J}{4N_0} \sum_{\abs{\vk_1}, \abs{\vk_2} \leq \Lambda_c} \sum_{\vq} \sum_{\substack{\alpha,\eta,s \\ a',\eta',s'}} \left[ \eta \eta' - (-1)^{\alpha + \alpha'} \right] \cre{f}{\vk_1 + \vq,\alpha',\eta',s'} \cre{f}{\vk_2 - \vq,\alpha,\eta,s} \des{c}{\vk_2,\alpha+2,\eta,s} \des{c}{\vk_1,\alpha'+2,\eta',s'} + \thc \label{app:eqn:THF_int:Jtilde} \\	
	H_{K} &= \frac{K}{N_0 \Omega_0} \sum_{\abs{\vk_1},\abs{\vk_2},\abs{\vk_3} \leq \Lambda_c} \sum_{\substack{\alpha,\eta,s \\ \eta',s'}} \eta \eta' \left(
	\cre{c}{\vk_1,\bar{\alpha},\eta,s} \des{c}{\vk_3,\alpha+2,\eta,s} \cre{f}{\vk_1 - \vk_2 - \vk_3,\alpha,\eta',s'} \des{c}{\vk_2,\alpha+2,\eta',s'} 
	\right. \nonumber \\
	&\left. -\cre{f}{\vk_2 + \vk_3 - \vk_1, \alpha, \eta', s'} \des{c}{\vk_2,\alpha+2,\eta',s'} \cre{c}{\vk_1,\bar{\alpha}+2,\eta,s} \des{c}{\vk_3,\alpha,\eta,s} 
	\right)
	+ \thc. \label{app:eqn:THF_int:K}
\end{align}
where $\bar{\alpha} = 3-\alpha$. These terms the represent various interactions within the THF model: onsite and nearest-neighbor repulsions between $f$-electrons ($H_{U_1}$ and $H_{U_2}$), Coulomb interactions among $c$-electrons ($H_V$), density-density and exchange interactions between $f$ and $c$ electrons ($H_W$ and $H_J$), and interactions involving hybridization ($H_{\tilde{J}}$ and $H_K$). $H_K$ is usually neglected as it does not conserve the $f$ number, which bears a large energy cost when $U>|\gamma|$. When $U<|\gamma|$, the formalism changes, and the Coulomb interaction must be projected into the flat bands, giving rise to the projected strong interaction picture~\cite{BER21a,BER21b,Kang2019Multipole,Zaletel2019Nov5MATBGIntegerFilling,Kang2018TBGFragile,LIA21}.  The normal ordering of an operator $\mathcal{O}$ subtracts its expectation value in the charge neutral ground state, $\ket{\mathrm{G}_0}$, which is defined such that
\begin{align}
\ev{\cre{f}{\vk,\alpha,\eta,s} \des{f}{\vk',\alpha',\eta',s'}}{\mathrm{G}_0} &= \frac{1}{2} \delta_{\vk, \vk'} \delta_{\alpha \alpha'} \delta_{\eta \eta'} \delta_{s s'}, \nonumber \\
	\ev{\cre{c}{\vk,a,\eta,s} \des{c}{\vk',a',\eta',s'}}{\mathrm{G}_0} &= \frac{1}{2} \delta_{\vk, \vk'} \delta_{a a'} \delta_{\eta \eta'} \delta_{s s'}, \nonumber \\
	\ev{\cre{f}{\vk,\alpha,\eta,s} \des{c}{\vk',a',\eta',s'}}{\mathrm{G}_0} &= \ev{\cre{c}{\vk,a,\eta,s} \des{f}{\vk',\alpha',\eta',s'}}{\mathrm{G}_0} = 0.
\end{align}
Additionally, in \cref{app:eqn:THF_int:U2}, $\left\langle \vec{R}, \vec{R}' \right\rangle$ denotes nearest-neighbor lattice sites. For $\theta = \SI{1.05}{\degree}$, $w_0 / w_1 = 0.8$, as well as the relative permittivity and screening length chosen around \cref{app:eqn:double_gate_interaction}, the six interaction parameters ($W_1 = W_2$ and $W_3 = W_4$ follows by symmetry~\cite{Song20211110MATBGHF}) are given by~\cite{Song20211110MATBGHF}
\begin{align}
	U_1 &= \SI{57.95}{\milli\electronvolt}, &
 J &= \SI{16.38}{\milli\electronvolt}, &
	W_1 &= W_2 = \SI{44.03}{\milli\electronvolt}, \nonumber \\ 
	U_2 &= \SI{2.239}{\milli\electronvolt}, &
	K &= \SI{4.887}{\milli\electronvolt}, &
	W_3 &= W_4 = \SI{50.2}{\milli\electronvolt}, \label{app:eqn:THF_interaction_params_TBG}
\end{align}
while the $c$-$c$ interaction potential is the same as in \cref{app:eqn:double_gate_interaction}.

\subsubsection{Berry Curvature of the TBG active bands}\label{app:sec:tgh:THF:berry}
For each spin $s$ and valley $\eta$, the active TBG bands near charge neutrality can be recombined into Chern-$(\pm 1)$ bands~\cite{Ahn2019TBGFragile,BER21a,Song2019TBGFragile}, whose operators are denoted by $\cre{d}{\vk,\zeta,\eta,s}$. Specifically, $\cre{d}{\vk, \zeta, \eta, s}$ creates a state of spin $s$ in valley $\eta$ within the Chern band with Chern number $\zeta \cdot \eta$. In terms of the TBG low-energy operators from \cref{app:eqn:BM_low_energy_ops_TBG}, the Chern-band operators are defined according to
\begin{equation}
    \label{app:eqn:def_TBG_chern_band_basis}
    \cre{d}{\vk, \zeta, \eta, s} = \sum_{\vQ,\beta}U^{\zeta}_{\vQ\beta;\eta}(\vk)\cre{c}{\vk,\vQ,\beta,\eta,s},
\end{equation}
with $U^{\zeta}_{\vQ\beta;\eta}(\vk)$ being the corresponding Chern-band wave function. 

The Berry curvature of the TBG active Chern bands can be defined in terms of the corresponding projector. Letting 
\begin{equation}
    \left[ P^{\zeta,\eta} ( \vk ) \right]_{\alpha \vQ; \beta \vQ' } = U^{\zeta}_{\vQ \alpha; \eta} ( \vk ) U^{*\zeta}_{\vQ' \beta; \eta} ( \vk )
\end{equation}
be the projector into the Chern band corresponding to $\cre{d}{\vk, \zeta, \eta, s}$, the associated Berry curvature is given by
\begin{equation} 
\Omega^{\zeta, \eta} (\vk)  =  -i\sum_{\mu,\nu} \epsilon^{\mu\nu} \text{Tr} \left[ P^{\zeta,\eta} ( \vk ) \partial_{k^\mu} P^{\zeta,\eta} ( \vk ) \partial_{k^\nu} P^{\zeta,\eta} ( \vk ) \right],
\end{equation} 
where $\epsilon^{\mu \nu}$ is the two-dimensional Levi-Civita symbol with $\epsilon^{xx}=\epsilon^{yy} = 0$ and $\epsilon^{xy}= - \epsilon^{yx} = 0$. 

\begin{figure}
    \centering
    \includegraphics[width=\textwidth]{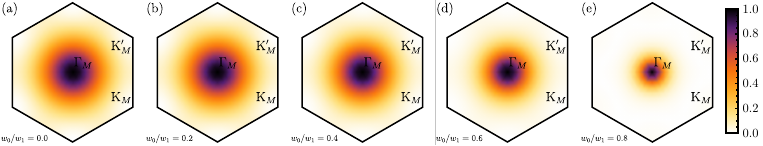}
    \caption{Berry curvature $\Omega^{+, +} (\vk)$ of the Chern-$(+1)$ band of TBG within valley $\eta = +$ at the magic angle $\theta = \SI{1.05}{\degree}$. The BM model parameters we employ are listed below \cref{app:eqn:BM_interlayerT}. (a)-(e) illustrate the effects of varying the tunneling amplitude ratio (indicated below each panel) from the unrealistic chiral limit $w_0 / w_1 = 0$~\cite{Tarnopolsky2019MagicAngleChiralLimit} to the more realistic value $w_0 / w_1 = 0.8$. The Berry curvature is plotted in arbitrary units, as indicated by the accompanying colorbar.}
    \label{fig:berry_curvature_dif_wRatios_THF}
\end{figure}

In \cref{fig:berry_curvature_dif_wRatios_THF}. we plot the distribution of the Berry curvature for the TBG Chern bands at the magic angle, across a range of tunneling ratios $0 \leq w_0/w_1 \leq 0.8$. For the realistic value $w_0 / w_1 = 0.8$, the Berry curvature is concentrated near the $\Gamma_M$ point. Conversely, as we approach the chiral limit ($w_0 / w_1 = 0$)~\cite{Tarnopolsky2019MagicAngleChiralLimit}, the Berry curvature becomes more evenly distributed across the moir\'e BZ. This uniform spread of the Berry curvature can be directly linked to the largely increased $f$-$c$ hybridization in the chiral limit~\cite{Song20211110MATBGHF,CAL23}.

\section{Analytical results within the THF model}\label{app:sec:anal}

In Ref.~\cite{CAL23}, it was shown that around the magic angle, the flat and closest remote bands of TBG within the single-particle BM model are extremely well-fitted by the THF model for a wide range of angles and tunneling amplitude ratios. It was, however, argued that the direct utility of the THF model is dependent on a separation of scales between the $f$-$c$ hybridization ($\gamma$) and the onsite Hubbard repulsion of the $f$-electrons ($U_1$)~\cite{CAL23}, where the magnitude of the latter should be larger than the former. In this case, a \emph{real-space} description of TBG is possible, in which the $f$-electrons form local moments \cite{Song20211110MATBGHF,LAU23,CHO23,HU23,YU23a,LI23a,RAI23a,HU23i,ZHO24,WAN24,CAL24}. 

Conversely, when $\abs{\gamma}>U_1$, a real-space description of the interaction terms is no longer useful, as the large $f$-$c$ hybridization does not allow the polarization of the $f$-fermions to minimize the onsite Hubbard repulsion $H_{U_1}$. Because in the $\abs{\gamma}>U_1$ limit, the energy scale of the interaction is smaller than the gap between the active and remote TBG, a flat-band-projected approach constitutes a more realistic formulation of the problem~\cite{Kang2018TBGFragile,LIA21}. Such a description involves the so-called TBG form factors -- overlaps between the active TBG band wave functions at different momenta~\cite{BER21a}. The TBG form factors can easily be computed numerically, but in order to obtain them analytically, the expressions of the active TBG wave functions are necessary. Obtaining the latter is an impossible task in the original BM model with a large number of bands, but becomes tractable within the THF formalism. 

In this appendix, we show that in the $\abs{\gamma} > U_1$ limit, the single-particle THF model can still be employed to derive analytical expressions for the TBG flat-band wave functions. We then show how the latter can be used to compute the TBG form factors analytically. In turn, this enables us to analytically obtain the charge-one excitations of TBG using the method introduced by Ref.~\cite{BER21b}. Finally, we also show that for $\abs{\gamma} > U_1$, a flat-band projected limit of the THF model can be defined, wherein the charge-one excitations of TBG around charge neutrality can also be obtained analytically. This confirms the conceptual utility of the THF model not just in the $\abs{\gamma} < U_1$ limit, but also in the opposite limit $\abs{\gamma}> U_1$.

\subsection{Flat-band wave functions}\label{app:sec:anal:wavf}

In the THF model, the wave functions of the flat bands can be obtained analytically within the zero-bandwidth limit, which is defined when $v_{\star}^\prime = v_{\star}^{\prime\prime} = 0$, and $M=0$~\cite{Song20211110MATBGHF}. In this limit, the model exhibits an additional symmetry, which Ref.~\cite{Song20211110MATBGHF} dubs ``chiral symmetry''. Notably, the chiral symmetry of the THF model within the zero-bandwidth limit is \emph{different} from the chiral sublattice symmetry of the TBG BM model in the $w_0 / w_1 = 0$ limit~\cite{Tarnopolsky2019MagicAngleChiralLimit}. This can be seen by noting that in the $w_0/w_1 =0$ limit, the active TBG Chern bands are eigenstates of the chiral sublattice operator and are, therefore, graphene sublattice-polarized. However, the active TBG Chern bands, obtained analytically from the THF model in its zero-bandwidth limit~\cite{Song20211110MATBGHF} are not polarized on the single-layer graphene sublattices. As a result, the analytical expression for the TBG flat bands obtained with the THF model in zero-bandwidth limit are still a good approximation of the realistic TBG bands. 

\subsubsection{Analytical expression for the TBG flat-band wave functions}
\label{app:sec:anal:wavf:wavefunctions}

Working in the zero-bandwidth limit, the single-particle Hamiltonian from \cref{app:eqn:single_part_THF_TBG} can be diagonalized analytically. The fermion operators $\cre{d}{\vk, \zeta, \eta, s}$ ($\zeta = \pm$) and $\cre{b}{\vk, j,\eta,s}$ ($j = \pm 1, \pm 2$), corresponding to the Chern band eigenstates of the active bands introduced in \cref{app:eqn:def_TBG_chern_band_basis} and to the remote-band eigenstates, respectively, are given in the THF model basis by
\begin{align}
\cre{d}{\vk, +, \eta, s} &= g_{\vk} 
     e^{i\eta \theta_\vk }\cre{f}{\vk,1,\eta,s} 
     - \eta h_{\vk} \cre{c}{\vk,3,\eta,s} , \nonumber\\
\cre{b}{\vk, +1, \eta, s} &= 
\frac{1}{\sqrt{2}}\left( 
h_\vk   \cre{f}{\vk,1,\eta,s}  - \cre{c}{\vk, 1,\eta,s} + \eta g_{\vk}e^{-i\eta \theta_\vk } \cre{c}{\vk,3,\eta,s} 
\right)
, \nonumber\\
\cre{b}{\vk, +2, \eta, s} &= 
\frac{1}{\sqrt{2}}\left( 
h_\vk   \cre{f}{\vk,1,\eta,s}  + \cre{c}{\vk, 1,\eta,s} + \eta g_{\vk}e^{-i\eta \theta_\vk } \cre{c}{\vk,3,\eta,s} 
\right)
, \nonumber\\
\cre{d}{\vk, -, \eta, s} &= g_{\vk} e^{-i\eta \theta_\vk }
     \cre{f}{\vk,2,\eta,s} 
     - \eta h_{\vk} \cre{c}{\vk,4,\eta,s} , \nonumber\\
\cre{b}{\vk, -1, \eta, s} &= 
\frac{1}{\sqrt{2}}\left( 
h_\vk   \cre{f}{\vk,2,\eta,s}  - \cre{c}{\vk, 2,\eta,s} + \eta g_{\vk}e^{i\eta \theta_\vk } \cre{c}{\vk,4,\eta,s} 
\right)
, \nonumber\\
\cre{b}{\vk, -2, \eta, s} &= 
\frac{1}{\sqrt{2}}\left( 
h_\vk   \cre{f}{\vk,2,\eta,s}  + \cre{c}{\vk, 2,\eta,s} + \eta g_{\vk}e^{i\eta \theta_\vk } \cre{c}{\vk,4,\eta,s} 
\right), \label{app:eqn:anal_of_THF_in_zero_band_limit}
\end{align}
where 
\begin{align}
    \label{app:eqn:aux_functions_for_ffs}
    g_\vk = \frac{ v_\star \abs{\vk } }{\sqrt{ \abs{v_\star \vk}^2 + \gamma^2e^{-|\vk|^2\lambda^2}} },\quad   h_\vk = \frac{  \gamma e^{-|\vk|^2\lambda^2/2}}{\sqrt{ \abs{v_\star \vk}^2 + \gamma^2e^{-|\vk|^2\lambda^2}} },\qq{and} e^{i\theta_\vk} = \frac{k_x+ik_y}{\abs{\vk}}.
\end{align}

In the resulting band-basis, the single-particle Hamiltonian can then be written simply as 
\begin{equation}
    H_0^{\tTBG} = \sum_{\vk,j, \eta,s} \sqrt{\gamma^2e^{-\abs{\vk}^2\lambda^2} + v_\star^2 \abs{\vk}^2 } \left( 2 \abs{j} - 3 \right) \cre{b}{\vk, j, \eta, s} \des{b}{\vk, j, \eta, s}.
\end{equation}
For simplicity, in \cref{app:eqn:anal_of_THF_in_zero_band_limit}, we restricted the $c$-electrons' momenta to the first moir\'e BZ. In the generic case that includes all the $c$-electrons with momenta $\abs{\vk}<\Lambda_c$, the $\cre{d}{\vk, \zeta, \eta, s}$ operators are given by 
\begin{align}
\cre{d}{\vk, +, \eta, s} &= \frac1{\sqrt{\mathcal{N}_\vk}} 
     \cre{f}{\vk,1,\eta,s} 
     - \frac1{\sqrt{\mathcal{N}_\vk}} \sum_{\substack{\vG \\ |\vk+\vG|<\Lambda_c}} \frac{\gamma e^{-\frac{1}{2}\lambda^2\abs{\vk+\vG}^2} /v_\star }{\eta (k_x+G_x) + i (k_y+G_y) } \cre{c}{\vk+\vG,3,\eta,s} , \\
\cre{d}{\vk, -, \eta, s} &= \frac1{\sqrt{\mathcal{N}_\vk}} 
     \cre{f}{\vk,2,\eta,s} 
     - \frac1{\sqrt{\mathcal{N}_\vk}} \sum_{\substack{\vG \\ |\vk+\vG|<\Lambda_c}} \frac{\gamma e^{-\frac{1}{2}\lambda^2\abs{\vk+\vG}^2}/v_\star}{\eta (k_x+G_x) - i (k_y+G_y) } \cre{c}{\vk+\vG,4,\eta,s} ,
\end{align}
where the normalization prefactor reads as
\begin{equation}
    \mathcal{N}_\vk = 1 + \sum_{\substack{\vG \\ |\vk+\vG|<\Lambda_c}} \frac{\gamma^2e^{-\lambda^2\abs{\vk+\vG}^2}/v_\star^2}{|\vk+\vG|^2}\ . 
\end{equation} 

We remind the reader that the operator $d_{\vk, \zeta, \eta, s}^\dagger$ corresponds to a Chern band with Chern number $\zeta \cdot \eta$. It is worth noting that, for the valley $\eta=+$, the $\cre{d}{\vk, \zeta, \eta, s}$ operator is the same as the Chern-band operator introduced by Ref.~\cite{BER21a}. However, in valley $\eta=-$, the $\cre{d}{\vk, \zeta, -, s}$ operator introduced here corresponds to the $\cre{d}{\vk, -\zeta, -, s}$ one of Ref.~\cite{BER21a}. 

The wave function of the TBG flat bands can, therefore, be expressed \emph{analytically} in terms of the $f$- and $c$-electron operators. The latter also admit analytical expressions given by \cref{app:eqn:wannier_states_I,app:eqn:wannier_states_II,app:eqn:anal_func_c_1,%
app:eqn:anal_func_c_2,app:eqn:anal_func_c_3,app:eqn:anal_func_c_4}. In turn, this enables one to obtain the active band TBG wave functions analytically within the plane wave basis from \cref{app:eqn:BM_low_energy_ops_TBG}, a result which is not otherwise directly available from the BM model. 

By combining \cref{app:eqn:f_fermions_mom_def,app:eqn:def:c_electrons,app:eqn:anal_of_THF_in_zero_band_limit}, we can express the electron operators of the flat TBG Chern bands in terms of the the low-energy TBG operators from \cref{app:eqn:BM_low_energy_ops_TBG} as 
\begin{align}
    &\cre{d}{\vk, +,\eta ,s } = g_{\vk} e^{i\eta \theta_{\vk}}\sum_{\vQ,\beta}v^{\eta}_{\vQ\beta;1}(\vk)\cre{c}{\vk,\vQ,\beta,\eta,s} 
    -\eta h_{\vk}\sum_{\vQ,\beta}\tilde{u}_{\vQ\beta;3}^\eta(\vec{0})\cre{c}{\vk,\vQ,\beta,\eta,s}, \nonumber\\ 
    &\cre{d}{\vk, -,\eta ,s } = g_{\vk} e^{-i\eta \theta_{\vk}}\sum_{\vQ,\beta}v^{\eta}_{\vQ\beta;2}(\vk)\cre{c}{\vk,\vQ,\beta,\eta,s} 
    -\eta h_{\vk}\sum_{\vQ,\beta}\tilde{u}_{\vQ\beta;4}^\eta(\vec{0})\cre{c}{\vk,\vQ,\beta,\eta,s}.
\end{align}
Comparing with \cref{app:eqn:def_TBG_chern_band_basis}, we can directly identify the Chern band wave functions  
\begin{equation}
\label{app:eq:eigen_fun_flat_band}
    U_{\vQ \beta;\eta}^{\zeta}(\vk) = e^{i\eta\zeta \theta_{\vk}}g_\vk v^{\eta}_{\vQ\beta;\zeta}(\vk) - \eta h_{\vk}\tilde{u}^{\eta}_{\vQ\eta;i_{\zeta}}(\vec{0}),
\end{equation}
where, for simplicity, we have denoted $i_{\zeta=+}=3$, $i_{\zeta=-}=4$. 

\subsubsection{Analytical expressions for the TBG form factors}\label{app:sec:anal:wavf:ff}

Another useful quantity which can be obtained analytically using \cref{app:eq:eigen_fun_flat_band} are the TBG form factors corresponding to the active bands. The latter arise when projecting the TBG density operator into the Chern-band basis. Specifically, the projected TBG density operator is given by
\begin{equation}
	\delta \rho \left( \vq  \right)  = \sum_{\vk, \eta, s} \sum_{\zeta, \zeta'} M_{\zeta \zeta'}^{(\eta)}(\vk,\vk + \vq) \left( \cre{d}{\vk + \vq, \zeta, \eta, s} \des{d}{\vk, \zeta', \eta, s} - \frac{1}{2} \delta_{\vq,\vec{0}} \right),
\end{equation}
where the corresponding form factors are defined by 
\begin{equation}
    \label{app:eqn:def_form_factors}
    M_{\zeta \zeta'}^{(\eta)}(\vk,\vk') = \sum_{\vQ,\beta}U^{*\zeta}_{\vQ\beta;\eta}(\vk)U^{\zeta'}_{\vQ\beta;\eta}(\vk').
\end{equation}

The form factors can be calculated via~\cref{app:eq:eigen_fun_flat_band}. Approximately, we find that 
\begin{align}
     M_{\zeta \zeta'}^{(\eta)}(\vk,\vk')  \approx \sum_{\vQ,\beta}\left( g_{\vk}g_{\vk'}e^{-i\theta_{\vk}\zeta-i\theta_{\vk'}\zeta'}v^{*\eta}_{\vQ\beta;\zeta}(\vk)v^{\eta}_{\vQ\beta;\zeta'}(\vk')
     +h_{\vk}h_{\vk'} \tilde{u}^{*\eta}_{\vQ\beta;i_{\zeta}}(\vec{0})\tilde{u}^{\eta}_{\vQ\beta;i_{\zeta'}}(\vec{0}) \right),
\end{align}
where we have dropped the cross-terms  $v^{*\eta}_{\vQ\beta;\zeta}(\vk)\tilde{u}^\eta_{\vQ\eta;i_{\zeta'}}(\vec{0})$ and $v^{\eta}_{\vQ\beta;\zeta}(\vk')\tilde{u}^{*\eta}_{\vQ\eta;i_{\zeta'}}(\vec{0})$. In general, the $f$- and $c$-fermions wave functions are orthogonal at any momentum $\vk$ in the moir\'e BZ
\begin{equation}
    \sum_{\vQ \beta} v^{*\eta}_{\vQ\beta;\zeta}(\vk) \tilde{u}^{\eta}_{\vQ\beta;i_\zeta}(\vk) = 0\, ,
\end{equation} 
Coupled with the approximation $\tilde{u}_{\vQ\beta,i}(\vk) \approx \tilde{u}_{\vQ\beta,i}(\vec{0})$ from \cref{app:eqn:approx_of_c_wavf}, this implies that the cross-terms should approximately vanish. Using also the orthonormal condition for the $c$-electron wave functions at the $\Gamma_M$ point, $\sum_{\vQ,\beta} \tilde{u}^{*\eta}_{\vQ\beta;i_{\zeta}}(\vec{0})\tilde{u}^{\eta}_{\vQ\beta;i_{\zeta'}}(\vec{0}) = \delta_{\zeta,\zeta'} $, we find that
\begin{equation}
\label{app:eq:form_factor}
     M_{\zeta \zeta'}^{(\eta)}(\vk,\vk')  \approx \sum_{\vQ,\beta}\left( g_{\vk}g_{\vk'} e^{-i(\theta_{\vk}-\theta_{\vk'})\zeta} v^{*\eta}_{\vQ\beta;\zeta}(\vk)v^{\eta}_{\vQ\beta;\zeta'}(\vk')
     +h_{\vk}h_{\vk'}\delta_{\zeta,\zeta'}\right) . 
\end{equation}

To further simplify the expression from \cref{app:eq:form_factor}, we now compute the sum $\sum_{\vQ,\beta}v^{\eta}_{\vQ\beta;\zeta}(\vk)v^{*\eta}_{\vQ\beta;\zeta'}(\vk')$ using the analytical formulae for the $f$-electron wave functions from \cref{eq:vQ-1,eq:vQ-2}. Additionally, we approximate the summation over $\vQ$ by an integral according to 
\begin{equation}
    \label{app:eqn:sum_to_int}
    \sum_{\vQ} \rightarrow \frac{\Omega_0}{2\pi^2}\int \text{d}^2 Q .
\end{equation}
This allows us to obtain 
\begin{align}
&\sum_{\vQ,\beta}v^{\eta}_{\vQ\beta;+}(\vk)v^{*\eta}_{\vQ\beta;+}(\vk') \nonumber \\ 
\approx & \frac{\Omega_0}{2\pi^2}
\frac{2\pi}{\Omega_0\sqrt{\mathcal{N}_{f,\vk}
\mathcal{N}_{f,\vk'}}}
\int \text{d}^2 Q 
\left\{ \lambda_1^2 \alpha_1^2 e^{-(\vk-\vQ)^2\lambda_1^2/2 -(\vk'-\vQ)^2\lambda_1^2/2 } \right. \nonumber\\ 
&\left. + \lambda_2^4 \alpha_2^2 e^{-(\vk-\vQ)^2\lambda_2^2/2 -(\vk'-\vQ)^2\lambda_2^2/2 }\left[i\eta(k_x-Q_x)-(k_y-Q_y)\right]\left[-i\eta(k_x'-Q_x)-(k_y'-Q_y)\right]
\right\} \nonumber\\ 
=&\frac{1}{\sqrt{\mathcal{N}_{f,\vk}\mathcal{N}_{f,\vk'}}}\left[
\alpha_1^2 e^{-\frac{\abs{\vk_1-\vk_2}^2\lambda_1^2}{4}} + \alpha_2^2 \left(
1 - \frac{\abs{\vk_1-\vk_2}^2 \lambda_1^2}{4}
\right)
e^{-\frac{\abs{\vk_1-\vk_2}^2\lambda_2^2}{4}}
\right], \\
&\sum_{\vQ,\beta}v^{\eta}_{\vQ\beta;-}(\vk)v^{*\eta}_{\vQ\beta;-}(\vk') \nonumber \\ 
\approx & \frac{\Omega_0}{2\pi^2}
\frac{2\pi}{\Omega_0\sqrt{\mathcal{N}_{f,\vk}
\mathcal{N}_{f,\vk'}}}
\int \text{d}^2 Q 
\left\{ \lambda_1^2 \alpha_1^2 e^{-(\vk-\vQ)^2\lambda_1^2/2 -(\vk'-\vQ)^2\lambda_1^2/2 } \right. \nonumber\\ 
&\left. + \lambda_2^4 \alpha_2^2 e^{-(\vk-\vQ)^2\lambda_1^2/2 -(\vk'-\vQ)^2\lambda_1^2/2 }\left[-i\eta(k_x-Q_x)-(k_y-Q_y)\right]\left[i\eta(k_x'-Q_x)-(k_y'-Q_y)\right]
\right\}\nonumber\\  
=&\frac{1}{\sqrt{\mathcal{N}_{f,\vk}\mathcal{N}_{f,\vk'}}}\left[
\alpha_1^2 e^{-\frac{\abs{\vk_1-\vk_2}^2\lambda_1^2}{4}} + \alpha_2^2 \left(
1 - \frac{\abs{\vk_1-\vk_2}^2 \lambda_1^2}{4}
\right)
e^{-\frac{\abs{\vk_1-\vk_2}^2\lambda_2^2}{4}}
\right],
\end{align}
and 
\begin{align}
&\sum_{\vQ,\beta}v^{\eta}_{\vQ\beta;+}(\vk)v^{*\eta}_{\vQ\beta;-}(\vk') \nonumber \\ 
\approx & \frac{\Omega_0}{2\pi^2}
\frac{2\pi}{\Omega_0\sqrt{\mathcal{N}_{f,\vk}
\mathcal{N}_{f,\vk'}}}
\int \text{d}^2 Q \alpha_1\alpha_2\lambda_1\lambda_2^2
\left[ 
-\eta\left(k_x'-Q_x\right)-i \left(k_y'-Q_y'\right)
\right] 
e^{ -\frac{1}{2}(\vk-\vQ)^2\lambda_1^2 -\frac{1}{2}(\vk'-\vQ)^2\lambda_2^2} \nonumber\\ 
&+\frac{\Omega_0}{2\pi^2}
\frac{2\pi}{\Omega_0\sqrt{\mathcal{N}_{f,\vk}
\mathcal{N}_{f,\vk'}}}
\int \text{d}^2 Q 
\alpha_1\alpha_2\lambda_1\lambda_2^2
\left[ -
\eta(k_x-Q_x) -i(k_y-Q_y)
\right] 
e^{ -\frac{1}{2}(\vk-\vQ)^2\lambda_2^2 -\frac{1}{2}(\vk'-\vQ)^2\lambda_1^2} =0, \\ 
&\sum_{\vQ,\beta}v^{\eta}_{\vQ\beta;-}(\vk)v^{*\eta}_{\vQ\beta;+}(\vk') =0.
\end{align} 
Moreover, the continuum limit from \cref{app:eqn:sum_to_int} enables us to approximate the normalization constant from \cref{app:eq:def_Nfk} and obtain
\begin{equation}
\mathcal{N}_{f,\vk} \approx \frac{\Omega_0}{2\pi^2}\int \text{d}^2Q\left[ \alpha_1^2 \frac{2\pi \lambda_1^2}{\Omega_0} 
     e^{-(\vk-\vQ)^2\lambda_1^2} 
+ \alpha_2^2 \frac{2\pi \lambda_2^4}{\Omega_0}  
    (\vk-\vQ)^2 e^{-(\vk-\vQ)^2 \lambda_2^2}\right] 
    = 1.
\end{equation}
As a result, we find that
\begin{equation}
\sum_{\vQ,\beta}v^{\eta}_{\vQ\beta;\zeta}(\vk)v^{*\eta}_{\vQ\beta;\zeta'}(\vk')\approx\delta_{\zeta,\zeta'} \frac{
1
}{\alpha_1^2+\alpha_2^2}\left[
\alpha_1^2 e^{-\frac{\abs{\vk-\vk'}^2\lambda_1^2}{4}} + \alpha_2^2 \left(
1 - \frac{\abs{\vk-\vk'}^2 \lambda_1^2}{4}
\right)
e^{-\frac{\abs{\vk-\vk'}^2\lambda_2^2}{4}}
\right].
\label{app:eq:sum_vqvq}
\end{equation}
From \cref{app:eq:form_factor} and \cref{app:eq:sum_vqvq}, the TBG form factors can then be written as 
\begin{equation}
    \label{app:eqn:approx_for_ff_using_THF}
    M_{\zeta \zeta'}^{(\eta)}(\vk,\vk') = \delta_{\zeta,\zeta'}g_{\vk}g_{\vk'}e^{-i(\theta_{\vk}-\theta_{\vk'})\zeta}
\left[
\alpha_1^2 e^{-\frac{\abs{\vk-\vk'}^2\lambda_1^2}{4}} + \alpha_2^2 \left(
1 - \frac{\abs{\vk-\vk'}^2 \lambda_1^2}{4}
\right)
e^{-\frac{\abs{\vk-\vk'}^2\lambda_2^2}{4}}
\right] + \delta_{\zeta,\zeta'}h_{\vk}h_{\vk'}.
\end{equation}
In the limit $\lambda_1 \approx \lambda_2$ (which holds approximately for TBG near magic angle), the form factors can be simplified to 
\begin{equation}
      M_{\zeta \zeta'}^{(\eta)}(\vk,\vk') \approx \delta_{\zeta,\zeta'}g_{\vk}g_{\vk'}e^{-i(\theta_{\vk}-\theta_{\vk'})\zeta}
 \left(
1 - \frac{\alpha_2^2\abs{\vk-\vk'}^2 \lambda_1^2}{4}
\right)
e^{-\frac{\abs{\vk-\vk'}^2\lambda_1^2}{4}}+ \delta_{\zeta,\zeta'}h_{\vk}h_{\vk'}\, .
\end{equation}

\subsection{The charge-one excitation spectrum of TBG in the $\abs{\gamma} > U_1$ limit}\label{app:sec:anal:ch_one}

Armed with analytical expressions of the active TBG bands in both the THF model basis and the TBG plane wave basis, we now show that the charge-one excitations of TBG can be obtained analytically in the $\abs{\gamma} > U_1$ limit. We illustrate two different approaches to do so: one which leverages the results of Ref.~\cite{BER21b} and employs the analytically derived TBG form factors from \cref{app:eqn:approx_for_ff_using_THF}, and another one in which we project the THF interaction Hamiltonian within the active TBG bands using the analytical eigenstates of the THF single-particle Hamiltonian obtained in \cref{app:eqn:anal_of_THF_in_zero_band_limit}, within the zero-bandwidth limit.

\subsubsection{Charge-one excitation matrices} \label{app:sec:anal:ch_one:from_rMat}

\newcommand{\hN}{\hat{N}}
\newcommand{\omegaTBG}{\Omega_0}
\newcommand{\Uncf}{U_{\text{non-chiral flats}}}
\newcommand{\UN}[1]{U(#1)} 
As shown in Ref.~\cite{BER21b}, for the integer-filled exact ground states of TBG derived by Ref.~\cite{LIA21}, the charge-one excitation spectrum of TBG can be calculated directly, as we will review below. For a given exact eigenstate $\ket{\varphi}$ of the projected interaction $H^{\tTBG}_I$~\cite{LIA21}, the charge-one excitation is obtained by evaluating the following commutators~\cite{BER21b}
\begin{align}
	\left[H^{\tTBG}_I -\mu \hN , \cre{d}{\vk,\zeta,\eta,s} \right] \ket{\varphi} &= \sum_{\zeta'} R^{\eta}_{\zeta' \zeta} \left( \vk \right) \cre{d}{\vk,\zeta',\eta,s} \ket{\varphi}, \label{app:eqn:r_mat_commut_+} \\
	\left[H^{\tTBG}_I -\mu \hN, \des{d}{\vk,\zeta,\eta,s} \right] \ket{\varphi} &= \sum_{\zeta'} \tilde{R}^{\eta}_{\zeta' \zeta} \left( \vk \right) \des{d}{\vk,\zeta',\eta,s} \ket{\varphi}, \label{app:eqn:r_mat_commut_-}
\end{align}
where $\mu$ is the chemical potential and $\hN$ is defined as
\begin{equation}
	\hN = \sum_{\vk,\zeta,\eta,s} \cre{d}{\vk,\zeta,\eta,s} \des{d}{\vk,\zeta,\eta,s}.
\end{equation} 

The matrices $R^{\eta}_{\zeta' \zeta} \left( \vk \right)$ and $\tilde{R}^{\eta}_{\zeta' \zeta} \left( \vk \right)$ in~\cref{app:eqn:r_mat_commut_+,app:eqn:r_mat_commut_-} are determined by the filling $\nu$ of the ground states $\ket{\varphi}$ and by the form factors of \cref{app:eqn:def_form_factors}. The explicit formulas are given below~\cite{BER21b}
\begin{align}
	R^{\eta}_{\zeta' \zeta} \left( \vk \right) &= \frac{1}{2 \omegaTBG N_0} \sum_{\vG \in \mathcal{Q}_0} \left[ \left( \sum_{\vq \in \MBZ, \zeta''} V\left( \vq + \vG \right) M^{* (\eta)}_{\zeta'' \zeta'} \left(\vk + \vq + \vG, \vk \right) M^{(\eta)}_{\zeta'' \zeta} \left(\vk + \vq + \vG, \vk \right) \right) \right. \nonumber \\
	&\left.+ 2 A_{-\vG} \sqrt{V \left(\vG \right)} M^{(\eta)}_{\zeta' \zeta} \left( \vk + \vG, \vk \right) \right] - \mu \delta_{\zeta', \zeta}, \label{app:eqn:def_r_+}\\
	\tilde{R}^{\eta}_{\zeta' \zeta} \left( \vk \right) &= \frac{1}{2 \omegaTBG N_0} \sum_{\vG \in \mathcal{Q}_0} \left[ \left( \sum_{\vq \in \MBZ, \zeta''} V\left( \vq + \vG \right) M^{(\eta)}_{\zeta'' \zeta'} \left(\vk + \vq + \vG, \vk \right) M^{*(\eta)}_{\zeta'' \zeta} \left(\vk + \vq + \vG, \vk \right) \right) \right. \nonumber \\
	&\left.- 2 A_{-\vG} \sqrt{V \left(\vG \right)} M^{*(\eta)}_{\zeta' \zeta} \left( \vk + \vG, \vk \right) \right] + \mu \delta_{\zeta', \zeta}, \label{app:eqn:def_r_-}
\end{align}
where $\MBZ$ denotes the moir\'e BZ and $A_{\vG}$ is defined as~\cite{BER21b} 
\begin{equation}
	\label{app:eqn:def_Afact}
	A_{\vG} = \sqrt{V \left( \vG \right)} \sum_{\substack{\vk \in \MBZ \\ \zeta,\eta}} \frac{\nu}{4} M^{(\eta)}_{\zeta \zeta} \left( \vk + \vG, \vk \right) = \sqrt{V \left( \vG \right)} \sum_{\vk \in \MBZ} \frac{\nu}{4} \Tr \left[ M \left( \vk + \vG, \vk \right)\right] .
\end{equation} 
Using the fact that the approximate TBG form factors from \cref{app:eqn:approx_for_ff_using_THF} are diagonal in the Chern band basis, we find that the corresponding charge-one excitations are also diagonal in the same basis, \ie{} $R^{\eta}_{\zeta' \zeta} \left(  \vk \right) = \delta_{\zeta', \zeta} R^{\eta}_{\zeta \zeta} \left(  \vk \right)$ and $\tilde{R}^{\eta}_{\zeta' \zeta} \left(  \vk \right) = \delta_{\zeta', \zeta} \tilde{R}^{\eta}_{\zeta \zeta} \left(  \vk \right)$ with
\begin{align}
	R^{\eta}_{\zeta \zeta} \left( \vk \right) &= \frac{1}{2 \omegaTBG N_0} \sum_{\vG \in \mathcal{Q}_0} \left[ \left( \sum_{\vq \in \MBZ} V\left( \vq + \vG \right) M^{* (\eta)}_{\zeta\zeta} \left(\vk + \vq + \vG, \vk \right) M^{(\eta)}_{\zeta \zeta} \left(\vk + \vq + \vG, \vk \right) \right) \right. \nonumber \\
	&\left.+ 2 A_{-\vG} \sqrt{V \left(\vG \right)} M^{(\eta)}_{\zeta \zeta} \left( \vk + \vG, \vk \right) \right] - \mu , \label{app:eqn:def_r_+_diag}\\
	\tilde{R}^{\eta}_{\zeta \zeta} \left( \vk \right) &= \frac{1}{2 \omegaTBG N_0} \sum_{\vG \in \mathcal{Q}_0} \left[ \left( \sum_{\vq \in \MBZ} V\left( \vq + \vG \right) M^{(\eta)}_{\zeta \zeta} \left(\vk + \vq + \vG, \vk \right) M^{*(\eta)}_{\zeta \zeta} \left(\vk + \vq + \vG, \vk \right) \right) \right. \nonumber \\
	&\left.- 2 A_{-\vG} \sqrt{V \left(\vG \right)} M^{*(\eta)}_{\zeta \zeta} \left( \vk + \vG, \vk \right) \right] + \mu . \label{app:eqn:def_r_-_diag}
\end{align}
At the same time, it is easy to see that $M^{*(\eta)}_{(-\zeta) (-\zeta)} \left( \vk, \vk' \right) = M^{(\eta)}_{\zeta \zeta} \left( \vk, \vk' \right)$, which implies that $A_{\vG}$ is real and that $R^{\eta}_{\zeta \zeta} \left(  \vk \right) = R^{*\eta}_{(-\zeta) (-\zeta)} \left(  \vk \right)$ and $\tilde{R}^{\eta}_{\zeta \zeta} \left(  \vk \right) = \tilde{R}^{*\eta}_{(-\zeta) (-\zeta)} \left(  \vk \right)$. However, while the first term in each of \cref{app:eqn:def_r_+_diag,app:eqn:def_r_-_diag} is real, the second term is not manifestly so. Generically, the TBG form factors obey the hermicitiy property $M^{(\eta)}_{\zeta \zeta'} \left( \vk, \vk' \right) = M^{*(\eta)}_{\zeta' \zeta} \left( \vk', \vk \right)$, which follows trivially from their definition in \cref{app:eqn:def_form_factors}. Together with the even symmetries of $A_{\vG} = A_{-\vG}$ and $V(\vG) = V(-\vG)$, the hermiticity property of the form factors coupled with the periodicity in reciprocal space $M^{(\eta)}_{\zeta \zeta'} \left( \vk + \vG, \vk' + \vG \right) = M^{(\eta)}_{\zeta \zeta'} \left( \vk + \vG, \vk' + \vG \right)$ is enough to guarantee the reality of the second term of each of \cref{app:eqn:def_r_+_diag,app:eqn:def_r_-_diag}. However, the approximation in \cref{app:eqn:approx_for_ff_using_THF} does not obey the periodicity property, which makes the charge-one excitation matrices nonhermitian at $\nu \neq 0$. For $\nu = 0$ (or alternatively when the so-called flat metric condition is imposed~\cite{BER21b} at generic $\nu \neq 0$), $A_{\vG} = 0$, which implies that the charge-one excitation matrices are proportional to the identity matrix (and are therefore \textit{bona-fide} hermitian matrices).

\subsubsection{Analytical expressions for the charge-one excitation matrices of TBG} \label{app:sec:anal:ch_one:super_anal}

We now provide asymptotic analytical expression for the charge-one excitation matrices at $\nu = 0$. The latter can be rewritten from \cref{app:eqn:def_r_+_diag,app:eqn:def_r_-_diag} as 
\begin{align}
    R^{\eta}_{\zeta \zeta} \left( \vk \right) = \tilde{R}^{\eta}_{\zeta \zeta} \left( \vk \right) =& \frac{1}{2 \Omega_0 N_0} \frac{\Omega_0 N_0}{(2 \pi)^2} \sum_{\vG \in \mathcal{Q}_0} \int_{\MBZ} V \left( \vq + \vG \right) \abs{M^{(\eta)}_{\zeta \zeta} \left(\vk + \vq + \vG, \vk \right)}^2 \text{d}^2 q \nonumber \\
    =& \frac{1}{8 \pi^2} \int_{-\infty}^{\infty} \int_{-\infty}^{\infty} V \left( \vq - \vk \right) \abs{M^{(\eta)}_{\zeta \zeta} \left(\vq, \vk \right)}^2 \text{d} q_x \text{d} q_y. \label{app:eqn:integral_of_rMat}
\end{align}
To move forward, we need to make two simplifying assumptions. First, we consider the $\sqrt{ \abs{v_\star \vk}^2 + \gamma^2 e^{-|\vk|^2\lambda^2}}$ denominator of the $g_{\vk}$ and $h_{\vk}$ functions from \cref{app:eqn:aux_functions_for_ffs}, which appears in the expression of the form factors from \cref{app:eqn:approx_for_ff_using_THF}. For small values of $\abs{\vk}$ the exponential $e^{-\abs{\vk}^2\lambda^2}$ is approximately 1. At large $\abs{\vk}$, $g_{\vk}$ is already multiplied by decaying exponentials in \cref{app:eqn:approx_for_ff_using_THF}, while $h_{\vk}$ itself contains a decaying exponential factor in the numerator. In other words, for large $\abs{\vq}$, the exact values of the $g_{\vq}$ and $h_{\vq}$ functions is unimportant, as they become exponentially suppressed in the form factors. As a result, we will ignore the exponential decay factor appearing in the denominators of the $g_{\vq}$ and $h_{\vq}$ functions in \cref{app:eqn:integral_of_rMat} and approximate the form factor as
\begin{equation}
    \label{app:eqn:approx_for_ff_using_THF}
    M_{\zeta \zeta}^{(\eta)}(\vk,\vq) = g_{\vk}\tilde{g}_{\vq}e^{-i(\theta_{\vk}-\theta_{\vq})\zeta}
\left[
\alpha_1^2 e^{-\frac{\abs{\vk-\vq}^2\lambda_1^2}{4}} + \alpha_2^2 \left(
1 - \frac{\abs{\vk-\vq}^2 \lambda_1^2}{4}
\right)
e^{-\frac{\abs{\vk-\vq}^2\lambda_2^2}{4}}
\right] + h_{\vk}\tilde{h}_{\vq},
\end{equation}
where we have defined
\begin{equation}
     \tilde{g}_\vq \approx \frac{ v_\star \abs{\vq } }{\sqrt{ \abs{v_\star \vq}^2 + \gamma^2 } } \qq{and}   \tilde{h}_\vq = \frac{  \gamma e^{-|\vq|^2\lambda^2/2}}{\sqrt{ \abs{v_\star \vq}^2 + \gamma^2} }.
\end{equation}
Additionally, we note that, while the form factor decays exponentially in $\abs{\vq}$, the screened Coulomb potential only decays polynomially over the same momentum scale. As a simple approximation, we can can therefore approximate the Coulomb potential to a constant
\begin{equation}
    V \left(\vq - \vk \right) \approx U_\xi \pi \xi^2.
\end{equation}
Finally, we note that the integrand in \cref{app:eqn:integral_of_rMat} only depends on the the relative angle between $\vk$ and $\vq$ and their magnitudes. Letting $\vk = \left(k, 0 \right)$ and $\vq = \left(q \cos \theta, q\sin \theta \right)$ without loss of generality, the charge-one excitation matrices become
\begin{align}
    R^{\eta}_{\zeta \zeta} \left( \vk \right) =& \frac{U_\xi \xi^2}{8\pi}\int_0^{\infty} q\text{d} q \int_{0}^{2 \pi} \text {d}\theta \left( g^2_{k}\tilde{g}^2_{q}  f^2 \left(k, q, \theta \right) + h^2_{k}\tilde{h}^2_{q} +2 g_{k} h_{k} \tilde{g}_{q} \tilde{h}_{q} f \left(k, q, \theta \right) \cos \theta \right), \label{app:eqn:integral_of_rMat_2}
\end{align}
with
\begin{equation}
    f (k, q, \theta) = \alpha_1^2 e^{-\frac{\lambda_1^2 \left( k^2 + q^2 - 2 k q \cos \theta \right) }{4}} + \alpha_2^2 \left( 1 - \frac{\lambda_1^2 \left( k^2 + q^2 - 2 k q \cos \theta \right)}{4} \right)
    e^{-\frac{\lambda_2^2 \left( k^2 + q^2 - 2 k q \cos \theta \right) }{4}}.
\end{equation}

The integral over the angle in \cref{app:eqn:integral_of_rMat_2} can be readily performed giving
\begin{align}
    R^{\eta}_{\zeta \zeta} \left( \vk \right) = \frac{U_\xi \xi^2}{8 \pi} \frac{1}{k^2 v_{\star}^2 + \gamma^2 e^{-\frac{k^2 \lambda^2}{2}}} \int_{0}^{\infty} \frac{q \text{d}q }{q^2 v_{\star}^2 + \gamma^2} F (k,q), \label{app:eqn:integral_of_rMat_3}
\end{align}
where we have defined
\begin{align}
    F(k,q) = &2 \pi  \gamma ^4 e^{-\lambda ^2 \left(k^2+q^2\right)}\nonumber \\ 
    +&2 \pi  \alpha _1^4 k^2 q^2 v_*^4 e^{-\frac{1}{2} \lambda _1^2 \left(k^2+q^2\right)}I_0\left(k \lambda _1^2 q\right)\nonumber \\ 
    +&\frac{1}{8} \pi  \alpha _2^4 k^2 q^2 v_*^4 e^{-\frac{1}{2} \lambda _2^2 \left(k^2+q^2\right)} \left(k^2 \lambda _1^2+\lambda _1^2 q^2-4\right){}^2I_0\left(k \lambda _2^2 q\right)\nonumber \\ 
    +&\pi  \alpha _1^2 \alpha _2^2 k^2 q^2 v_*^4 \left(-e^{-\frac{1}{4} \left(\lambda _1^2+\lambda _2^2\right) \left(k^2+q^2\right)}\right) \left(k^2 \lambda _1^2+\lambda _1^2 q^2-4\right)I_0\left(\frac{1}{2} k \left(\lambda _1^2+\lambda _2^2\right) q\right)\nonumber \\ 
    +&4 \pi  \alpha _1^2 \gamma ^2 k q v_*^2 e^{-\frac{1}{4} \left(2 \lambda ^2+\lambda _1^2\right) \left(k^2+q^2\right)}I_1\left(\frac{1}{2} k \lambda _1^2 q\right)\nonumber \\ 
    -&\frac{\pi  \alpha _2^2 \gamma ^2 k q v_*^2 e^{-\frac{1}{4} \left(2 \lambda ^2+\lambda _2^2\right) \left(k^2+q^2\right)} \left(\lambda _1^2 \left(k^2 \lambda _2^2+\lambda _2^2 q^2-4\right)-4 \lambda _2^2\right)}{\lambda _2^2}I_1\left(\frac{1}{2} k \lambda _2^2 q\right)\nonumber \\ 
    -&\frac{\pi  \alpha _2^4 k^3 \lambda _1^2 q^3 v_*^4 e^{-\frac{1}{2} \lambda _2^2 \left(k^2+q^2\right)} \left(\lambda _1^2 \left(k^2 \lambda _2^2+\lambda _2^2 q^2-1\right)-4 \lambda _2^2\right)}{2 \lambda _2^2}I_1\left(k \lambda _2^2 q\right)\nonumber \\ 
    +&2 \pi  \alpha _1^2 \alpha _2^2 k^3 \lambda _1^2 q^3 v_*^4 e^{-\frac{1}{4} \left(\lambda _1^2+\lambda _2^2\right) \left(k^2+q^2\right)}I_1\left(\frac{1}{2} k \left(\lambda _1^2+\lambda _2^2\right) q\right)\nonumber \\ 
    +&2 \pi  \alpha _2^2 \gamma ^2 k^2 \lambda _1^2 q^2 v_*^2 e^{-\frac{1}{4} \left(2 \lambda ^2+\lambda _2^2\right) \left(k^2+q^2\right)}I_2\left(\frac{1}{2} k \lambda _2^2 q\right)\nonumber \\ 
    +&\frac{1}{2} \pi  \alpha _2^4 k^4 \lambda _1^4 q^4 v_*^4 e^{-\frac{1}{2} \lambda _2^2 \left(k^2+q^2\right)}I_2\left(k \lambda _2^2 q\right),
\label{app:eqn:def_of_F_for_ff}
\end{align}
where $I_n \left( z \right)$ is the $n$-th order modified Bessel function of the first kind. For the first term of $F(k,q)$, the integral over $q$ can be directly performed
\begin{equation}
    \int_{0}^{\infty} 2 \pi  \gamma ^4 e^{-\lambda ^2 \left(k^2+q^2\right)} \frac{q \text{d}q }{q^2 v_{\star}^2 + \gamma^2} = \frac{\pi  \gamma ^4 e^{-\lambda ^2 \left(k^2-\frac{\gamma ^2}{v_*^2}\right)} \Gamma \left(0,\frac{\gamma ^2 \lambda ^2}{v_*^2}\right)}{v_*^2}.
\end{equation}
For the other terms, we apply the following strategy. First, we expand the Bessel functions according to 
\begin{equation}
    \label{app:eqn:series_bessel}
    I_n (z) = \sum_{m=0}^{\infty} \frac{1}{m! (m+n)!} \left( \frac{z}{2} \right)^{2m+n}, \qq{for} n \in \mathbb{N},
\end{equation}
which is equivalent to a power series expansion in $q$ of $F(k,q)$. We find it easier to perform this step for each term of $F(k,q)$ separately. Consider the typical term in \cref{app:eqn:def_of_F_for_ff}, which has the form $\mathcal{C}(k) q^{n_1} e^{-\frac{\Lambda_1^2}{2} q^2} I_{n_2} \left( \frac{k q \Lambda_2^2}{2} \right)$, where $\mathcal{C}(k)$ is a function of $k$, $n_1, n_2 \in \mathbb{N}$, while $\Lambda_1$ and $\Lambda_2$ are combinations of $\lambda$, $\lambda_1$, or $\lambda_2$ having dimension of length. For example, the second-to-last term of \cref{app:eqn:def_of_F_for_ff} has this form with
\begin{align}
    \mathcal{C}(k) &= 2 \pi \alpha _2^2 \gamma ^2 k^2 \lambda _1^2 v_*^2 e^{-\frac{1}{4} \left(2 \lambda ^2+\lambda _2^2\right) k^2},  \\
    n_1 = 2,  \quad n_2 &= 2, \quad \Lambda_1 = \sqrt{2\lambda^2 + \lambda^2_2}, \quad \Lambda_2 = \lambda^2_2. 
\end{align}
Such a term can be integrated by first expanding the Bessel function according to \cref{app:eqn:series_bessel} and then integrating the resulting series in a term-by-term fashion
\begin{align}
   &\int_{0}^{\infty} \frac{q \text{d}q }{q^2 v_{\star}^2 + \gamma^2} \mathcal{C}(k) q^{n_1} e^{-\frac{\Lambda_1^2}{2} q^2} I_{n_2} \left( \frac{k q \Lambda_2^2}{2} \right) = \nonumber \\
   =&\frac{1}{v^2_{\star}} \mathcal{C}(k) \abs{\frac{\gamma}{v_{\star}}}^{n_1} \int_{0}^{\infty} \frac{z \text{d}z }{1+z^2}  z^{n_1} e^{-\frac{\Lambda_1^2 \gamma^2}{2 v^2_{\star}} z^2} I_{n_2} \left( k z \Lambda_2^2 \abs{\frac{\gamma}{2 v_{\star}}}  \right) \nonumber \\
   =&\frac{1}{v^2_{\star}} \mathcal{C}(k) \abs{\frac{\gamma}{v_{\star}}}^{n_1} \sum_{m=0}^{\infty} \frac{1}{m! (m+n_2)!} \left( k \Lambda_2^2 \abs{\frac{\gamma}{4 v_{\star}}}  \right)^{2m+n_2} \int_{0}^{\infty} \frac{z^{n_1 + n_2 + 2m +1} \text{d}z }{1+z^2}  e^{-\frac{\Lambda_1^2 \gamma^2}{2 v^2_{\star}} z^2} \nonumber \\
   =&\frac{1}{v^2_{\star}} \mathcal{C}(k) \abs{\frac{\gamma}{v_{\star}}}^{n_1} \sum_{m=0}^{\infty} \frac{\Gamma\left( \frac{n_1 + n_2}{2} + m + 1 \right)}{m! (m+n_2)!} \left( k \Lambda_2^2 \abs{\frac{\gamma}{4 v_{\star}}}  \right)^{2m+n_2} \frac{1}{2} e^{\frac{\Lambda_1^2 \gamma^2}{2 v^2_{\star}}} \Gamma\left(- \frac{n_1 +n_2}{2} - m, \frac{\Lambda_1^2 \gamma^2}{4 v^2_{\star}}\right).  \label{app:eqn:integral_of_rMat_3}
\end{align}
We now note that for all the terms in \cref{app:eqn:def_of_F_for_ff}, $\frac{\Lambda_1^2 \gamma^2}{2 v^2_{\star}} \ll 1$, suggesting that the incomplete Gamma function can be expanded in the second argument. Moreover, we find that for all the terms in \cref{app:eqn:def_of_F_for_ff} proportional to a Bessel function, $n_1+n_2 \geq 2$ is even. As such, we can employ the following representation of the incomplete Gamma function 
\begin{equation}
    \label{app:incomplete_gamma_expansion_1}
    \Gamma (-m,z) = \frac{(-1)^m}{m!} \left[ \Gamma (0,z) - e^{-z} \sum_{j=0}^{m-1} \frac{j! (-1)^j}{z^{j+1}} \right],\qq{for} m \in \mathbb{N}.
\end{equation}
For small $z$, the incomplete Gamma function has the following asymptotic form 
\begin{equation}
    \Gamma (-m,z) = \frac{1}{m z^m} + \begin{cases}
        \mathcal{O} \left( z^{-m + 1} \right), & \quad m \geq 1 \\
        \mathcal{O} \left( \log z \right), & \quad m = 1 \\
    \end{cases} 
\end{equation}
which enables us to rewrite the integral in \cref{app:eqn:integral_of_rMat_3} as
\begin{align}
   &\int_{0}^{\infty} \frac{q \text{d}q }{q^2 v_{\star}^2 + \gamma^2} \mathcal{C}(k) q^{n_1} e^{-\frac{\Lambda_1^2}{2} q^2} I_{n_2} \left( \frac{k q \Lambda_2^2}{2} \right)= \nonumber\\
   = & \frac{1}{v^2_{\star}} \mathcal{C}(k) \abs{\frac{\gamma}{v_{\star}}}^{n_1} \sum_{m=0}^{\infty} \frac{\Gamma\left( \frac{n_1 + n_2}{2} + m + 1 \right)}{m! (m+n_2)!} \left( k \Lambda_2^2 \abs{\frac{\gamma}{4 v_{\star}}}  \right)^{2m+n_2} \nonumber \\
   & \times \left[ \frac{\left( \frac{\Lambda_1^2 \gamma^2}{2 v^2_{\star}} \right)^{- \frac{n_1 +n_2}{2} - m}}{n_1 +n_2 + 2 m} + \mathcal{O} \left( \left( \frac{\Lambda_1^2 \gamma^2}{2 v^2_{\star}} \right)^{- \frac{n_1 +n_2}{2} - m + 1} + \log \left( \frac{\Lambda_1^2 \gamma^2}{2 v^2_{\star}} \right) \right) \right]   \nonumber\\
   = & \frac{1}{v^2_{\star}} \mathcal{C}(k) \abs{\frac{\gamma}{v_{\star}}}^{n_1} \sum_{m=0}^{\infty} \frac{\Gamma\left( \frac{n_1 + n_2}{2} + m + 1 \right)}{m! (m+n_2)!} \left( k \Lambda_2^2 \abs{\frac{\gamma}{4 v_{\star}}}  \right)^{n_2} \left( \frac{k^2 \Lambda_2^2}{2 \sqrt{2}} \right)^{m} \left( \frac{\Lambda_2}{\Lambda_1} \right)^{m} \left( \frac{\Lambda_1^2 \gamma^2}{2 v^2_{\star}} \right)^{m} \nonumber \\
   & \times \left[ \frac{\left( \frac{\Lambda_1^2 \gamma^2}{2 v^2_{\star}} \right)^{- \frac{n_1 +n_2}{2} - m}}{n_1 +n_2 + 2 m} + \mathcal{O} \left( \left( \frac{\Lambda_1^2 \gamma^2}{2 v^2_{\star}} \right)^{- \frac{n_1 +n_2}{2} - m + 1} + \log \left( \frac{\Lambda_1^2 \gamma^2}{2 v^2_{\star}} \right) \right) \right]. \label{app:eqn:integral_of_rMat_4}
\end{align}
Keeping only the leading contribution in the small parameter $\frac{\Lambda_1^2 \gamma^2}{2 v^2_{\star}}$, we can approximate the integral from \cref{app:eqn:integral_of_rMat_3} as 
\begin{align}
   &\int_{0}^{\infty} \frac{q \text{d}q }{q^2 v_{\star}^2 + \gamma^2} \mathcal{C}(k) q^{n_1} e^{-\frac{\Lambda_1^2}{2} q^2} I_{n_2} \left( \frac{k q \Lambda_2^2}{2} \right) \approx \nonumber\\
   & \approx \frac{1}{v^2_{\star}} \mathcal{C}(k) \abs{\frac{\gamma}{v_{\star}}}^{n_1} \sum_{m=0}^{\infty} \frac{\Gamma\left( \frac{n_1 + n_2}{2} + m + 1 \right)}{m! (m+n_2)!} \left( k \Lambda_2^2 \abs{\frac{\gamma}{4 v_{\star}}}  \right)^{n_2} \left( \frac{k^2 \Lambda_2^2}{2\sqrt{2}} \right)^{m} \left( \frac{\Lambda_2}{\Lambda_1} \right)^{m} \frac{\left( \frac{\Lambda_1^2 \gamma^2}{2 v^2_{\star}} \right)^{- \frac{n_1 +n_2}{2}}}{n_1 +n_2 + 2 m}. \label{app:eqn:integral_of_rMat_5}
\end{align}
Finally, the series in \cref{app:eqn:integral_of_rMat_5} can be re-summed into an elementary function. Performing this approximation for each term in \cref{app:eqn:def_of_F_for_ff}, we arrive at the following result
\begin{align}
     &\frac{8 \pi}{U_\xi \xi^2} \left( k^2 v_{\star}^2 + \gamma^2 e^{-\frac{k^2 \lambda^2}{2}} \right) R^{\eta}_{\zeta \zeta} \left( \vk \right) \approx \nonumber \\
     \approx & \frac{\pi  \gamma ^4 e^{-\lambda ^2 \left(k^2-\frac{\gamma ^2}{v_*^2}\right)} \Gamma \left(0,\frac{\gamma ^2 \lambda ^2}{v_*^2}\right)}{v_*^2} \nonumber \\
     +&\frac{\pi  \alpha _2^4 k^2 \lambda _1^4 \left(k^4 \lambda _2^4-8 k^2 \lambda _2^2+8\right) v_*^2}{8 \lambda _2^6}\nonumber \\ 
    -&\frac{\pi  \alpha _2^4 k^2 \lambda _1^2 \left(\lambda _1^2 \lambda _2^2 k^4-4 \left(\lambda _1^2+\lambda _2^2\right) k^2+8\right) v_*^2}{4 \lambda _2^4}\nonumber \\ 
    +&\frac{2 \pi  \alpha _2^2 \gamma ^2 \lambda _1^2 e^{-\frac{1}{4} k^2 \left(2 \lambda ^2+\lambda _2^2\right)} \left(e^{\frac{k^2 \lambda _2^4}{8 \lambda ^2+4 \lambda _2^2}} \left(4 \lambda _2^2 \lambda ^2 \left(k^2 \lambda _2^2-8\right)+\lambda _2^4 \left(k^2 \lambda _2^2-8\right)-32 \lambda ^4\right)+8 \left(2 \lambda ^2+\lambda _2^2\right){}^2\right)}{\lambda _2^4 \left(2 \lambda ^2+\lambda _2^2\right){}^2}\nonumber \\ 
    +&\frac{2 \pi  \alpha _1^2 \alpha _2^2 k^2 \lambda _1^2 \left(k^2 \left(\lambda _1^2+\lambda _2^2\right)-4\right) v_*^2}{\left(\lambda _1^2+\lambda _2^2\right){}^2}\nonumber \\ 
    +&\frac{2 \pi  \alpha _1^4 k^2 v_*^2}{\lambda _1^2}\nonumber \\ 
    +&\frac{8 \pi  \alpha _1^2 \gamma ^2 e^{-\frac{1}{4} k^2 \left(2 \lambda ^2+\lambda _1^2\right)} \left(e^{\frac{k^2 \lambda _1^4}{8 \lambda ^2+4 \lambda _1^2}}-1\right)}{\lambda _1^2}\nonumber \\ 
    +&\frac{\pi  \alpha _2^4 k^2 \left(k^2 \lambda _1^2-4\right){}^2 v_*^2}{8 \lambda _2^2}\nonumber \\ 
    -&\frac{2 \pi  \alpha _2^2 \gamma ^2 e^{-\frac{1}{4} k^2 \left(2 \lambda ^2+\lambda _2^2\right)} \left(e^{\frac{k^2 \lambda _2^4}{8 \lambda ^2+4 \lambda _2^2}}-1\right) \left(\lambda _1^2 \left(k^2 \lambda _2^2-4\right)-4 \lambda _2^2\right)}{\lambda _2^4}\nonumber \\ 
    -&\frac{2 \pi  \alpha _1^2 \alpha _2^2 k^2 \left(k^2 \lambda _1^2-4\right) v_*^2}{\lambda _1^2+\lambda _2^2}. \label{app:eqn:ch_1_exc_complicated}
\end{align}
To simplify the expression in \cref{app:eqn:ch_1_exc_complicated}, we now take the approximation $\lambda_1 \approx \lambda_2 \approx \frac{\lambda}{\sqrt{2}}$, which holds approximately for TBG around the magic angle and realistic interlayer tunneling amplitudes $w_0/w_1 = 0.8$. Coupled with the normalization condition $\alpha_1^2 + \alpha_2^2=1$, this enables us to obtain the following approximate expression for the charge-one excitation matrix
\begin{align}
     &\frac{k^2 v_{\star}^2 + \gamma^2 e^{-\frac{k^2 \lambda^2}{2}}}{U_\xi \xi^2} R^{\eta}_{\zeta \zeta} \left( \vk \right) \approx \nonumber \\
     \approx & \frac{\pi  \gamma ^4 e^{-\lambda ^2 \left(k^2-\frac{\gamma ^2}{v_*^2}\right)} \Gamma \left(0,\frac{\gamma ^2 \lambda ^2}{v_*^2}\right)}{v_*^2} \nonumber \\
     +& \frac{-25 \gamma ^2 e^{-\frac{5}{8} k^2 \lambda ^2} \left(\alpha _1^2 \left(k^2 \lambda ^2+8\right)-k^2 \lambda ^2\right)+8 \gamma ^2 e^{-\frac{3}{5} k^2 \lambda ^2} \left(\alpha _1^2 \left(2 k^2 \lambda ^2+25\right)-2 k^2 \lambda ^2\right)+25 \left(\alpha _1^4+1\right) k^2 v_*^2}{100 \lambda ^2}.  \label{app:eqn:ch_1_simplified}
\end{align}
The simplified expression in \cref{app:eqn:ch_1_simplified} readily explains the shape of the charge-one excitation spectrum around charge neutrality
\begin{equation}
    \frac{8 R^{\eta}_{\zeta \zeta} \left( \vk \right)}{U_{\xi} \xi^2} =  \begin{cases}
       \frac{2   \left(\alpha _1^4+1\right)}{\lambda ^2}, \qq{for} k \to \infty \\
    k^2 \left(\frac{2}{25}  \left(-4 \alpha _1^2+\frac{25 \left(\alpha _1^4+1\right) v_*^2}{\gamma ^2 \lambda ^2}+9\right)-  e^{\frac{\gamma ^2 \lambda ^2}{v_*^2}} \Gamma \left(0,\frac{\gamma ^2 \lambda ^2}{v_*^2}\right)\right)+\frac{  \gamma ^2 e^{\frac{\gamma ^2 \lambda ^2}{v_*^2}} \Gamma \left(0,\frac{\gamma ^2 \lambda ^2}{v_*^2}\right)}{v_*^2}, \qq{for} k \approx 0
    \end{cases}.
\end{equation}

\begin{figure}
    \centering
    \includegraphics[width=0.5\textwidth]{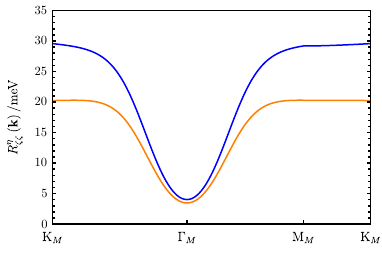}
    \caption{Charge-one excitations of TBG above the $\nu = 0$ correlated insulators computed with the analytical form-factor expression from \cref{app:eqn:approx_for_ff_using_THF}. We take $\theta = \SI{1.05}{\degree}$ and $w_0/w_1 = 0.5$ for which the relevant THF model parameters are given by $\alpha_1=0.9208 $, $\alpha_2 = 0.3901$, $\lambda = 0.533\abs{\vec{a}_{M1}}$, $\lambda_1 = 0.2268 \abs{\vec{a}_{M1}}$, $\lambda_2 = 0.2089 \abs{\vec{a}_{M1}}$, $\abs{\gamma} = \SI{65.78}{\milli\electronvolt}$, and $U_1 = \SI{37.61}{\milli\electronvolt}$ (such that the condition $\abs{\gamma} > U_1$ is satisfied). The blue line show the dispersion computed using the formula from \cref{app:eqn:def_r_+_diag,app:eqn:def_r_-_diag}, while the orange line corresponds to its analytical asymptotic expression from \cref{app:eqn:ch_1_simplified}. }
    \label{app:fig:ch_one_from_rMat}
\end{figure}

We benchmark the analytical expressions for the TBG form factors from \cref{app:eqn:approx_for_ff_using_THF} by computing the charge-one excitation dispersion of TBG. We focus on charge neutrality, where the dispersion is identical for both the electron- and hole-excitations and compute the corresponding charge-one excitation matrices both exactly using \cref{app:eqn:def_r_+_diag,app:eqn:def_r_-_diag} and analytically using the asmyptotic expression derived in \cref{app:eqn:ch_1_simplified}.
The resulting charge-one excitation dispersion is shown in \cref{app:fig:ch_one_from_rMat} and qualitatively agrees with the numerical results of Ref.~\cite{BER21b}.

\subsubsection{Flat-band projected limit of the THF model} \label{app:sec:anal:ch_one:fb_projected}

The THF model has been used in Refs.~\cite{Song20211110MATBGHF,LAU23,CHO23,HU23,YU23a,LI23a,RAI23a,HU23i,ZHO24,WAN24,CAL24} to obtain many of the properties of TBG in the limit $U_1>\abs{\gamma}$. In this section, we discuss the opposite limit -- the flat-band projected limit of the THF model -- which is valid for $U_1<\abs{\gamma}$. Within this parameter regime, the remote bands are effectively gapped degrees of freedom and the interaction THF Hamiltonian can be projected into the active TBG bands. We will show that in the limit $U_1 < \abs{\gamma}$, the  THF model remains a reliable mapping of the TBG problem, which captures the essential physics of the system, even though the interaction cannot be treated locally anymore, due to the heavy mixing between the $c$- and $f$-electrons. 

In what follows, we analytically obtain the charge-$(\pm 1)$ excitation of the system around integer fillings, in the limit of $v_{\star}^\prime = v_{\star}^{\prime\prime} = 0$, $M=0$ and $U_1<\abs{\gamma}$. As already discussed near \cref{app:eqn:anal_of_THF_in_zero_band_limit}, the single-particle Hamiltonian of the system can be solved analytically. For $U_1<\abs{\gamma}$, the remote bands corresponding to $\cre{b}{\vk, j, \eta, s}$ are high-energy degrees of freedom, being separated from the active TBG bands by $\abs{\gamma}$. At integer fillings $-4 \leq \nu \leq +4$, the ground states of the system were shown to be Slater determinants obtained by fully filling the TBG Chern bands~\cite{LIA21}
\begin{equation}
\label{app:eqn:gnd_state_flat_limit}
    |\Psi_\nu \rangle = \prod_{\vk}\prod_{n=1}^{\nu+4}\cre{d}{\vk, \zeta_n ,\eta_n,s_n}|\text{RB}\rangle.
\end{equation}
where $\{\xi_n,\eta_n,s_n\}_n$ are $(\nu + 4)$ flavor indices arbitrarily sorted in $n$ and $|\text{RB}\rangle$ is the Slater determinant state corresponding to the fully occupied remote bands with negative energy 
\begin{equation}
    |\text{RB}\rangle  = \prod_\vk \prod_{\eta,s,\abs{j}=1} \cre{b}{\vk, j,\eta,s}|0\rangle \, . 
\end{equation}

For the given ground state, we can then obtain the charge $\pm 1$ excitation via the Hartree-Fock approximation~\cite{RAI23a,Song20211110MATBGHF}. In the Hartree-Fock approach, we can rewrite the interacting Hamiltonian as
\begin{equation}
	\label{app:eqn:hartree_fock_interaction}
	H_{I,\text{MF}} = \sum_{\substack{i,\eta,s \\ i',\eta',s'}} h^{I,\text{MF}}_{i \eta s; i' \eta' s'} \left( \vk \right) \cre{\gamma}{\vk,i,\eta,s} \des{\gamma}{\vk,i',\eta',s'} ,
\end{equation}
where we have introduced the following operators to simplify the notations
\begin{equation}
\label{app:eqn:shorthand_gamma_not}
\cre{\gamma}{\vk,\eta,i,s} \equiv \begin{cases}
	\cre{c}{\vk,\eta,i,s}, & \qq{for} 1 \leq i \leq 4 \\
	\cre{f}{\vk,\eta,i-4,s}, & \qq{for} 5 \leq i \leq 6
\end{cases}.
\end{equation}
We consider the full THF interacting Hamiltonain from \cref{app:eqn:THF_interaction_TBG}, but follow Ref.~\cite{SON22} and only consider the Hartree contribution stemming from $H_{V}$, but otherwise include both the Hartree and Fock contributions stemming from the other terms. The Hartree-Fock Hamiltonian is then characterized by the following matrix~\cite{RAI23a}
\begin{align}
	h^{I,\text{MF}}_{i \eta s; i' \eta' s'} \left( \vk \right) &= 
	\sum_{\alpha=1}^{2} \left( U_1\nu_f+3U_2\nu_f + \frac{1}{N_0} \sum_{a=1}^{4} W_a \sum_{\vk'} \sum_{\eta'',s''} \varrho_{a \eta'' s''; a \eta'' s''} \left( \vk' \right) \right) \delta_{(\alpha+4) i} \delta_{(\alpha+4) i'} \delta_{\eta \eta'} \delta_{s s'} \nonumber \\ 
	&+ \sum_{a=1}^{4} \left( V(\vec{0}) \nu_c + W_a \nu_f \right) \delta_{a i} \delta_{a i'} \delta_{\eta \eta'} \delta_{s s'}  \nonumber \\
	&- \frac{U}{N_0} \sum_{\alpha,\alpha'=1}^{2} \delta_{(\alpha+4) i} \delta_{(\alpha'+4) i'}\sum_{\vk'}  \varrho_{i' \eta' s'; i \eta s} \left( \vk' \right) \nonumber \\
  & + \frac{U_2}{N_0}\sum_{\vk'}\sum_{n=0}^5\cos\left((\vk-\vk')\cdot C_{6z}^n\vec{a}_{M_1}\right)\varrho_{i'\eta's';i\eta s}(\vk')\nonumber \\ 
	&- \sum_{\alpha=1}^{2} \sum_{a'=1}^{4} \frac{W_{a'}}{N_0} \left[ \delta_{(\alpha+4) i} \delta_{a' i'} + \delta_{(\alpha+4) i'} \delta_{a' i} \right] \sum_{\vk'} \varrho_{i' \eta' s'; i \eta s} \left( \vk' \right) \nonumber \\
	& + \frac{1}{N_0} \sum_{\substack{\iesC{1} \\ \iesC{2}}} \left( \mathcal{J}_{\ies{1};\ies{2}; i \eta s; i' \eta' s'} + \mathcal{J}_{i \eta s; i' \eta' s'; \ies{1};\ies{2}} \right)  \sum_{\vk'}  \varrho_{\ies{1}; \ies{2}} \left( \vk' \right) \nonumber \\
	& - \frac{1}{N_0} \sum_{\substack{\iesC{1} \\ \iesC{2}}} \left( \mathcal{J}_{\ies{1}; i' \eta' s' ; i \eta s; \ies{2}} + \mathcal{J}_{i \eta s; \ies{2}; \ies{1} i' \eta' s'} \right)  \sum_{\vk'}  \varrho_{\ies{1}; \ies{2}} \left( \vk' \right),	\label{app:eqn:hartree_fock_interaction_matrix}
\end{align}
where the interaction tensor is defined as 
\begin{align}
&\mathcal{J}_{\ies{1};\ies{2};\ies{3};\ies{4}} = - \frac{J}{2} \sum_{\substack{\alpha,\alpha'=1 \\ \eta,\eta'}}^{2} \left[ \eta \eta' + \left( -1 \right)^{\alpha+\alpha'} \right] \delta_{(\alpha+4)i_1} \delta_{(\alpha'+4)i_2} \delta_{(\alpha'+2)i_3} \delta_{(\alpha+2)i_4} \delta_{\eta \eta_1} \delta_{\eta' \eta_2} \delta_{\eta' \eta_3} \delta_{\eta \eta_4} \delta_{s_1 s_4}  \delta_{s_2 s_3} \nonumber \\
	-& \frac{J}{4} \sum_{\substack{\alpha,\alpha'=1 \\ \eta,\eta'}}^{2} \left[ \eta \eta' - \left( -1 \right)^{\alpha+\alpha'} \right] \delta_{(\alpha+4)i_1} \delta_{(\alpha'+2)i_2} \delta_{(\alpha'+4)i_3} \delta_{(\alpha+2)i_4} \delta_{\eta \eta_1} \delta_{\eta' \eta_2} \delta_{\eta' \eta_3} \delta_{\eta \eta_4} \delta_{s_1 s_4}  \delta_{s_2 s_3} \nonumber \\
	-& \frac{J}{4} \sum_{\substack{\alpha,\alpha'=1 \\ \eta,\eta'}}^{2} \left[ \eta \eta' - \left( -1 \right)^{\alpha+\alpha'} \right] \delta_{(\alpha+4)i_2} \delta_{(\alpha'+2)i_1} \delta_{(\alpha'+4)i_4} \delta_{(\alpha+2)i_3} \delta_{\eta \eta_1} \delta_{\eta' \eta_2} \delta_{\eta' \eta_3} \delta_{\eta \eta_4} \delta_{s_1 s_4}  \delta_{s_2 s_3}  \nonumber\\ 
  +&\frac{K}{\Omega_0}\sum_{\alpha=1}^2 \sum_{\eta\eta'}\eta\eta' [ \delta_{\bar{\alpha} i_1}\delta_{(\alpha+2) i_2}\delta_{(\alpha+4)i_3}\delta_{(\alpha+2)i_4}\delta_{\eta \eta_1}\delta_{\eta \eta_2}\delta_{\eta'\eta_3}\delta_{\eta'\eta_4}\delta_{s_1s_2}\delta_{s_3s_4}\nonumber\\ 
   -& \delta_{(\alpha+4)i_1}\delta_{(\alpha+2)i_2}\delta_{(\bar{\alpha}+2)i_3}\delta_{\alpha i_4}\delta_{\eta\eta_1}\delta_{\eta\eta_2}\delta_{\eta'\eta_3}\delta_{\eta'\eta_4}\delta_{s_1s_2}\delta_{s_3s_4}] \nonumber \\ 
  +&\frac{K}{\Omega_0}\sum_{\alpha=1}^2 \sum_{\eta\eta'}\eta\eta' [ \delta_{\bar{\alpha} i_4}\delta_{(\alpha+2) i_3}\delta_{(\alpha+4)i_2}\delta_{(\alpha+2)i_1}\delta_{\eta \eta_4}\delta_{\eta \eta_3}\delta_{\eta'\eta_2}\delta_{\eta'\eta_1}\delta_{s_1s_2}\delta_{s_3s_4}\nonumber\\ 
   -& \delta_{(\alpha+4)i_4}\delta_{(\alpha+2)i_3}\delta_{(\bar{\alpha}+2)i_2}\delta_{\alpha i_1}\delta_{\eta\eta_4}\delta_{\eta\eta_3}\delta_{\eta'\eta_2}\delta_{\eta'\eta_1}\delta_{s_1s_2}\delta_{s_3s_4}]
 \label{app:eqn:j_tensor} 
\end{align}
The Hartree-Fock Hamiltonian also depends on the density matrix of the system 
\begin{equation}
	\label{app:eqn:def_rho_HF}
	\varrho_{i \eta s; i' \eta' s'} \left(\vk \right) = \expec{ \normord{\cre{\gamma}{\vk, i, \eta, s} \des{\gamma}{\vk, i' \eta' s'}}}\, ,
\end{equation}
and the filling of the $f$-electrons ($\nu_f$) and $c$-electrons ($\nu_c$)
\begin{equation}
	\nu_c = \frac{1}{N_0} \sum_{i=1}^{4} \sum_{\vk,\eta,\sigma} \varrho_{i \eta \sigma; i \eta \sigma} \left(\vk \right) \qq{and}
	\nu_f = \frac{1}{N_0} \sum_{i=5}^{6} \sum_{\vk,\eta,\sigma} \varrho_{i \eta \sigma; i \eta \sigma} \left(\vk \right)\, .
\end{equation} 
The total filling of the system is $\nu = \nu_f+\nu_c$.

For the Slater-determinant ground states from \cref{app:eqn:gnd_state_flat_limit}, we candirectly compute the density matrix using \cref{app:eqn:def_rho_HF} and then diagonalize the Hartree-Fock Hamiltonian to get the charge-($\pm 1$) excitation spectra. 
Here, we also provide the density matrix for different integer fillings. We find that the density matrix is diagonal in the valley and spin spaces with 
\begin{equation}
    \varrho_{i \eta s; i' \eta' s'} \left(\vk \right) =\delta_{\eta,\eta'}\delta_{s,s'}\varrho_{\eta, s, ij}
\end{equation}
where $\varrho_{\eta,s,ij}$ is the density matrix of valley $\eta$ spin $s$ block. For each block, we have four cases corresponding to four different density matrices. The four cases are: (1) filling zero $\cre{d}{\vk,\zeta,\eta,s}$ electrons ($\varrho^{\text{empty}}_{\eta,s,ij}(\vk)$); (2) filling one $\cre{d}{\vk,+,\eta,s}$ electron for each $\vk$ ($\varrho^{+}_{\eta,s,ij}(\vk)$); (3) filling one $\cre{d}{\vk,-,\eta,s}$ electron for each $\vk$ ($\varrho^{-}_{\eta,s,ij}(\vk)$); (4) filling one $\cre{d}{\vk,+,\eta,s}$ electron and one  $\cre{d}{\vk,-,\eta,s}$ electron for each $\vk$ ($\varrho^{+-}_{\eta,s,ij}(\vk)$). The corresponding density matrices are
\begin{align}
    \varrho_{\eta,s}^{\text{empty}}(\vk) &=\left(
\begin{array}{cccccc}
0 & 0 & -\frac{1}{2} \eta  g_{\vk} e^{-i \eta  \theta_{\vk} } & 0 & -\frac{h_{\vk}}{2} & 0 \\
 0 & 0 & 0 & -\frac{1}{2} \eta  g_{\vk} e^{i \eta  \theta_{\vk} } & 0 & -\frac{h_{\vk}}{2} \\
 -\frac{1}{2} \eta  g_{\vk} e^{i \eta  \theta_{\vk} } & 0 & \frac{g_{\vk}^2-1}{2} & 0 & \frac{1}{2} \eta  g_{\vk} h_{\vk}
   e^{i \eta  \theta_{\vk} } & 0 \\
 0 & -\frac{1}{2} \eta  g_{\vk} e^{-i \eta  \theta_{\vk} } & 0 & \frac{ g_{\vk}^2-1}{2} & 0 & \frac{1}{2} \eta  g_{\vk}
   h_{\vk} e^{-i \eta  \theta_{\vk} } \\
 -\frac{h_{\vk}}{2} & 0 & \frac{1}{2} \eta  g_{\vk} h_{\vk} e^{-i \eta  \theta_{\vk} } & 0 & \frac{h_{\vk}^2-1}{2} & 0 \\
 0 & -\frac{h_{\vk}}{2} & 0 & \frac{1}{2} \eta  g_{\vk} h_{\vk} e^{i \eta  \theta_{\vk} } & 0 & \frac{h_{\vk}^2-1}{2} \\
\end{array}
\right)\\
    \varrho_{\eta,s}^{+}(\vk) &=
\left(
\begin{array}{cccccc}
0 & 0 & -\frac{1}{2} \eta  g_{\vk} e^{-i \eta  \theta_{\vk} } & 0 & -\frac{h_{\vk}}{2} & 0 \\
 0 & 0 & 0 & -\frac{1}{2} \eta  g_{\vk} e^{i \eta  \theta_{\vk} } & 0 & -\frac{h_{\vk}}{2} \\
 -\frac{1}{2} \eta  g_{\vk} e^{i \eta  \theta_{\vk} } & 0 & \frac{1}{2} h_{\vk}^2 & 0 &
   -\frac{1}{2} \eta  g_{\vk} h_{\vk} e^{i \eta  \theta_{\vk} } & 0 \\
 0 & -\frac{1}{2} \eta  g_{\vk} e^{-i \eta  \theta_{\vk} } & 0 & \frac{ g_{\vk}^2-1}{2} & 0 & \frac{1}{2} \eta  g_{\vk}
   h_{\vk} e^{-i \eta  \theta_{\vk} } \\
 -\frac{h_{\vk}}{2} & 0 & -\frac{1}{2} \eta  g_{\vk} h_{\vk} e^{-i \eta  \theta_{\vk} } & 0 &\frac{g_{\vk}^2}{2} &
   0 \\
 0 & -\frac{h_{\vk}}{2} & 0 & \frac{1}{2} \eta  g_{\vk} h_{\vk} e^{i \eta  \theta_{\vk} } & 0 & \frac{h_{\vk}^2-1}{2} \\
\end{array}
\right)\\
    \varrho_{\eta,s}^{-}(\vk) &=\left(
\begin{array}{cccccc}
0 & 0 & -\frac{1}{2} \eta  g_{\vk} e^{-i \eta  \theta_{\vk} } & 0 & -\frac{h_{\vk}}{2} & 0 \\
 0 & 0& 0 & -\frac{1}{2} \eta  g_{\vk} e^{i \eta  \theta_{\vk} } & 0 & -\frac{h_{\vk}}{2} \\
 -\frac{1}{2} \eta  g_{\vk} e^{i \eta  \theta_{\vk} } & 0 & \frac{ g_{\vk}^2-1}{2} & 0 & \frac{1}{2} \eta  g_{\vk} h_{\vk}
   e^{i \eta  \theta_{\vk} } & 0 \\
 0 & -\frac{1}{2} \eta  g_{\vk} e^{-i \eta  \theta_{\vk} } & 0 & \frac{1}{2}  h_{\vk}^2 & 0 &
   -\frac{1}{2} \eta  g_{\vk} h_{\vk} e^{-i \eta  \theta_{\vk} } \\
 -\frac{h_{\vk}}{2} & 0 & \frac{1}{2} \eta  g_{\vk} h_{\vk} e^{-i \eta  \theta_{\vk} } & 0 & \frac{h_{\vk}^2-1}{2} & 0 \\
 0 & -\frac{h_{\vk}}{2} & 0 & -\frac{1}{2} \eta  g_{\vk} h_{\vk} e^{i \eta  \theta_{\vk} } & 0 & \frac{ g_{\vk}^2}{2}
   \\
\end{array}
\right)\\
    \varrho_{\eta,s}^{+-}(\vk) &=\left(
\begin{array}{cccccc}
 0 & 0 & -\frac{1}{2} \eta  g_{\vk} e^{-i \eta  \theta_{\vk} } & 0 & -\frac{h_{\vk}}{2} & 0 \\
 0 & 0 & 0 & -\frac{1}{2} \eta  g_{\vk} e^{i \eta  \theta_{\vk} } & 0 & -\frac{h_{\vk}}{2} \\
 -\frac{1}{2} \eta  g_{\vk} e^{i \eta  \theta_{\vk} } & 0 & \frac{1}{2}  h_{\vk}^2 & 0 &
   -\frac{1}{2} \eta  g_{\vk} h_{\vk} e^{i \eta  \theta_{\vk} } & 0 \\
 0 & -\frac{1}{2} \eta  g_{\vk} e^{-i \eta  \theta_{\vk} } & 0 & \frac{1}{2} h_{\vk}^2& 0 &
   -\frac{1}{2} \eta  g_{\vk} h_{\vk} e^{-i \eta  \theta_{\vk} } \\
 -\frac{h_{\vk}}{2} & 0 & -\frac{1}{2} \eta  g_{\vk} h_{\vk} e^{-i \eta  \theta_{\vk} } & 0 &\frac{1}{2} g_{\vk}^2 &
   0 \\
 0 & -\frac{h_{\vk}}{2} & 0 & -\frac{1}{2} \eta  g_{\vk} h_{\vk} e^{i \eta  \theta_{\vk} } & 0 & \frac{1}{2} g_{\vk}^2
   \\
\end{array}
\right)
\end{align} 
Using the above density matrices, one can directly obtain the charge-($\pm 1$) excitations by diagaonalizing the corresponding Hartree-Fock Hamiltonian obtained using \cref{app:eqn:hartree_fock_interaction_matrix}.

\begin{figure}
    \centering
    \includegraphics[width=\textwidth]{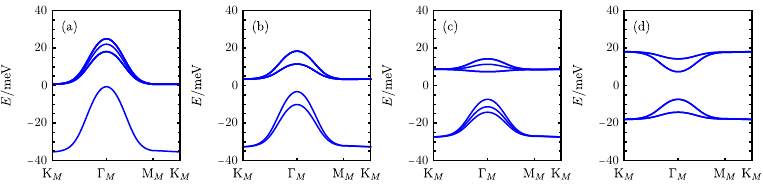}
    \caption{Charge-($\pm 1$) excitations at $w_0/w_1=0.5$ with $\nu=-3$ (a), $\nu=-2$ (b), $\nu=-1$ (c) and $\nu=0$ (d).}
    \label{app:fig:bs_proj_thf}
\end{figure}

In \cref{app:fig:bs_proj_thf}, we show the charge-($\pm 1$) excitation at fillings integer filling $-3 \leq \nu \leq 0$ of THF model for $w_0/w_1=0.5$. For this tunneling ratio, the Hubbard interaction $U_1 = \SI{37.61}{\milli\electronvolt}$ is smaller than the $f$-$c$ hybridization strength $\abs{\gamma} = \SI{65.78}{\milli\electronvolt}$. For the other parameters, we employed~\cite{Song20211110MATBGHF}
\begin{align}
&v_{\star} = \SI{-4.774}{\electronvolt\angstrom},\quad 
v_\star^\prime =0,\quad M=0,\quad \lambda = 0.5330\abs{\vec{a}_{M1}}\nonumber\\ 
&
W_1=W_2= 44.06\text{meV},\quad W_3=W_4= {48.35}\text{meV},\quad J= 23.75\text{meV},\quad K=5.11\text{meV},\quad U_2=3.709\text{meV}
\end{align}
The ground states we considered in~\cref{app:fig:bs_proj_thf} are 
\begin{align}
    &|\Psi_{\nu=-3}\rangle =\prod_{\vk} \cre{d}{\vk,+,+,\uparrow}|\text{RB}\rangle ,\quad \quad |\Psi_{\nu=-2}\rangle =\prod_{\vk} \cre{d}{\vk,+,+,\uparrow} \cre{d}{\vk,-,+,\uparrow}|\text{RB}\rangle, \nonumber\\
   & |\Psi_{\nu=-1}\rangle =\prod_{\vk} \cre{d}{\vk,+,+,\uparrow} \cre{d}{\vk,-,+,\uparrow}\cre{d}{\vk,+,-,\uparrow}|\text{RB}\rangle ,\quad\quad |\Psi_{\nu=0}\rangle =\prod_{\vk} \cre{d}{\vk,+,+,\uparrow} \cre{d}{\vk,-,+,\uparrow}\cre{d}{\vk,+,-,\uparrow}\cre{d}{\vk,-,-,\uparrow}|\text{RB}\rangle. 
\end{align}

\section{Chern bands with localized Berry curvature}\label{app:sec:gap_chern_bands}

In this appendix, we discuss the heavy fermion description of a model with gapped Chern bands with concentrated Berry curvature. Our simple $sp$-orbital model consists of one $s$ orbital and one $p_z$ orbital both located at the $1a$ Wyckoff position of the square lattice. We take the following non-interacting Hamiltonian featuring both inversion symmetry and time-reversal symmetry\cite{bhz_model} 
\begin{align} 
   & H_0 = \sum_{\vk,\sigma } \psi_{\vk,\sigma}^\dagger  h^\sigma_{\vk} \psi_{\vk,\sigma}, \qq{with} \psi_{\vk,\sigma} =\begin{bmatrix}
        \des{f}{\vk, \sigma} & \des{c}{\vk, \sigma} 
    \end{bmatrix}^T \nonumber\\ 
   & h_{\vk}^\sigma = \begin{bmatrix}
       \epsilon_{s} &  t_{sp}\left[\sigma\sin(k_y) -i\sin(k_x)\right] \\ 
       t_{sp}\left[\sigma\sin(k_y) +i\sin(k_x)\right] & \epsilon_p - t_{p}\left[\cos(k_x)+\cos(k_y)\right]
   \end{bmatrix},
   \label{app:eq:sp_signle_particle_ham}
\end{align}
where $\cre{f}{\vk,\sigma}$ ($\cre{c}{\vk,\sigma}$) creates a $s$($p_z$)-orbital electron at momentum $\vk$ with spin $\sigma$. $\epsilon_s, \epsilon_p$ denote the onsite energy of the $s$ and $p$ orbitals, respectively. At the same time, $t_{sp}$ is the hopping between two orbitals and $t_p$ denotes the hopping of between the $p_z$ orbitals. As a result of the time-reversal symmetry, the spin-$\uparrow$ and spin-$\downarrow$ bands are degenerate in this model.  

\begin{figure}
    \centering
    \includegraphics{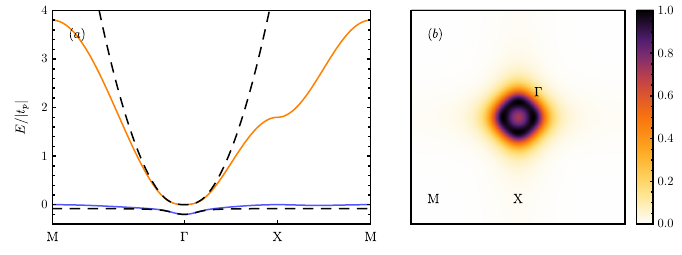}
    \caption{Heavy-fermion model for Chern bands with localized Berry curvature. (a) shows the dispersion of the $sp$-orbital model defined in \cref{app:eq:sp_signle_particle_ham} for $t_p = -1$, $t_sp=0.1$, $\epsilon_s = 0$, and $\epsilon_p =1.8$.  The band structure features a doubly-degenerate narrow Chern band (blue) and a doubly-degenerate dispersive band (orange). The dashed lines correspond to the band structure of the corresponding approximate Hamiltonian from \cref{app:eq:sp_signle_particle_ham_near_Gamma}. The berry curvature distributions of the narrow Chern band in arbitrary units is plotted in (b). }
    \label{app:fig:chern_band_THF}
\end{figure}

We consider the situation where $|t_p| \gg |t_{sp}|$, $\epsilon_p = 2t_p - \Delta\epsilon$ with $|\Delta\epsilon| \ll |t_p|$, and $\epsilon_s =0 $. The resulting band structure is shown in \cref{app:fig:chern_band_THF}(a) for $\Delta\epsilon/t_p = -0.2$ and $t_{sp}/t_p = 0.1$ and features a flat band and a dispersive band, both being doubly-degenerate. 

We now discuss the band structure of the system. In the limit $t_{sp} \to 0$, the $p$ orbitals develop a dispersive band with the band minimum at the $\Gamma$ point and dispersion 
\begin{equation}
    \epsilon_{p,\vk} = t_p\left[2-\cos(k_x)-\cos(k_y)\right] -\Delta\epsilon  \approx -\Delta \epsilon + \frac{t_p \abs{\vk}^2}{2}
\end{equation}
In this limit, the $s$ orbitals form a perfectly flat trivial band. After introducing a non-zero $t_{sp}$, the atomic $s$ orbitals and the dispersive $p$ orbitals hybridize near the $\Gamma$ point and a band inversion happens therein. Performing a leading-order expansion in $\vk$ of the $p$-orbital dispersion and of the $s$-$p$ orbital hybridization, the single-particle Hamiltonian from \cref{app:eq:sp_signle_particle_ham} can be approximated to 
\begin{equation}
    H_0  \approx \sum_{|\vk|<\Lambda_c,\sigma} \left(-\Delta\epsilon + \frac{t_p|\vk|^2}{2} \right) \cre{c}{\vk,\sigma}\des{c}{\vk,\sigma} +\frac{t_{sp}}{\sqrt{N}} \sum_{|\vk|<\Lambda_c,\vec{R},\sigma}\left[e^{-i\vk\cdot\vec{R}}(\sigma k_y+ik_x)\cre{c}{\vk,\sigma} \des{f}{\vec{R},\sigma}+\text{h.c.}\right] \, .
    \label{app:eq:sp_signle_particle_ham_near_Gamma}
\end{equation}
where $N$ is the number of unit cells. We have also introduced a momentum cutoff $\Lambda_c$ for the $c$ electrons, since the$c$ electron with large momenta are high-energy degrees of freedom and can be dropped. In practice, we can set the cutoff to the edge of the BZ. 
From \cref{app:eq:sp_signle_particle_ham_near_Gamma}, it becomes evident that the system is equivalent to a heavy-fermion model, characterized by the hybridization between atomic $f$-electrons ($s$ orbitals) and dispersive $c$ electrons ($p$ orbitals).

The eigenvalues of the single-particle Hamiltonian near the $\Gamma$ point are 
\begin{equation} 
    E^\sigma_{\vk, 1} \approx -\Delta\epsilon+ \frac{t_p \Delta\epsilon+ 2t_{sp}^2}{2\Delta\epsilon} \abs{\vk}^2,\quad E_{\vk,2} \approx  \frac{t_{sp}^2}{\Delta\epsilon} \abs{\vk}^2, 
\end{equation}
with the corresponding eigenvectors being given by
\begin{align}
    &U^\sigma_{\vk, 1} \approx \frac{1}{
    \sqrt{ 
    2\sqrt{\frac{s_{\vk}^2}{4} +t_{sp}^2\abs{\vk}^2 }
    \left[ \sqrt{\frac{s_{\vk}^2}{4} +t_{sp}^2\abs{\vk}^2 }-\frac{s_{\vk}}{2}
    \right]
    }
    }\begin{bmatrix}
        \frac{s_{\vk}}{2} -\sqrt{ \frac{s_{\vk}^2}{4} +t_{sp^2}\abs{\vk}^2}  & t_{sp}\left(ik_x+\sigma k_y \right)
    \end{bmatrix} \nonumber\\ 
   & U^\sigma_{\vk, 2} \approx\frac{1}{
    \sqrt{ 
    2\sqrt{\frac{s_{\vk}^2}{4} +t_{sp}^2\abs{\vk}^2 }
    \left[ \sqrt{\frac{s_{\vk}^2}{4} +t_{sp}^2\abs{\vk}^2 }+\frac{s_{\vk}}{2}
    \right]
    }}\begin{bmatrix}
        \frac{s_{\vk}}{2} +\sqrt{ \frac{s_{\vk}^2}{4} +t_{sp^2}\abs{\vk}^2}  & t_{sp}\left(ik_x+\sigma k_y \right)
    \end{bmatrix} 
    \label{app:eq:spmodel_eigenfun_Gamma}
\end{align}
where $s_{\vk} =-\Delta \epsilon + t_p |\vk|^2/2 $. It is easy to see that at the $\Gamma$ point, the lowest bands are formed by the $p_z$ orbital with $U_{\vk=\vec{0},1}^\sigma = [0,1]$. As we move away $\Gamma$ point, the lowest bands acquire more and more $s$-orbital weight. At large momenta $\abs{\vk}$, the $p_z$ orbitals form high-energy bands, and the lowest-energy bands are mostly formed by the $s$ orbitals.

We can also estimate the bandwidths of the bands. We first consider the lowest-energy bands. At the high-symmetry points, we have
\begin{equation}
    E_{\vec{0}, 1}^\sigma = -\Delta \epsilon ,\quad E_{(0,\pi),1}^\sigma = E_{( \pi,0),1}^\sigma \approx 0,\quad E_{( \pi, \pi), 1}^\sigma \approx 0
\end{equation}
where, away from $\Gamma$ point, the lowest-energy bands are mostly formed by $s$ orbitals with energy $\epsilon_s = 0$. Therefore the bandwidth of the lowest-energy bands are $\sim |\Delta\epsilon| $. For the high-energy bands, the energies at high-symmetry points are
\begin{equation} 
    E_{\vec{0},2}^\sigma = 0,\quad E_{(0, \pi),2} = E_{( \pi,0),2} \approx t_p,\quad vE_{( \pi,  \pi),2} \approx  2t_p
\end{equation}
where away from $\Gamma$ point, the high-energy bands are mostly formed by $p_z$ orbitals with energy $-\Delta \epsilon +t_p[2-\cos(k_x)-\cos(k_y)] \approx t_p[2-\cos(k_x)-\cos(k_y)] $. Therefore, the high-energy bands have bandwidth $\sim 2|t_p|$. Thus, our current model comprises a two-fold degenerate narrow band with bandwidth $\sim |\Delta \epsilon|$ and a two-fold degenerate dispersive band with bandwidth $\sim 2|t_p|$. 

We now show that the two-fold degenerate low-energy narrow bands have Chern number $\pm 1$ for spin-$\uparrow/\downarrow $. Since the system has inversion symmetry, it is sufficient to check the inversion eigenvalues of the lowest-energy band at high symmetry points $(0,0), (0, \pi), (\pi,0), (\pi,\pi)$. At the $\Gamma=(0,0)$ point, the narrow bands are formed by $p_z$ orbitals and have $-1$ inversion eigenvalues. At the $(0,\pi),(\pi,0),(\pi,\pi)$ points, the narrow bands are formed by $s$ orbitals with $+1$ inversion eigenvalue. Therefore, the two-fold degenerate narrow bands have Chern number $\pm 1$ due to the band inversion at the $\Gamma$ point.

Since the band inversion happens near the $\Gamma$ point, we also analyze the Berry curvature near $\Gamma$ point. We can calculate the Berry curvature of the lowest bands near $\Gamma$ point using~\cref{app:eq:spmodel_eigenfun_Gamma}, which gives
\begin{equation}
    \Omega(\vk) \approx \frac{ 8(2\Delta \epsilon + \abs{\vk}^2 t_p) t_{sp}^2}{\left[\left(-2 \Delta \epsilon + t_p^2 \abs{\vk}^2\right)^2+16 \abs{\vk}^2 t_{sp}^2 \right]^{3/2}}
\end{equation}
We can observe that, at large momentum the Berry curvature decays as $\frac{t_{sp}^2}{t_p^2\abs{\vk}}$. Since $|t_{sp}/t_p| \ll  1 $, the Berry curvature quickly goes to zero at large momenta. In~\cref{app:fig:chern_band_THF}(b), we also show the numerically computed Berry curvature distribution, where we observe the Berry curvature concentrates near $\Gamma$ point. 

We can also introduce interactions to our current model. Here, we consider onsite Hubbard interactions for the $s$ ($U$) and $p$ ($V$) orbitals, as well as onsite repulsion between the two orbitals ($W$). The interaction term can be written as
\begin{align}
    H_I = &\sum_{\vec{R}}\frac{U}{2} \left( \sum_\sigma :\cre{f}{\vec{R},\sigma}\des{f}{\vec{R},\sigma}:\right)^2+\sum_{\vec{R},|\vec{k}|<\Lambda_c,|\vec{k}+\vec{q}|<\Lambda_c} \frac{We^{i\vec{q}\cdot\vec{R}}}{{N}}\left( \sum_\sigma :\cre{f}{\vec{R},\sigma}\des{f}{\vec{R},\sigma}:\right) 
    \left(\sum_{\sigma'}:\cre{c}{\vk,\sigma'}\des{c}{\vk+\vq,\sigma'}:\right)
      \nonumber\\ 
    &+ \sum_{|\vec{k}|<\Lambda_c,|\vec{k}+\vec{q}|<\Lambda_c,|\vec{k'}|<\Lambda_c,|\vec{k'}-\vec{q}|<\Lambda_c}\frac{V}{2N}\left( :\sum_\sigma \cre{c}{\vec{k},\sigma}\des{c}{\vec{k}+\vec{q},\sigma} :\right)\left(
    :\sum_{\sigma'} \cre{c}{\vec{k'},\sigma'}\des{c}{\vec{k'}-\vec{q},\sigma'} :
    \right)
\end{align}
which takes a similar form to the interaction term of the THF model of TBG from \cref{app:eqn:THF_interaction_TBG}. 
For a given operator $\hat{O}$, we have also defined normal ordering as $:\hat{O}:=\hat{O}-\langle \Phi_0|\hat{O}|\Phi_0\rangle$ with $|\Phi_0\rangle$ the ground state of the non-interacting Hamiltonian. Finally, we mention that one could also consider the spinless case, where the only interaction term is the term proportional to $W$. 

In summary, our current model yields two-fold degenerate narrow Chern bands near the Fermi energy, exhibiting concentrated Berry curvature near the $\Gamma$ point. By including both the narrow Chern bands and the dispersive bands, the entire system can be effectively described by a heavy fermion model, as depicted by the single-particle Hamiltonian in~\cref{app:eq:sp_signle_particle_ham} (or~\cref{app:eq:sp_signle_particle_ham_near_Gamma}). Here, the $s$ orbitals correspond to localized $f$-electrons, while the $p_z$ orbitals represent dispersive $c$-electrons. The flatness of the narrow bands arises from the localized nature of $s$ orbitals ($f$-fermions), while the non-trivial topology emerges from the hybridization between $s$ and $p_z$ orbitals. Furthermore, after introducing interaction terms to the $s$ and $p$ orbitals, we could obtain a conventional interacting heavy fermion model. Thus we conclude that, in the case of flat Chern bands with concentrated Berry curvature, a heavy fermion description is still feasible. 

\section{Chiral-Symmetry-Protected Stable Anomaly}
\label{app:chiral_symm_anomaly}

We now present the general discussion for chiral-symmetry-protected stable anomaly.
Given a generic 2D single-particle matrix Hamiltonian $h(\bsl{k})$ with chiral symmetry $\mathcal{C}$, \ie, $\{\mathcal{C},h(\bsl{k})\}=0 $ and $\mathcal{C}^2 = 1$.
Suppose 
\eq{
h(\bsl{k}+\bsl{G}) = V_{\bsl{G}}^\dagger h(\bsl{k}) V_{\bsl{G}}
}
and
\eq{
\label{eq:embedding_commutes_with_chiral}
\left[V_{\bsl{G}}, \C \right] = 0 \ .
}
Let us focus on an isolated set of $2N$ bands whose energies are symmetric with respect to the zero energy. 
The eigenvectors ($U_{n,\bsl{k}}$ with $n=1,...,2N$) of the isolated set of bands are compatible with $\mathcal{C}$, \ie, the eigenvectors of the isolated set of bands furnish a rep of $\mathcal{C}$: $\mathcal{C} U_{n,\bsl{k}} = \sum_{n'} U_{n',\bsl{k}} \left[ D_{\mathcal{C}}(\bsl{k})\right]_{n'n}$ with $D_{\mathcal{C}}(\bsl{k})$ unitary for any $n$ and $\bsl{k}$.
Equivalently, the projection matrix $P(\bsl{k})$ of the isolated set of 2N bands commutes with $\mathcal{C}$:
\eq{
\label{eq:chiral_commuting_projection_matrix}
[\mathcal{C} , P(\bsl{k}) ] = 0 \ .
}

We can diagonalize $D_{\mathcal{C}}(\bsl{k})$, which gives a ``$+1$" subspace and a ``$-1$" subspace, each of which has a dimension $N$.
Specifically, we denote 
\eq{
D_{\mathcal{C}}(\bsl{k}) v_{\pm,i,\bsl{k}} = \pm v_{\pm,i,\bsl{k}} \text{ for i=1,...,N}\ ,
}
and then we have 
\eq{
U_{\pm,i,\bsl{k}} = \sum_{n} U_{n,\bsl{k}} \left[v_{\pm,i,\bsl{k}}\right]_n \ ,
}
where
\eq{
\mathcal{C} U_{\pm,i,\bsl{k}} = \pm U_{\pm,i,\bsl{k}}\ .
}
The projection matrix of the ``$ \pm 1 $" subspace of the isolated set of $2N$ bands reads
\eq{
P_{\pm}(\bsl{k}) = \sum_{i=1,...,N}  U_{\pm,i,\bsl{k}} U_{\pm,i,\bsl{k}}^\dagger\ .
}

Owing to \eqnref{eq:embedding_commutes_with_chiral}, $P_{\pm}(\bsl{k}+\bsl{G}) = V_{\bsl{G}}^\dagger P_{\pm}(\bsl{k}) V_{\bsl{G}}$, which has the same behavior as $P(\bsl{k})$.
Combined with the fact that they are projection matrices of the smooth $P(\bsl{k})$ for chiral eigenvalues $+$ and $-$, respectively, both $P_{+}(\bsl{k}) $ and $P_{-}(\bsl{k})$ are globally smooth in the momentum space.
Therefore, the Berry curvature of each subspace reads
\eq{
F_{\pm}(\bsl{k}) = \Tr\left[ P_{+}(\bsl{k}) \left[\partial_{k_x} P_{+}(\bsl{k}) , \partial_{k_y} P_{+}(\bsl{k}) \right] \right]
}
which is smooth with $F_{\pm}(\bsl{k}+\bsl{G})=F_{\pm}(\bsl{k})$, leading to well-defined Chern numbers for $P_{\pm}(\bsl{k})$:
\eq{
\Ch_\pm = \frac{1}{2\pi}\int d^2 k\ F_{\pm}(\bsl{k}) \ .
}
The total Chern number of the entire isolated set of $2N$ bands is
\eq{
\Ch = \Ch_+ + \Ch_-\ .
}
Note that \eqnref{eq:embedding_commutes_with_chiral} makes sure that $\Ch_{\pm}$ are well defined.
If \eqnref{eq:embedding_commutes_with_chiral} is violated by having $\{ \C , V(\bsl{b}_1) \} = 0$ with $\bsl{b}_1$ one primitive reciprocal lattice vector, we have $\C P_{+}(\bsl{k}-\bsl{b}_1) = \C V_{\bsl{b}_1} P_+(\bsl{k}) V_{\bsl{b}_1}^\dagger = -  P_{+}(\bsl{k}-\bsl{b}_1)$ and thus $F_{+}(\bsl{k}+\bsl{b}_1)=F_{-}(\bsl{k})$, which cannot guarantee well-defined $\Ch_{\pm}$.

If $\Ch \neq 0$, it is impossible to construct exponentially-localized Wannier functions for them, which is the classic obstruction from the total Chern number~\cite{Brouder2007Wannier}.
Nevertheless, if $\Ch_{\mathcal{C}} = \Ch_+ - \Ch_- \neq 0$, it is also impossible to construct exponentially-localized Wannier functions for them while keeping the rep of $\mathcal{C}$ strictly-onsite in the real space or momentum-independent in the momentum space, even if $\Ch=0$.
To see this, let us assume we can rotate $U_{n,\bsl{k}}$ to a set of globally smooth vectors $\zeta_{a,\bsl{k}}$ with $a=1,2,...,2N$, \ie,
\eq{
\zeta_{a,\bsl{k}} = \sum_{n=1}^{2N} U_{n,\bsl{k}} R_{na}(\bsl{k})
}
and all $\zeta_{a,\bsl{k}}$ are globally smooth while $V_{\bsl{G}}^\dagger\zeta_{a,\bsl{k}} = \zeta_{a,\bsl{k}+\bsl{G}}$.
We further assume the rep of $\C$ furnished by $\zeta_{a,\bsl{k}}$ is independent of $\bsl{k}$ (which means that the rep of $\mathcal{C}$ is strictly-onsite in the real space), \ie,
\eq{
\C \zeta_{a,\bsl{k}} = \sum_{a'} \zeta_{a',\bsl{k}} \left[ \widetilde{D}_{\C} \right]_{a' a}\ .
}
Then, we can diagonalize $\widetilde{D}_{\C}$: 
\eq{
\widetilde{D}_{\mathcal{C}} \widetilde{v}_{\pm,i} = \pm \widetilde{v}_{\pm,i} \text{ for } i=1,...,N\ ,
}
and have
\eqa{
\zeta_{\pm,i,\bsl{k}} = \sum_{a} \zeta_{a,\bsl{k}} \left[ \widetilde{v}_{\pm,i} \right]_{a}\ .
}
Since $\zeta_{\pm,i,\bsl{k}}$ are globally smooth and $V_{\bsl{G}}^\dagger\zeta_{\pm,i,\bsl{k}} = \zeta_{\pm,i,\bsl{k}+\bsl{G}}$, we have $\Ch_+ = \Ch_- =0$.
As globally smooth $\zeta_{a,\bsl{k}}$ give exponentially localized Wannier functions, we know nonzero $\Ch_{\mathcal{C}} = \Ch_+ - \Ch_- \neq 0$ forbids exponentially-localized Wannier functions with the rep of $\mathcal{C}$ being strictly-onsite in the real space or momentum-independent in the momentum space, even if $\Ch=0$.

Now we present an alternative way to evaluate $\Ch_+ - \Ch_-$.
Suppose $h(\bsl{k})$ has several isolated band-touching points at zero energy, labeled as $\bsl{k}_l$.
Then, for each node $\bsl{k}_l$, we can define the chiral winding number $W_l$ as 
\eq{
W_l = \frac{\ii}{2\pi } \int_{\partial D_l } d\bsl{k}\cdot \Tr\left[ Q^{-1}_{\bsl{k}} \nabla_{\bsl{k}} Q_{\bsl{k}} \right]\ ,
}
where
\eq{
\left[ Q_{\bsl{k}} \right]_{i i'} = U_{+,i,\bsl{k}}^\dagger h(\bsl{k}) U_{-,i',\bsl{k}} \ ,
}
$D_l $ is a disk-like region that only includes $\bsl{k}_l$, and $\partial D_l $ is the boundary (with direction) of $D_l$.
We now prove 
\eq{
\label{eq:winding_ch_chiral}
\sum_l W_l = \Ch_{\C} = \Ch_+ - \Ch_-\ .
}
The proof is analugous to the proof for monopole Cooper pairing in \refcite{Li2018WSMObstructedPairing}.
As the first step of the proof, we define 
\eqa{
\label{eq:v_k_chiral}
\bsl{v}_{\bsl{k}} & = \Tr\left[ \left( U_{-,\bsl{k}} Q_{\bsl{k}}^{-1} U_{+,\bsl{k}}^\dagger \right) \ii \nabla_{\bsl{k}}  \left( U_{+,\bsl{k}} Q_{\bsl{k}} U_{-,\bsl{k}}^\dagger \right) \right] \\
& = \Tr\left[  U_{+,\bsl{k}}^\dagger \ii \nabla_{\bsl{k}} U_{+,\bsl{k}}\right] + \Tr\left[ Q_{\bsl{k}}^{-1}\ii \nabla_{\bsl{k}}  Q_{\bsl{k}}  \right] - \Tr\left[ U_{-,\bsl{k}}^\dagger \ii \nabla_{\bsl{k}}   U_{-,\bsl{k}}  \right]\\
& = \Tr\left[ Q_{\bsl{k}}^{-1}\ii \nabla_{\bsl{k}}  Q_{\bsl{k}}  \right] -  \bsl{A}_+(\bsl{k})  +  \bsl{A}_-(\bsl{k})  \ ,
}
where
\eq{
U_{\pm,\bsl{k}} = \mat{ U_{\pm,1,\bsl{k}} & U_{\pm,2,\bsl{k}} & ... & U_{\pm,N,\bsl{k}}   } \ ,
}
and 
\eq{
 \bsl{A}_\pm (\bsl{k}) = -\ii \Tr\left[  U_{\pm,\bsl{k}}^\dagger \nabla_{\bsl{k}} U_{\pm,\bsl{k}}\right] \ .
}
$\bsl{v}_{\bsl{k}}$ is smooth except at $\bsl{k}_l$'s, since both  $U_{+,\bsl{k}}^\dagger U_{+,\bsl{k}}$ and $U_{-,\bsl{k}}^\dagger U_{-,\bsl{k}}$ are smooth, and $Q_{\bsl{k}}^{-1} = U_{-,\bsl{k}}^\dagger h^{-1}(\bsl{k}) U_{+,\bsl{k}} $.
Moreover, $\bsl{v}_{\bsl{k}+\bsl{G}} = \bsl{v}_{\bsl{k}}$.
The next step is to choose $D_l$ such that (i) two $D_l$'s either have no intersection or intersect only on the boundaries, and (ii) $\cup_l D_l = 1BZ$.
Then, we have
\eq{
\frac{1}{2\pi } \sum_l \int_{\partial D_l } d\bsl{k} \cdot \bsl{v}_{\bsl{k}} = 0\ ,
} 
since the integration always cancels along a line shared by two neighboring $D_l$'s.
Combined with \eqnref{eq:v_k_chiral}, we have
\eqa{
& \frac{1}{2\pi } \sum_l \int_{\partial D_l } d\bsl{k} \cdot \left[ \Tr\left[ Q_{\bsl{k}}^{-1}\ii \nabla_{\bsl{k}}  Q_{\bsl{k}}  \right] -  \bsl{A}_+(\bsl{k})  +  \bsl{A}_-(\bsl{k}) \right] \\
& = \sum_l W_l - \frac{1}{2\pi } \sum_l \int_{\partial D_l } d\bsl{k} \cdot \bsl{A}_+(\bsl{k}) + \frac{1}{2\pi } \sum_l \int_{\partial D_l } d\bsl{k} \cdot\bsl{A}_-(\bsl{k}) \\
& =  \sum_l W_l - \frac{1}{2\pi } \sum_l \int_{  D_l } dk^2 F_+(\bsl{k}) + \frac{1}{2\pi } \sum_l \int_{  D_l } d k^2 F_-(\bsl{k}) \\
& = \sum_l W_l - \Ch_+ + \Ch_- = 0\ ,
}
leading to \eqnref{eq:winding_ch_chiral}.
In the derivation, we have used $F_\pm(\bsl{k}) = \nabla_{\bsl{k}}\times\bsl{A}_\pm(\bsl{k})$, and used the fact that we can always choose $
U_{\pm,\bsl{k}}$ to be smooth within one single $D_l$ (as it is disk-like) to use Stoke's theorem for the second equality. 
We note that $U_{\pm,\bsl{k}}$ cannot be made globally smooth if $\Ch_\pm$ are nonzero.

Based on \eqnref{eq:winding_ch_chiral}, we know that as long as $\sum_l W_l$ is nonzero, any isolated set of $2N$ chiral-symmetric bands have chiral-protected Wannier obstruction, leading to a stable chiral-protected anomaly.

\section{Twisted Checkerboard Model}
\label{app:twisted_chekerboard}

This section contained more details on the twisted checkerboard model in \refcite{Yao2022TwistedCheckorboard}, and the construction of the heavy femrion model of it.
We specify that \refcite{KaiSun2023THF} studied the heavy fermion framework for a special quadratic band touching model clearly related to the twisted checkerbaord model. We have here complemented their analysis by uncovering the stable chiral-symmetric anomaly and obtaining the interaction Hamiltonian.

The twisted checkerboard model in \refcite{Yao2022TwistedCheckorboard} is generated by stacking two checkerboard lattices with relative twisted angle $\theta$.
For each checkerboard lattice, the primitive lattice vectors read
\eqa{
& \bsl{a}_1 = (1,0) \\
& \bsl{a}_2 = (0,1) \ ,
}
where we have chosen the lattice constant to be 1, and the primitive reciprocal lattice vectors are
\eqa{
& \bsl{b}_1 = (2\pi,0) \\
& \bsl{b}_2 = (0,2\pi) \ .
}
The Hamiltonian reads
\eq{
H_{CBL} = \sum_{\bsl{k}} \mat{ c^{\dagger}_{\bsl{k},A} & c^{\dagger}_{\bsl{k},B}} h(\bsl{k}) \otimes s_0 \mat{ c_{\bsl{k},A} \\ c_{\bsl{k},B}}\ ,
}
where
\eq{
h(\bsl{k}) =  - 4 t \cos(k_x/2) \cos(k_y/2) \sigma_x + 2 t' (\cos(k_x)-\cos(k_y)) \sigma_z\ ,
}
\eq{
c^{\dagger}_{\bsl{k},A/B} = \left( c^{\dagger}_{\bsl{k},A/B,\uparrow}, c^{\dagger}_{\bsl{k},A/B,\downarrow} \right)
}
with $\sigma_{0,x,y,z}$ are identity and Pauli matrices for the sublattice index, and $s_{0,x,y,z}$ are identity and Pauli matrices for the spin index.
Furthermore, for any reciprocal lattice vector $\bsl{G}$, we have
\eqa{
& c^{\dagger}_{\bsl{k}+\bsl{G},A} = c^{\dagger}_{\bsl{k},A} \\
& c^{\dagger}_{\bsl{k}+\bsl{G},B} = c^{\dagger}_{\bsl{k},B} e^{-\ii \bsl{G}\cdot (\bsl{a}_1+\bsl{a}_2)/2}\\
}

With a basis transformation
\eq{
\mat{ \widetilde{c}^{\dagger}_{\bsl{k}, + } & \widetilde{c}^{\dagger}_{\bsl{k},-}} = \mat{ c^{\dagger}_{\bsl{k},A} & c^{\dagger}_{\bsl{k},B}} \frac{1}{\sqrt{2}}\mat{ 1 & -\ii \\ -\ii & 1}\otimes s_0\ ,
}
we get 
\eq{
\label{eq:CBL_model}
H_{CBL} = \sum_{\bsl{k}} \mat{ \widetilde{c}^{\dagger}_{\bsl{k},+} & \widetilde{c}^{\dagger}_{\bsl{k},-}} \widetilde{h}(\bsl{k}) \otimes s_0 \mat{ \widetilde{c}_{\bsl{k},+} \\ \widetilde{c}_{\bsl{k},-}}\ ,
}
where
\eq{
\label{eq:CBL_htilde}
\widetilde{h}(\bsl{k}) =  4 t \cos(k_x/2) \cos(k_y/2) \sigma_x + 2 t' (\cos(k_x)-\cos(k_y)) \sigma_y\ .
}

The symmetry group of the checkerboard lattice model in \eqnref{eq:CBL_model} is spanned by spin-charge $\U(2)$, spinless $C_{4z}$, spinless $C_{2x}$, $\TR$ and lattice translations, where representations (reps) of $C_{4z}$, $C_{2x}$ and $\TR$ are
\eqa{
& C_{4z}  \widetilde{c}^{\dagger}_{\bsl{k}} C_{4z}^{-1} =  \widetilde{c}^{\dagger}_{C_{4z}\bsl{k}} \sigma_x s_0 \\
& C_{2x}  \widetilde{c}^{\dagger}_{\bsl{k}} C_{2x}^{-1} =   \widetilde{c}^{\dagger}_{C_{2x}\bsl{k}} \sigma_0 s_0\\
& \TR \widetilde{c}^{\dagger}_{\bsl{k}} \TR^{-1} =  \widetilde{c}^{\dagger}_{-\bsl{k}} (-\sigma_y) \ii s_y\ .
}
At $\bsl{k}=(\pi,\pi)=$M, the basis furnishes a 2D irrep of $C_{4z}$ and $C_{2x}$ symmetries:
\eqa{
& C_{4z}  \widetilde{c}^{\dagger}_{\M} C_{4z}^{-1} =  \widetilde{c}^{\dagger}_{M-\bsl{b}_1} \sigma_x s_0  =  c^{\dagger}_{M} \sigma_z  \frac{1}{\sqrt{2}}\mat{ 1 & -\ii \\ -\ii & 1} \sigma_x s_0=  \widetilde{c}^{\dagger}_{\M}\frac{1}{2}  \mat{ 1 & \ii \\ \ii & 1} \sigma_z  \mat{ 1 & -\ii \\ -\ii & 1} \sigma_x s_0  = \widetilde{c}^{\dagger}_{\M}(-\ii) \sigma_z s_0 \\
& C_{2x}  \widetilde{c}^{\dagger}_{\M} C_{2x}^{-1} =  \widetilde{c}^{\dagger}_{M-\bsl{b}_2} \sigma_0 s_0  =  c^{\dagger}_{M} \sigma_z  \frac{1}{\sqrt{2}}\mat{ 1 & -\ii \\ -\ii & 1} \sigma_0 s_0=  \widetilde{c}^{\dagger}_{\M}\frac{1}{2}  \mat{ 1 & \ii \\ \ii & 1} \sigma_z  \mat{ 1 & -\ii \\ -\ii & 1} \sigma_0 s_0  = \widetilde{c}^{\dagger}_{\M}\sigma_y s_0 \ .
}
Therefore, $\widetilde{h}(\M)$ must be proportional to an identity matrix, which indeed is the case: $\widetilde{h}(\M) = 0$ according to \eqnref{eq:CBL_htilde}.
Away from $\M$, $\widetilde{h}(\M+\bsl{p})$ reads
\eq{
\label{eq:CBL_htilde_M}
\widetilde{h}(\M+\bsl{p}) =  t p_x p_y\sigma_x + t' (p_x^2 + p_y^2) \sigma_y + O(p^4)\ .
}
The moir\'e model in the following is built from \eqnref{eq:CBL_htilde_M} to the second order of $p$.

The basis of the moir\'e model is $\psi^\dagger_{\bsl{r},l,\sigma,s}$, where $l=t,b$ labels the layer, $\sigma=A,B$ labels the sublattice, and $s=\uparrow,\downarrow$ labels the spin.
Specifically, 
\eqa{
 &  \psi^\dagger_{\bsl{r},t,\sigma,s} = \frac{1}{\sqrt{N \Omega}}\sum_{\bsl{p}} e^{\ii \bsl{p}\cdot\bsl{r}}\ \widetilde{c}^{\dagger}_{C_{\theta/2} M +\bsl{p},\sigma,s} \\
 & \psi^\dagger_{\bsl{r},b,\sigma,s} = \frac{1}{\sqrt{N \Omega}}\sum_{\bsl{p}} e^{\ii \bsl{p}\cdot\bsl{r}}\ \widetilde{c}^{\dagger}_{C_{-\theta/2} M +\bsl{p},\sigma,s}
}
where $N$ is the number of moir\'e unit cells, and $\Omega$ is the area of the moir\'e unit cell.
The model has moire lattice translations generated by 
\eqa{
& \bsl{a}_{M,1} = \frac{\sqrt{2} \pi}{k_\theta} (1,0)^T \\
& \bsl{a}_{M,2} = \frac{\sqrt{2} \pi}{k_\theta} (0,1)^T\ ,
}
where $k_\theta = 2\sqrt{2} \pi \sin(\frac{\theta}{2})$ and recall that the lattice constant of the underlying checkerboard lattice is set to 1.
The moir\'e Hamiltonian in the real space reads
\eq{
H =\int d^2 r \mat{ \psi^\dagger_{\bsl{r},t} , \psi^\dagger_{\bsl{r},b} } \mat { -t \partial_x \partial_y \sigma_x - t' (\partial_x^2 - \partial_y^2) \sigma_y &  W(\bsl{r}) \\
 W^\dagger(\bsl{r}) & -t \partial_x \partial_y \sigma_x - t' (\partial_x^2 - \partial_y^2) \sigma_y} \otimes s_0 \mat{ \psi_{\bsl{r},t} \\ \psi_{\bsl{r},b} }\ ,
}
where 
\eq{
\psi^\dagger_{\bsl{r},l} = (\psi^\dagger_{\bsl{r},l,+,\uparrow}, \psi^\dagger_{\bsl{r},l,+,\downarrow}, \psi^\dagger_{\bsl{r},l,-,\uparrow}, \psi^\dagger_{\bsl{r},l,-,\downarrow})\ ,
}
\eq{
 W(\bsl{r})=  2 \sum_{i=1}^2 T_i \cos(\bsl{r}\cdot\bsl{q}_i) + 2 \sum_{i=1}^4 T_i'\cos(\bsl{r}\cdot\bsl{g}_i) \ ,
 }
$\bsl{q}_n = C_{4z}^{n-1} \frac{k_\theta}{\sqrt{2}} (1,1)^T$ with $n=1,2,3,4$ (see $\bsl{q}_1$ in \cref{fig:CherkerBoard}(c)), $\bsl{g}_i=\bsl{b}_{M,i}+\bsl{q}_1$ with $i=1,2$, $\bsl{g}_i=C_{4z} \bsl{g}_{i-2}$ with $i=3,4$.
The interlayer tunnelling chosen in \refcite{Yao2022TwistedCheckorboard} reads
\eq{
T_1 = \sigma_z T_2 \sigma_z  = w_1 \sigma_x\ ,
}
and we added 
\eq{
T_1'=T_2'= w_1' \sigma_x\ ,\ T_3'=T_4'= \sigma_z T_1' \sigma_z
}
to allow more tunability of the model.
The symmetry group of the model is spanned by spin $\SU(2)$, spinless $C_{4z}$, spinless $C_{2x}$, $\TR$ and moir\'e-lattice translations.
It also has an effective $m_z$ that exchanges the two layers, and chiral symmetry $\C$. (We neglect the identity term in $T_1$ and $T_2$ to achieve this.)
The representations (reps) of the symmetries furnished by the basis are
\eqa{
& C_{4z} \psi^\dagger_{\bsl{r},l} C_{4z}^{-1} = \psi^\dagger_{C_{4z}\bsl{r},l} (-\ii \sigma_z s_0) \\
& C_{2x} \psi^\dagger_{\bsl{r},l} C_{2x}^{-1} = \psi^\dagger_{C_{2x}\bsl{r},\bar{l}} \sigma_y s_0\\
& \TR \psi^\dagger_{\bsl{r},l} \TR^{-1} = \psi^\dagger_{\bsl{r},l} \ii\sigma_x \ii s_y\\
& T_{\bsl{R}_M}  \psi^\dagger_{\bsl{r},l} T_{\bsl{R}_M}^{-1} = e^{-\ii \bsl{R}_M\cdot M_{l}} \psi^\dagger_{\bsl{r}+\bsl{R}_M,l} \\
& m_z  \psi^\dagger_{\bsl{r},l} m_z^{-1} = \psi^\dagger_{\bsl{r},\bar{l}} \\
& \C  \psi^\dagger_{\bsl{r},l} \C^{-1} =  \psi^\dagger_{\bsl{r},l} \sigma_z s_0\ ,
}
where $\bar{l} = b/t$ if $l=t/b$, $M_t = C_{\theta/2} (\pi,\pi)^T$, and $M_b = C_{-\theta/2} (\pi,\pi)^T$.

In practical calculation, we always need to convert the model to the momentum spaces.
To do so, we first define 
\eq{
\psi^\dagger_{\bsl{p},l,\sigma,s} = \frac{1}{\sqrt{N \Omega}} \int d^2 r e^{\ii \bsl{p}\cdot \bsl{r} } \psi^\dagger_{\bsl{r},l,\sigma,s}\ ,
}
where $\sigma=\pm$.
The moir\'e reciprocal lattice vectors are $\{ \bsl{G}_M \} = \mat{ \bsl{b}_{M,1} & \bsl{b}_{M,2}} \dsZ^2$, where $\bsl{b}_{M,1} = \sqrt{2} k_\theta (1,0)^T$ and $\bsl{b}_{M,2} = \sqrt{2} k_\theta (0,1)^T$.
The Q vectors for the two layers are 
\eqa{
\label{eq:Q_lattices}
& \Q_t = \{ \bsl{G}_M \}  + \bsl{q}_1 \\
& \Q_b = \{ \bsl{G}_M \}  \\
& \Q = \Q_t \cup \Q_b\ ,
}
and then we can define
\eq{
\psi^\dagger_{\bsl{k},\bsl{Q},\sigma,s} = \psi^\dagger_{\bsl{k}-\bsl{Q},l_{\bsl{Q}},\sigma,s}
}
with $l_{\bsl{Q}} = l$ iff $\bsl{Q}\in \bsl{Q}_l$.

With those definitions, we have
\eq{
\label{eq:moire_SP}
H= \sum_{\bsl{k}\in \MBZ} \sum_{\bsl{Q} \bsl{Q}'} \psi^\dagger_{\bsl{k},\bsl{Q}} \left[ h_0(\bsl{k}-\bsl{Q})\delta_{\bsl{Q}\bsl{Q}'} + \sum_{i=1}^2 T_i (\delta_{\bsl{Q},\bsl{Q}'+\bsl{q}_i} + \delta_{\bsl{Q},\bsl{Q}'-\bsl{q}_i}) + \sum_{i'=1}^4 T_{i'} (\delta_{\bsl{Q},\bsl{Q}'+\bsl{g}_{i'}} + \delta_{\bsl{Q},\bsl{Q}'-\bsl{g}_{i'}}) \right]\otimes s_0 \psi_{\bsl{k},\bsl{Q}'} \ ,
}
where 
\eq{
h_0(\bsl{p}) = t p_x p_y \tau_x + t' (p_x^2 - p_y^2) \tau_y\ .
}
The symmetry reps in the momentum space read
\eqa{
\label{eq:sym_rep_moire_momentum_basis}
& C_{4z} \psi^\dagger_{\bsl{k},\bsl{Q}} C_{4z}^{-1} = \psi^\dagger_{C_{4z}\bsl{k},C_{4z}\bsl{Q}} (-\ii \tau_z) s_0 \\
& C_{2x} \psi^\dagger_{\bsl{k},\bsl{Q}} C_{2x}^{-1} = \psi^\dagger_{\C_{2x}\bsl{k}-C_{2x}\bsl{Q},\bar{l}_{\bsl{Q}}} \tau_y s_0= \psi^\dagger_{\C_{2x}\bsl{k} + \bsl{q}_1 -C_{2x}\bsl{Q}-\bsl{q}_1 ,l_{C_{2x}\bsl{Q}+\bsl{q}_1}}\tau_y s_0= \psi^\dagger_{\C_{2x}\bsl{k} + \bsl{q}_1,C_{2x}\bsl{Q}+\bsl{q}_1}\tau_y s_0\\
& \TR \psi^\dagger_{\bsl{k},\bsl{Q}} \TR^{-1} = \psi^\dagger_{-\bsl{k},-\bsl{Q}} \ii\tau_x \ii s_y \\
& m_z  \psi^\dagger_{\bsl{k},\bsl{Q}} m_z^{-1} = \psi^\dagger_{\bsl{k}+\bsl{q}_1,\bsl{Q}+\bsl{q}_1}\\
& \C  \psi^\dagger_{\bsl{k},\bsl{Q}} \C^{-1} =  \psi^\dagger_{\bsl{k},\bsl{Q}} \sigma_z s_0\ .
}
In this basis, we see that the representation for $C_{2x}$ changes the Bloch momentum.
The reason is that $C_{2x}$ changes $\bsl{k}-\bsl{Q}_t$ in the top layer to $C_{2x} \bsl{k} - C_{2x} \bsl{Q}_t$ in the second layer. However, $\Q_t$ and $\Q_b$ are both $C_{2x}$-invariant, and thus we have to shift $C_{2x} \bsl{Q}_t$ by $\bsl{q}_1$ to enter $\Q_b$, leading to the new seperation $C_{2x} \bsl{k}+\bsl{q}_1 - C_{2x} \bsl{Q}_t-\bsl{q}_1$ in \eqnref{eq:sym_rep_moire_momentum_basis}.
The same reasoning holds for $m_z$.
The argument extends to any twisted bilayer rectangular/square systems (i) whose low-energy physics comes from M point and (ii) which has in-plane $C_2$ symmetries that flip two layers.
In contrast, in TBG, the two $\bsl{Q}$ lattices of two layers are $C_{2x}$ partners, and thus there is no momentum shift for $C_{2x}$ there.

\subsection{Chiral-Symmetry Protected Stable Anomaly}

In this section, we discuss the stable anomaly of the moir\'e single-particle Hamiltonian in \eqnref{eq:moire_SP} that is protected by the chiral symmetry.
If we set $w_1=w_1'=0$, we have two quadratic touching points (from two layers) at the zero energies at $\Gamma_M$ and $M_M$ in the moir\'e Brillouin zone, and each of it has chiral winding number 2, adding up to a total chiral winding number of $\sum_{l}W_l = 4$.
The nonzero total chiral winding number indicates a stable anomaly, which holds for nonzero $w_1$ and $w_1'$ due to the presence of chiral symmetry.

\begin{figure}
    \centering
    \includegraphics[width=\columnwidth]{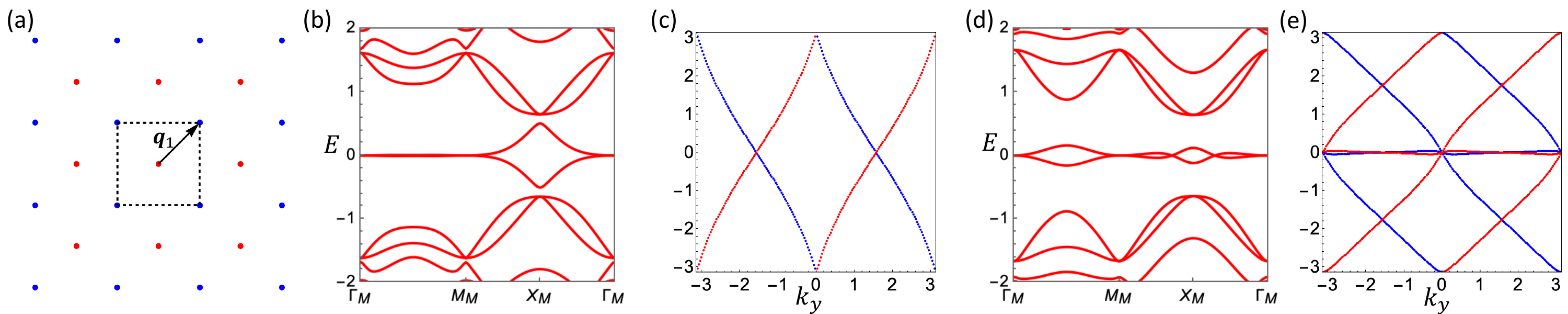}
    \caption{(a) The red and blue dots are $\Q_b$ and $\Q_t$, respectively, in \figref{eq:Q_lattices}. 
    (b) and (d) show the band structures for the parameter set 1 in \cref{eq:para_1} and set 2 in \cref{eq:para_2}, respectively.
    (c) and (e) show the chiral-resolved Wilson loop spectrum for the isolated chiral-symmetric set of 2 bands in (b) around zero energies and the isolated chiral-symmetric set of 6 bands in (d) around zero energies, respectively.
    The Wilson spectrum in the chiral-even (chiral-odd) subspace is in red (blue). 
    }
    \label{fig:CheckerBoard_app_1}
\end{figure}

The chiral-symmetry-protected anomaly should be reflected as nontrivial $\Ch_+-\Ch_-$ for any chiral-symmetric set of even number of bands.
To show this, we first choose 
\eq{
\label{eq:para_1}
t = 4/k_\theta^2,\ t' = 1.26/k_\theta^2,\ w_1 = 0.66\ ,w_1' = -0.4\ ,
}
whose band structure is shown in \cref{fig:CheckerBoard_app_1}(b).
We have $\Ch_+-\Ch_-=4$ for the isolated chiral-symmetric set of 2 bands in \cref{fig:CheckerBoard_app_1}(b) around zero energies, as indicated by the Wilson loop spectrum in \cref{fig:CheckerBoard_app_1}(c).
We also plot the band structure for 
\eq{
\label{eq:para_2}
t = 4/k_\theta^2,\ t' = 1.26/k_\theta^2,\ w_1 = 0.4\ ,w_1' = -0.4
}
in \cref{fig:CheckerBoard_app_1}(d), which has $\Ch_+-\Ch_-=4$ for the isolated chiral-symmetric set of 6 bands around zero energies, as indicated by the Wilson loop spectrum in \cref{fig:CheckerBoard_app_1}(d).

The nonzero $\Ch_+-\Ch_-$ can be indicated by the symmetries.
Owing to the $C_{2z}\TR$ symmery, any isolated set of bands have zero total Chern number, and thus we always have $\Ch_+ = -\Ch_-$, which means we only need to the indicate the nonzero $\Ch_+$.
To do so, we look at $C_{4z}$ eigenvalues at $\Gamma_M$ and $M_M$, and $C_{2z}$ eigenvalues at $X_M$.
As a result, for the isolated chiral-symmetric set of 2 bands for \eqnref{eq:para_1}, the chrial-even subspace has $C_{4z}\dot{=}-\ii$ at $\Gamma_M$, $C_{4z}\dot{=}-\ii$ at $M_M$ and $C_{2z}\dot{=}1$ at $X_M$, which gives $\ii^{\Ch_+} = (-\ii) (-\ii) 1 = -1$; then we know $\Ch_+ = 2 \mod 4$, which is consistent with the Wilson loop in \cref{fig:CheckerBoard_app_1}(c).
Similarly, for the isolated chiral-symmetric set of 6 bands for \eqnref{eq:para_2}, the chrial-even subspace has $C_{4z}\sim\text{diag}(-\ii, \ii, -\ii)$ at $\Gamma_M$, $C_{4z}\sim\text{diag}(-\ii, \ii , -\ii)$ at $M_M$ and $C_{2z}\sim\text{diag}(-1, -1, 1)$ at $X_M$, which gives $\ii^{\Ch_+} = (-\ii) (-\ii) 1 = -1 \Rightarrow \Ch_+ = 2 \mod 4$, which is consistent with the Wilson loop in \cref{fig:CheckerBoard_app_1}(e).

\subsection{Heavy Fermion Model: Single-Particle}

The stable anomaly of the model forbids any local-chiral Wannierization for any chiral-symmetric set of bands.
However, for the case of the nearly flat bands with concentrated curvature, we can resolve the issue via the construction of topological heavy fermion model~\cite{KaiSun2023THF}.

To show this, we focus on the parameter set in \eqnref{eq:para_1}, where the near flat bands has concentrated $F_{+}(\bsl{k})-F_{-}(\bsl{k})$ as shown in \figref{fig:CherkerBoard}(c).
Using \emph{Wannier90}, we can construct the exponentially-localized Wannier functions by combining most states of nearly flat bands near the zero energy and the states near $X_M$ and $Y_M$ from the remote bands.
The resultant $f$ modes have the form
\eq{
f^\dagger_{\bsl{k},\alpha,s} =\sum_{\bsl{Q}\sigma} \psi^\dagger_{\bsl{k},\bsl{Q},\sigma,s} \left[ U_{f,\alpha}(\bsl{k}) \right]_{\bsl{Q}\sigma}
}
with $\alpha=1,2$.
The trial functions for the Wannierization were chosen as 
\eqa{
& \left[ U_{f,1}(\bsl{k}) \right]_{\bsl{Q}\sigma} = \frac{1}{\sqrt{\Omega}}\sqrt{\pi\lambda^2} (1,\ii)_\alpha e^{- |\bsl{k}-\bsl{Q}|^2\lambda^2/2} (-)^{\bsl{Q}}\\
& \left[ U_{f,2}(\bsl{k}) \right]_{\bsl{Q}\sigma} = \frac{1}{\sqrt{\Omega}}\sqrt{\pi\lambda^2} (-\ii,-1)_\alpha e^{- |\bsl{k}-\bsl{Q}|^2\lambda^2/2} (-)^{\bsl{Q}}\ ,
}
where $(-)^{\bsl{Q}} = 1 $ for $\bsl{Q}\in\Q_t$ and $(-)^{\bsl{Q}} = -1 $ for $\bsl{Q}\in\Q_b$.
The resultant Wannier function have the forms
\eq{
w_{\alpha l \sigma}(\bsl{r})  = \frac{1}{N \sqrt{\Omega}} \sum_{\bsl{Q}\in \Q_l} e^{\ii (\bsl{k}-\bsl{Q})\cdot\bsl{r}} \left[ U_{f,\alpha}(\bsl{k})\right]_{\bsl{Q}\sigma}\ ,
}
which has spread $(0.4285 a_M)^2$ with $a_M$ the moir\'e lattice constant.
The symmetry reps furnished by $f^\dagger_{\bsl{k},\alpha,s}$ are
\eqa{
\label{eq:sym_rep_moire_momentum_basis}
& C_{4z} f^\dagger_{\bsl{k}} C_{4z}^{-1} = f^\dagger_{C_{4z}\bsl{k}} (-\ii \tau_z) s_0 \\
& C_{2x} f^\dagger_{\bsl{k}} C_{2x}^{-1} = f^\dagger_{\C_{2x}\bsl{k} + \bsl{q}_1 }\tau_x s_0\\
& \TR f^\dagger_{\bsl{k}} \TR^{-1} = f^\dagger_{-\bsl{k}} \tau_x \ii s_y \\
& m_z  f^\dagger_{\bsl{k}} m_z^{-1} = f^\dagger_{\bsl{k}+\bsl{q}_1} (-)\\
& \C  f^\dagger_{\bsl{k}} \C^{-1} =  f^\dagger_{\bsl{k}} \sigma_z s_0\ ,
}
indicating they have the same symmetry reps as $p_x \pm p_y$ at 1a position of the moir\'e unit cell.

The remaining $c$ modes are localized around $X_M$ and $Y_M$, which have the form
\eq{
c^\dagger_{X_M+\bsl{q},\beta,s} = \sum_{\bsl{Q}\sigma} \psi^\dagger_{X_M+\bsl{q},\bsl{Q},\sigma,s} \left[ U_{c,\beta}(X_M+\bsl{q})\right]_{\bsl{Q}\sigma}
}
and
\eq{
c^\dagger_{Y_M+\bsl{q},\beta,s} = \sum_{\bsl{Q}\sigma} \psi^\dagger_{Y_M+\bsl{q},\bsl{Q},\sigma,s} \left[ U_{c,\beta}(Y_M+\bsl{q})\right]_{\bsl{Q}\sigma} \ .
}
Their symmetry reps are
\eqa{
& C_{4z} m_z c^\dagger_{X_M+\bsl{q}} (C_{4z} m_z)^{-1} = c^\dagger_{X_M+C_{4z}\bsl{q}} \left(
\begin{array}{cccc}
 -1 & 0 & 0 & 0 \\
 0 & -1 & 0 & 0 \\
 0 & 0 & -i & 0 \\
 0 & 0 & 0 & i \\
\end{array}
\right) s_0 \ ,\  m_z C_{2x} c^\dagger_{X_M+\bsl{q}} ( m_z C_{2x})^{-1} = c^\dagger_{X_M+\bsl{q}}\left(
\begin{array}{cccc}
 0 & -1 & 0 & 0 \\
 -1 & 0 & 0 & 0 \\
 0 & 0 & 0 & i \\
 0 & 0 & -i & 0 \\
\end{array}
\right) s_0\\
& C_{2z} c^\dagger_{X_M+\bsl{q}} C_{2z}^{-1} = c^\dagger_{X_M-\bsl{q}} \left(
\begin{array}{cccc}
 1 & 0 & 0 & 0 \\
 0 & 1 & 0 & 0 \\
 0 & 0 & -1 & 0 \\
 0 & 0 & 0 & -1 \\
\end{array}
\right) s_0  \ ,\ \TR c^\dagger_{X_M+\bsl{q}} \TR^{-1} = c^\dagger_{X_M-\bsl{q}} \left(
\begin{array}{cccc}
 0 & 1 & 0 & 0 \\
 1 & 0 & 0 & 0 \\
 0 & 0 & 0 & 1 \\
 0 & 0 & 1 & 0 \\
\end{array}
\right) \ii s_y \\
& \C  c^\dagger_{X_M+\bsl{q}} \C^{-1} =  c^\dagger_{X_M+\bsl{q}} \left(
\begin{array}{cccc}
 1 & 0 & 0 & 0 \\
 0 & -1 & 0 & 0 \\
 0 & 0 & 1 & 0 \\
 0 & 0 & 0 & -1 \\
\end{array}
\right) s_0 \ ,\ C_{4z}  c^\dagger_{X_M+\bsl{q}} C_{4z}^{-1} =  c^\dagger_{Y_M+C_{4z}\bsl{q}} \ .
}

With the symmetry representations, we can obtain the single-particle part of the Heavy fermion model.
The resulting heavy fermion model consists of three parts: the $f$-mode part which we choose to have zero Hamiltonian, the $c$-mode part, and the $f-c$ coupling.
The $c$-mode part reads
\eq{
H_c = \sum_{\bsl{p}}^{\Lambda} c^\dagger_{X_M+\bsl{p}} h_{cc}^{X_M}(\bsl{p})\otimes s_0 c_{X_M+\bsl{p}} + \text{$C_{4z}$ partner}\ ,
}
where $c^\dagger_{X_M+\bsl{p}}$ $=(c^\dagger_{X_M+\bsl{p},1}$, $c^\dagger_{X_M+\bsl{p},2}$, $c^\dagger_{X_M+\bsl{p},3}$, $c^\dagger_{X_M+\bsl{p},4})$ $\otimes (\uparrow,\downarrow)$, 
\eq{
h_{cc}^{X_M}(\bsl{p}) =
\mat{ 
m \sigma_x & \ii v_2 \left[ (p_x + p_y) \sigma_x + (p_x - p_y) \sigma_y \right]\\
h.c. & 0_{2\times 2} \ .
}
}
to the linear order of $\bsl{p}$, and the $C_{4z}$ partner term stands for the $Y_M$ Hamiltonian:
\eq{
h_{cc}^{Y_M}(\bsl{p}) = 
h_{cc}^{X_M}(C_{4z}^{-1}\bsl{p})\ .
}
The $f-c$ coupling reads
\eqa{
H_{fc} & = \sum_{\bsl{p}}^{\Lambda} f^\dagger_{\bsl{p}} h_{fc}^{X_M}(\bsl{p}) c_{X_M+\bsl{p}} e^{-\frac{|\bsl{p}|^2 \lambda^2 }{2}}  + (C_{4z}\text{ partner}) \\
& \quad+ h.c.\ .
}
where $\lambda=0.4254 a_M$ is square root of the Wannier spread of the $f$ modes with $a_M$ the moir\'e lattice constant, and 
\eq{
h_{fc}^{X_M}(\bsl{p}) = \mat{ \ii v_1 (p_x \sigma_x + p_y \sigma_y) & \gamma (\sigma_x + \sigma_y)}\ .
}
From the tight-binding parameters and the Wannier states, we compute $v_1 = -0.52/k_\theta$, $\gamma=0.45$, $m = 0.51$ and $v_2 =1.53/k_\theta$. The heavy fermion model matches the continuum model remarkably well(\figref{fig:CherkerBoard}(d)).

\subsection{Heavy Fermion Model: Interaction}

In this part, we show the interaction of the Heavy fermion model.
The microscopic Hamiltonian is the following density-density interaction 
\eq{
\label{eq:H_int_gen}
H_{int} = \int d^2 r \int d^2 r' V(\bsl{r}-\bsl{r}') :\rho(\bsl{r}): :\rho(\bsl{r}'):\ ,
}
where 
\eq{
\rho(\bsl{r}) = \sum_{l,\sigma,s} \psi^\dagger_{\bsl{r},l,\sigma,s} \psi_{\bsl{r},l,\sigma,s} \ ,
}
and $ :\rho(\bsl{r}):  =\rho(\bsl{r}) - \bra{G_0} \rho(\bsl{r}) \ket{G_0} $ for certain reference state $\ket{G_0}$.
The idea is to project the density-density interaction to the $f$ and $c$ modes.
In the following, we will use the general form in \eqnref{eq:H_int_gen} to do the analytical derivations, and use double-gate screened Coulomb interaction for numerical calculations:
\eq{
\label{eq:Coulomb_V}
V(\bsl{r}) = \frac{1}{N\Omega} \sum_{\bsl{p}} e^{ -\ii \bsl{p}\cdot \bsl{r} } V(\bsl{p})\ ,
}
where  
\eq{
V(\bsl{p}) = \pi \xi^2 V_\xi \frac{ \tanh( \xi |\bsl{p}|/2) }{ \xi |\bsl{p}|/2 }\ ,
}
$\xi=10/k_{\theta}$ is the distance between two gates, and $V_\xi = \frac{e^2}{4\pi \epsilon \xi} = 2$ with $e$ the elementary charge and $\epsilon$ the dielectric constant.
We will always consider the $f$ and $c$ modes derived from the single-particle parameters in \cref{eq:para_1}.

To do the projection, we first consider $f$ and $c$ modes in the real space:
\eq{
f^\dagger_{\bsl{R},\alpha,s} = \frac{1}{\sqrt{N}}\sum_{\bsl{k}} e^{ -\ii \bsl{k}\cdot \bsl{R} } f^\dagger_{\bsl{k},\alpha,s} \ ,
}
and
\eq{
c^\dagger_{\eta,\bsl{r},\beta} = \frac{1}{\sqrt{V}} \sum_{\bsl{q}}^{\Lambda} e^{-\ii (\eta+\bsl{q})\cdot\bsl{r}} c^\dagger_{\bsl{k}_0+\bsl{q},\beta}\ .
}
Then, we can approximate $\psi^\dagger_{\bsl{r},l,\sigma,s}$ as 
\eq{
\psi^\dagger_{\bsl{r},l,\sigma,s}\approx \sum_{\alpha}\sum_{\bsl{R}} f^\dagger_{\bsl{R},\alpha,s} e^{\ii \bsl{R}\cdot\Delta K_l} w^*_{\alpha l \sigma}(\bsl{r}-\bsl{R}) + \sum_{\eta=X,Y}\sum_{\beta}c^\dagger_{\eta,\bsl{r},\beta,s} g_{c,\eta,\beta,l,\sigma}(\bsl{r}) \ ,
}
where
\eq{
g_{c,\eta,\beta,l,\sigma
} (\bsl{r}) = \sum_{\bsl{Q}\in\Q_l} e^{\ii \bsl{Q}\cdot \bsl{r}} \left[ U_{c,\beta}(\eta) \right]_{\bsl{Q}\sigma}^*
}
with $\eta=X,Y$, and the high energy modes are neglected.
Therefore, the density operator can now be expanded into
\eqa{
\rho(\bsl{r}) & = \sum_{l \sigma}\psi^\dagger_{\bsl{r},l,\sigma} \psi_{\bsl{r},l,\sigma} 
\approx \sum_{l \sigma} \sum_{\alpha_1 \alpha_2} \sum_{\bsl{R}_1 \bsl{R}_2} f^\dagger_{\bsl{R}_1,\alpha_1} f_{\bsl{R}_2,\alpha_2} e^{\ii (\bsl{R}_1-\bsl{R}_2)\cdot \Delta K_l } w^*_{\alpha_1 l \sigma} (\bsl{r}-\bsl{R}_1) w_{\alpha_2 l \sigma}(\bsl{r}-\bsl{R}_2) \\
& \qquad + \left( \sum_{l\sigma} \sum_{\eta_1=X,Y} \sum_{\beta_1} c^\dagger_{\eta_1,\bsl{r},\beta_1} g_{c,\eta_1,\beta_1,l,\sigma}(\bsl{r}) \sum_{\alpha_2 \bsl{R}_2} f_{\bsl{R}_2 \alpha_2} e^{-\ii \bsl{R}_2\cdot\Delta K_l} w_{\alpha_2 l \sigma}(\bsl{r}-\bsl{R}_2) + h.c. \right) \\
& \qquad + \sum_{\eta_1=X,Y} \sum_{\beta_1} \sum_{\eta_2=\X,Y} \sum_{\beta_2} c^\dagger_{\eta_1,\bsl{r},\beta_1} c_{\eta_2,\bsl{r},\beta_2} g_{c,\eta_1,\beta_1,l,\sigma}(\bsl{r}) g^*_{c,\eta_2,\beta_2,l,\sigma}(\bsl{r})\ .
}
With the expression of $\rho(\bsl{r}) $, we can project $H_{int}$ in \eqnref{eq:H_int_gen} to $f$ and $c$ modes, and split the results into seven parts:
\eq{
H_{int} = \int d^2 d^2 r' V(\bsl{r}-\bsl{r}') :\rho(\bsl{r}): :\rho(\bsl{r}'): = \sum_{i=1,...,7} H_{int,i}\ .
}

In the following, we will specify the form of each of $H_{int,i}$.
To do so, we will use the following three functions
\eq{
z(\bsl{r}-\bsl{R}_1,\bsl{r}-\bsl{R}_2) = \frac{1}{\sqrt{\Omega} N^2} \sum_{\bsl{k}_1}^\MBZ\sum_{\bsl{p}_2}^{\dsR^2} e^{-\ii \bsl{k}_1\cdot(\bsl{r}-\bsl{R}_1)}  e^{ \ii \bsl{p}_2\cdot(\bsl{r}-\bsl{R}_2)} \frac{1}{2} \Tr\left[ U_f^\dagger(\bsl{k}_1) U_{f}(\bsl{p}_2) \right]\ ,
}
\eq{
\chi_{\eta_1\beta_1\alpha_2}(\bsl{r}-\bsl{R}) = \sum_{l\sigma} g_{c,\eta_1,\beta_1,l,\sigma}(\bsl{r}0 e^{-\ii \bsl{R}\cdot \Delta K_l} w_{\alpha_2 l \sigma}(\bsl{r}-\bsl{R}) = \frac{1}{N\sqrt{\Omega}} \sum_{\bsl{p}}^{\dsR^2} e^{\ii \bsl{p}\cdot(\bsl{r}-\bsl{R})} U^\dagger_{c,\beta_1}(\eta_1) U_{f,\alpha_2}(\bsl{p})\ ,
}
and
\eq{
y_{\eta_1\beta_1,\eta_2\beta_2}(\bsl{r}) = \sum_{l\sigma} g_{c,\eta_1,\beta_1,l,\sigma}(\bsl{r}) g_{c,\eta_2,\beta_2,l,\sigma}^*(\bsl{r}) = \sum_{\bsl{G}} e^{\ii \bsl{G}\cdot \bsl{r}} U^\dagger_{c,\beta_1}(\eta_1) U_{c,\beta_2}(\eta_2+\bsl{G})
}

\subsubsection{$H_{int,1}$}

\eq{
H_{int,1} = \int d^2 r d^2 r' V(\bsl{r}-\bsl{r}') \sum_{\bsl{R}_1 \bsl{R}_2 }  :f^\dagger_{\bsl{R}_1} f_{\bsl{R}_2}: z(\bsl{r}-\bsl{R}_1,\bsl{r}-\bsl{R}_2) \sum_{\bsl{R}_3 \bsl{R}_4} :f^\dagger_{\bsl{R}_3} f_{\bsl{R}_4}: z(\bsl{r}'-\bsl{R}_4,\bsl{r}'-\bsl{R}_4)
}

Owing to the effective $m_z$ symmetry, we have
\eqa{
& m_z \sum_{\bsl{R}_1 \bsl{R}_2} f_{\bsl{R}_1\bsl{R}_2}^\dagger f_{\bsl{R}_2} z(\bsl{r}-\bsl{R}_1,\bsl{r}-\bsl{R}_2) m_z^{-1} = \sum_{\bsl{R}_1 \bsl{R}_2} f_{\bsl{R}_1\bsl{R}_2}^\dagger f_{\bsl{R}_2} z(\bsl{r}-\bsl{R}_1,\bsl{r}-\bsl{R}_2) \\
& \Rightarrow e^{\ii \bsl{q}_1\cdot (\bsl{R}_1-\bsl{R}_2)} z(\bsl{r}-\bsl{R}_1,\bsl{r}-\bsl{R}_2) = z(\bsl{r}-\bsl{R}_1,\bsl{r}-\bsl{R}_2) \\
& \Rightarrow z(\bsl{r}-\bsl{R}_1,\bsl{r}-\bsl{R}_2) = 0\text{ for }|\bsl{R}_1-\bsl{R}_2| = a_M \\
& \Rightarrow z(\bsl{r}-\bsl{R}_1,\bsl{r}-\bsl{R}_2) \approx \delta_{\bsl{R}_1 \bsl{R}_2 } n_f(\bsl{r}-\bsl{R}_1)\ ,
}
where we use the fact that the Wannier functions have a localization length (square root of the Wannier spread) about $0.43a_M$,
\eq{
n_f(\bsl{r}) = \frac{1}{\V} \sum_{\bsl{p}} e^{-\ii \bsl{p}\cdot\bsl{r}} \widetilde{n}_f(\bsl{p}) \ ,
}
and 
\eq{
\widetilde{n}_f(\bsl{p}) = \frac{1}{2 N} \sum_{\bsl{k}}^{MBZ} \Tr\left[ U^\dagger(\bsl{p}+\bsl{k}) U_f(\bsl{k}) \right] \ .
}

As a result, we have
\eqa{
H_{int,1} & \approx  \sum_{\bsl{R} \bsl{R}' } \int d^2 r d^2 r' V(\bsl{r}-\bsl{r}') n_f(\bsl{r}-\bsl{R}) n_f(\bsl{r}'-\bsl{R}')  :\rho_f(\bsl{R}): :\rho(\bsl{R}'): \\
& = \sum_{\bsl{R} \bsl{R}' } U_f(\bsl{R}-\bsl{R}')  :\rho_f(\bsl{R}): :\rho_f(\bsl{R}'): \ ,
}
where
\eqa{
 U_f(\bsl{R}-\bsl{R}') & = \int d^2 r d^2 r' V(\bsl{r}-\bsl{r}') n_f(\bsl{r}-\bsl{R}) n_f(\bsl{r}'-\bsl{R}')  = \frac{1}{N \Omega} \sum_{\bsl{p}} V(\bsl{p}) |\widetilde{n}_f(\bsl{p})|^2 e^{-\ii \bsl{p}\cdot (\bsl{R}-\bsl{R}')} \ .
}
Numerically, we have
\eq{
 U_f(\bsl{0})=11.27\ ,\   U_f(\bsl{a}_{M,1})=2.55\ ,
}
and thus $H_{int,1}$ can further be approximated as
\eqa{
H_{int,1} & \approx H_U =   \sum_{\bsl{R} } U :\rho_f(\bsl{R}): :\rho_f(\bsl{R}): 
}
with $U=11.27$.

\subsubsection{$H_{int,2}$}

\eqa{
H_{int,2} & = \int d^2 r d^2 r' V(\bsl{r}-\bsl{r}')  :c^\dagger_{\bsl{r}} y(\bsl{r}) c_{\bsl{r}}: :c^\dagger_{\bsl{r}'} y(\bsl{r}') c_{\bsl{r}'}: \\
& = \frac{1}{\mathcal{V}} \sum_{\bsl{G}\bsl{G}'}\sum^{\Lambda}_{\bsl{q}_1\bsl{q}_2\bsl{q}_3 \bsl{q}_4} \sum_{\eta_1 ... \eta_4} \sum_{\beta_1 ...\beta_4} :c^\dagger_{\eta_1 +\bsl{q}_1,\eta_1} c_{\eta_2+\bsl{q}_2,\beta_2}: :c^\dagger_{\eta_3+\bsl{q}_3,\beta_3} c_{\eta_4+\bsl{q}_4,\beta_4}: V(\eta_3 + \bsl{q}_3 - \eta_4 - \bsl{q}_4 - \bsl{G}')    \\
& \qquad \times U^\dagger_{c,\beta_1} (\eta_1) U_{c,\beta_2}(\eta_2 + \bsl{G}) U^\dagger_{c,\beta_3} (\eta_3) U_{c,\beta_4} (\eta_4 + \bsl{G}') \delta(-\eta_1-\bsl{q}_1 + \eta_2 + \bsl{q}_2 + \bsl{G} -\eta_3 -\bsl{q}_3 + \eta_4 + \bsl{q}_4 + \bsl{G}') \\
& \approx \frac{1}{\mathcal{V}} \sum_{\bsl{G}\bsl{G}'}\sum^{\Lambda}_{\bsl{q}_1\bsl{q}_3 } \sum_{\eta_1 ... \eta_4} \sum_{\beta_1 ...\beta_4} :c^\dagger_{\eta_1 +\bsl{q}_1,\eta_1} c_{\eta_2+\bsl{q}_1,\beta_2}: :c^\dagger_{\eta_3+\bsl{q}_3,\beta_3} c_{\eta_4+\bsl{q}_3,\beta_4}: V(\eta_3 - \eta_4  - \bsl{G}')    \\
& \qquad \times U^\dagger_{c,\beta_1} (\eta_1) U_{c,\beta_2}(\eta_2 + \bsl{G}) U^\dagger_{c,\beta_3} (\eta_3) U_{c,\beta_4} (\eta_4 + \bsl{G}') \delta(-\eta_1 + \eta_2  + \bsl{G} -\eta_3  + \eta_4  + \bsl{G}') \\
& = \frac{1}{N \Omega} \sum_{\bsl{q}\bsl{q}'}^{\Lambda} \sum_{\eta \eta', \beta \beta'} :c^\dagger_{\eta+\bsl{q},\beta} c_{\eta+\bsl{q},\beta}: :c^\dagger_{\eta'+\bsl{q}',\beta'} c_{\eta'+\bsl{q}',\beta'}:  V(\bsl{p}=0) \\
& \qquad + \frac{1}{N\Omega} \sum_{\bsl{q}\bsl{q}'}^{\Lambda} \sum_{\eta \eta'} \sum_{\beta_1 \beta-2 \beta_3 \beta_4} :c^\dagger_{\eta+\bsl{q},\beta_1} c_{\eta+\bsl{q},\beta_2}: :c^\dagger_{\eta'+\bsl{q}',\beta_3} c_{\eta'+\bsl{q}',\beta_4}: V'_{\beta_1 \beta_2 \beta_3 \beta_4}(\eta,\eta')\ ,
}
where
\eq{
V'_{\beta_1 \beta_2 \beta_3 \beta_4}(\eta,\eta') = \sum_{\bsl{G}'} V(\bsl{G}') U^\dagger_{c,\beta_1} (\eta) U_{c,\beta_2}(\eta + \bsl{G}') U^\dagger_{c,\beta_3} (\eta') U_{c,\beta_4} (\eta' + \bsl{G}') \ .
}

Numerically, we have $V(\bsl{p}=0)=628.32(1/k_\theta)^2$ and $V'_{\beta_1 \beta_2 \beta_3 \beta_4}(\eta,\eta')$ have dominating values of the same order.

\subsubsection{$H_{int,3}$}

\eq{
H_{int,3} = \int d^2 r d^2 r' V(\bsl{r}-\bsl{r}') \sum_{\bsl{R}_1 \bsl{R}_2 }  :f^\dagger_{\bsl{R}_1} f_{\bsl{R}_2}: z(\bsl{r}-\bsl{R}_1,\bsl{r}-\bsl{R}_2) :c^\dagger_{\bsl{r}'} y(\bsl{r}') c_{\bsl{r}'}: + h.c.
}

\eqa{
H_{int,3} & = \int d^2 r d^2 r' V(\bsl{r}-\bsl{r}') \sum_{\bsl{R}_1 \bsl{R}_2 }  :\rho_f(\bsl{R}): n_f(\bsl{r}-\bsl{R}) :c^\dagger_{\bsl{r}'} y(\bsl{r}') c_{\bsl{r}'}: + h.c.\\
& = \sum_{\bsl{R}} :\rho_f(\bsl{R}): \frac{1}{N \Omega} \sum_{\bsl{q}_1 \bsl{q}_2 }^{\Lambda} :c^\dagger_{\eta_1+\bsl{q}_1,\beta_1} c_{\eta_2+\bsl{q}_2,\beta_2}: \sum_{\bsl{G}} e^{\ii (-\eta_1 - \bsl{q}_1 +\eta_2 + \bsl{q}_2 + \bsl{G})\cdot\bsl{R}} V(\eta_1+\bsl{q}_1-\eta_2 - \bsl{q}_2 - \bsl{G}) \\
& \qquad \times \widetilde{n}_f(-\eta_1-\bsl{q}_1 +\eta_2 +\bsl{q}_2 +\bsl{G}) U^\dagger_{c,\beta_1}(\eta_1) U_{c,\beta_2} (\eta_2+\bsl{G}) + h.c.\\
& \approx \sum_{\bsl{R}} :\rho_f(\bsl{R}): :c^\dagger_{\eta_1,\bsl{R},\beta_1} c_{\eta_2,\bsl{R},\beta_2}: \sum_{\bsl{G}} V(\eta_1-\eta_2 - \bsl{G}) \widetilde{n}_f(-\eta_1 +\eta_2 +\bsl{G}) U^\dagger_{c,\beta_1}(\eta_1) U_{c,\beta_2} (\eta_2+\bsl{G}) + h.c. \\
& = \Omega \sum_{\bsl{R}} :\rho_f(\bsl{R}): :c^\dagger_{\bsl{R}} \widetilde{W} c_{\bsl{R}}: + h.c. \ ,
}
where
\eq{
\widetilde{W}_{\eta_1 \beta_1,\eta_2 \beta_2} = \frac{1}{\Omega} \sum_{\bsl{G}} V(\eta_1-\eta_2 - \bsl{G}) \widetilde{n}_f(-\eta_1 +\eta_2 +\bsl{G}) U^\dagger_{c,\beta_1}(\eta_1) U_{c,\beta_2} (\eta_2+\bsl{G}) \ .
}

Numerically, we find that 
\eq{
\widetilde{W}_{\eta_1 \beta_1,\eta_2 \beta_2} \approx W \delta_{\eta_1 \eta_2} \delta_{\beta_1 \beta_2} 
}
with $W=32$, leading to
\eq{
H_{int,3} \approx \Omega W \sum_{\bsl{R}} :\rho_f(\bsl{R}): :c^\dagger_{\bsl{R}} c_{\bsl{R}}: + h.c.\ .
}

\subsubsection{$H_{int,4}$}

\eqa{
H_{int,4} & = \int d^2 r d^2 r' V(\bsl{r}-\bsl{r}') \left\{ \sum_{\bsl{R} }  c^\dagger_{\bsl{r}} \chi(\bsl{r}-\bsl{R})f_{\bsl{R}} , \sum_{\bsl{R}' }  f_{\bsl{R}'}^\dagger \chi^\dagger(\bsl{r}'-\bsl{R}') c_{\bsl{r}'} \right\} \\
& = \sum_{\bsl{R}\bsl{R}'} \sum_{\eta_1 \beta_1} \sum_{\alpha_2 \alpha_3} \sum_{\eta_4 \beta_4} \sum_{\bsl{q}_1 \bsl{q}_3} \left\{ c^\dagger_{\eta_1+\bsl{q}_1,\beta_1} f_{\bsl{R},\alpha_2} \ ,\ f^\dagger_{\bsl{R}',\alpha_4} c_{\eta_3+\bsl{q}_3,\beta_3} \right\} \frac{1}{N\Omega} \int d^2r d^2 r' e^{-\ii (\eta_1 +\bsl{q}_1)\cdot \bsl{r} } e^{\ii (\eta_3 + \bsl{q}_3 ) \cdot \bsl{r}'} V(\bsl{r}-\bsl{r}') \\
& \qquad \times \chi_{\eta_1 \beta_1,\alpha_2} (\bsl{r}-\bsl{R}) \chi_{\eta_3 \beta_3,\alpha_4}^*(\bsl{r}'-\bsl{R}') 
}

\eqa{
& \frac{1}{N\Omega} \int d^2r d^2 r' e^{-\ii (\eta_1 +\bsl{q}_1)\cdot \bsl{r} } e^{\ii (\eta_3 + \bsl{q}_3 ) \cdot \bsl{r}'} V(\bsl{r}-\bsl{r}')  \chi_{\eta_1 \beta_1,\alpha_2} (\bsl{r}-\bsl{R}) \chi_{\eta_3 \beta_3,\alpha_4}^*(\bsl{r}'-\bsl{R}') \\
& = \frac{1}{N^2 \Omega} \sum_{\bsl{p}} e^{-\ii (\bsl{q}_1 + \eta_1 + \bsl{p}) \cdot \bsl{R} } e^{-\ii (\bsl{q}_3 + \eta_3 + \bsl{p}) \cdot \bsl{R}' }  V(\bsl{p}) U^\dagger_{c,\beta_1}(\eta_1) U_{f,\alpha_2}(\bsl{q}_1+\eta_1 + \bsl{p}) U^\dagger_{f,\alpha_4}(\bsl{q}_3+\eta_3+\bsl{p}) U_{c,\beta_3}(\eta_3) \\
& \approx \frac{1}{N^2 \Omega} \sum_{\bsl{p}} e^{-\ii (\bsl{q}_1 + \eta_1 +\bsl{p}) \cdot \bsl{R} } e^{ +\ii (\bsl{q}_3 + \eta_3 +\bsl{p}) \cdot \bsl{R}' } V(\bsl{p}) U^\dagger_{c,\beta_1}(\eta_1) U_{f,\alpha_2}(\eta_1+\bsl{p}) U^\dagger_{f,\alpha_4}(\eta_3+\bsl{p}) U_{c,\beta_3 }(\eta_3) \\
& = \frac{1}{N} e^{-\ii (\bsl{q}_1+\eta_1)\cdot\bsl{R}} e^{\ii (\bsl{q}_3+\eta_3)\cdot\bsl{R}'}  \widetilde{J}_{\eta_1\beta_1\alpha_2,\eta_3\beta_3\alpha_4}(\bsl{R}-\bsl{R}') \ ,
}
where
\eq{
\label{eq:J_tilde}
\widetilde{J}_{\eta_1\beta_1\alpha_2,\eta_3\beta_3\alpha_4}(\bsl{R}-\bsl{R}') = \frac{1}{N \Omega} \sum_{\bsl{p}} e^{-\ii \bsl{p}\cdot(\bsl{R}-\bsl{R}')} V(\bsl{p})  U^\dagger_{c,\beta_1}(\eta_1) U_{f,\alpha_2}(\eta_1+\bsl{p}) U^\dagger_{f,\alpha_4}(\eta_3+\bsl{p}) U_{c,\beta_3 }(\eta_3) \ .
}

It means that
\eq{
H_{int,4} \approx  \sum_{\bsl{R}\bsl{R}'} \sum_{\eta_1 \beta_2} \sum_{\alpha_2} \sum_{\eta_3 \beta_3} \sum_{\alpha_4} \left\{ c^\dagger_{\eta_1,\bsl{R},\beta_1} f_{\bsl{R},\alpha_2}, f^\dagger_{\bsl{R}',\alpha_4} c_{\eta_3,\bsl{R}',\beta_3} \right\} \Omega \widetilde{J}_{\eta_1\beta_1\alpha_2,\eta_3\beta_3\alpha_4}(\bsl{R}-\bsl{R}') 
}

Numerically, we find that $\widetilde{J}_{\eta_1\beta_1\alpha_2,\eta_3\beta_3\alpha_4}(\bsl{R}-\bsl{R}')\approx \widetilde{J}_{\eta_1\beta_1\alpha_2,\eta_3\beta_3\alpha_4}(0) \delta_{\bsl{R},\bsl{R}'}\approx J \delta_{\beta_1 \alpha_2}  \delta_{\beta_3 \alpha_2}  \delta_{\alpha_4 \alpha_2} \delta_{\bsl{R},\bsl{R}'} $ with $J\approx 4.5$.
As a result, we have
\eq{
H_{int,4} \approx  \Omega J  \sum_{\bsl{R}} \sum_{\eta \eta'} \sum_{\alpha} \left\{ c^\dagger_{\eta,\bsl{R},\alpha} f_{\bsl{R},\alpha}, f^\dagger_{\bsl{R},\alpha} c_{\eta',\bsl{R},\alpha} \right\} \ .
}

\subsubsection{$H_{int,5}$}

\eqa{
H_{int,5} &  = \int d^2 r d^2 r' V(\bsl{r}-\bsl{r}')\sum_{\bsl{R} }  c^\dagger_{\bsl{r}} \chi(\bsl{r}-\bsl{R})f_{\bsl{R}} \sum_{\bsl{R}' }  c^\dagger_{\bsl{r}'} \chi(\bsl{r}'-\bsl{R}')f_{\bsl{R}'} + h.c. \\
& = \sum_{\bsl{q}_1,\bsl{q}_3}^{\Lambda} \sum_{\eta_1 \beta_1} \sum_{\alpha_2}  \sum_{\eta_3 \beta_3} \sum_{\alpha_4} \sum_{\bsl{R}\bsl{R}'} c^\dagger_{\eta_1+\bsl{q}_1,\beta_1} f_{\bsl{R},\alpha_2} c^\dagger_{\eta_3+\bsl{q}_3,\beta_3} f_{\bsl{R}',\alpha_4} \\
& \qquad \times \frac{1}{N\Omega} \int d^2 r d^2 r' V(\bsl{r}-\bsl{r}') e^{-\ii (\eta_1 + \bsl{q}_1) \cdot\bsl{r}} e^{-\ii (\eta_3+\bsl{q}_3)\cdot\bsl{r}'} \chi_{\eta_1\beta_1,\alpha_2}(\bsl{r}-\bsl{R}) \chi_{\eta_3\beta_3,\alpha_4}(\bsl{r}'-\bsl{R}') + h.c.\ ,
}
where
\eqa{
& \frac{1}{N\Omega} \int d^2 r d^2 r' V(\bsl{r}-\bsl{r}') e^{-\ii (\eta_1 + \bsl{q}_1) \cdot\bsl{r}} e^{-\ii (\eta_3+\bsl{q}_3)\cdot\bsl{r}'} \chi_{\eta_1\beta_1,\alpha_2}(\bsl{r}-\bsl{R}) \chi_{\eta_3\beta_3,\alpha_4}(\bsl{r}'-\bsl{R}')  \\
    & = \frac{1}{N^2 \Omega} \sum_{\bsl{p}} e^{-\ii (\bsl{p} + \eta_1 + \bsl{q}_1)\cdot\bsl{R}} e^{-\ii (-\bsl{p}+\eta_3+\bsl{q}_3)\cdot\bsl{R}'} V(\bsl{p})  U^\dagger_{c,\beta_1}(\eta_1) U_{f,\alpha_2} (\bsl{p}+\eta_1 +\bsl{q}_1) U^\dagger_{c,\beta_3}(\eta_3) U_{f,\alpha_4}(-\bsl{p}+\eta_3+\bsl{q}_3) \\
    & \approx \frac{1}{N \Omega} e^{-\ii (\eta_1 + \bsl{q}_1) \cdot \bsl{R}} e^{-\ii (\eta_3+\bsl{q}_3)\cdot \bsl{R}'} \Omega  \widetilde{J}'_{\eta_1\beta_1\alpha_2,\eta_3\beta_4 \alpha_4}(\bsl{R}-\bsl{R}')\ ,
}
and
\eq{
\widetilde{J}'_{\eta_1\beta_1\alpha_2,\eta_3\beta_4 \alpha_4}(\bsl{R}-\bsl{R}') = \frac{1}{N \Omega} \sum_{\bsl{p}} e^{-\ii \bsl{p}\cdot(\bsl{R}-\bsl{R}')} V(\bsl{p}) U^\dagger_{c,\beta_1}(\eta_1) U_{f,\alpha_2}(\bsl{p}+\eta_1) U^\dagger_{c,\beta_3}(\eta_3) U_{f,\alpha_4}(-\bsl{p}+\eta_3)\ .
}
As a result, we have
\eq{
H_{int,5} \approx \sum_{\eta_1 \beta_1} \sum_{\eta_3 \beta_3} \sum_{\alpha_2 \alpha_4} \sum_{\bsl{R}\bsl{R}'} c^\dagger_{\eta_1,\bsl{R},\beta_1} f_{\bsl{R},\alpha_2} c^\dagger_{\eta_3,\bsl{R}',\beta_3} f_{\bsl{R}',\alpha_4} \Omega \widetilde{J}'_{\eta_1\beta_1\alpha_2,\eta_3\beta_3\alpha_4}(\bsl{R}-\bsl{R}')+ h.c. 
}
Owing to the TR symmetry, 
\eq{
\widetilde{J}'_{\eta_1\beta_1\alpha_2,\eta_3\beta_3\alpha_4}(\bsl{R}-\bsl{R}') = \sum_{\beta_3' a_4'} 
\mat{ 
 & 1 & & \\
  1 & & & \\
   & & & 1 \\
   & & 1 &
}_{\beta_3' \beta_3} \mat{ & 1 \\ 1 & }_{a_4' a_4} \widetilde{J}_{\eta_1 \beta_1 a_2,\eta_3 \beta_3 a_4'}(\bsl{R}-\bsl{R}') \ ,
}
where $\widetilde{J}_{\eta_1 \beta_1 a_2,\eta_3 \beta_3 a_4'}(\bsl{R}-\bsl{R}')$ is in \eqnref{eq:J_tilde}.
Therefore, we numerically have 
\eq{
H_{int,5} \approx \Omega  J  \sum_{\eta \eta'} \sum_{\bsl{R} \alpha} c^\dagger_{\eta,\bsl{R}, \alpha} f_{\bsl{R},\alpha} c^\dagger_{\eta',\bsl{R},\bar{\alpha}} f_{\bsl{R},\bar{\alpha}} + h.c. 
}
with $J\approx 4.5$ and $\bar{\alpha}=2,1$ for $\alpha=1,2$.

\subsubsection{$H_{int,6}$}

\eqa{
H_{int,6} &  = \int d^2 r d^2 r' V(\bsl{r}-\bsl{r}') \left\{ \sum_{\bsl{R}_1 \bsl{R}_2 }  :f^\dagger_{\bsl{R}_1} f_{\bsl{R}_2}: z(\bsl{r}-\bsl{R}_1,\bsl{r}-\bsl{R}_2)\ ,\  \sum_{ \bsl{R}'} :c^\dagger_{\bsl{r}'} \chi(\bsl{r}'-\bsl{R}') f_{\bsl{R}'} : \right\}  + h.c. \\
& \approx \int d^2 r d^2 r' V(\bsl{r}-\bsl{r}') \sum_{\bsl{R}\bsl{R}' } \left\{  :\rho_f(\bsl{R}): n_f(\bsl{r}-\bsl{R}) \ ,\  :c^\dagger_{\bsl{r}'} \chi(\bsl{r}'-\bsl{R}') f_{\bsl{R}'} : \right\}  + h.c. \\
& = \sum_{\bsl{R}\bsl{R}' } \sum_{\eta'\beta'\alpha'} \sum_{\bsl{q}'}^{\Omega} \left\{  :\rho_f(\bsl{R}):  \ ,\  : c^\dagger_{\eta'+\bsl{q}',\beta'} f_{\bsl{R}',\alpha'} : \right\} \\
& \qquad \times \frac{1}{\sqrt{N\Omega}} \int d^2 r d^2 r' V(\bsl{r}-\bsl{r}') e^{-\ii (\eta'+\bsl{q}')\cdot\bsl{r}'} n_f(\bsl{r}-\bsl{R}) \chi_{\eta'\beta',\alpha'} \chi(\bsl{r}'-\bsl{R}')  + h.c. \ ,
}
where
\eqa{
& \frac{1}{\sqrt{N\Omega}} \int d^2 r d^2 r' V(\bsl{r}-\bsl{r}') e^{-\ii (\eta'+\bsl{q}')\cdot\bsl{r}'} n_f(\bsl{r}-\bsl{R}) \chi_{\eta'\beta',\alpha'} \chi(\bsl{r}'-\bsl{R}') \\
& = \frac{1}{\sqrt{N}} e^{-\ii (\eta'+\bsl{q}')\cdot\bsl{R}'} \frac{1}{\N \Omega} \sum_{\bsl{p}} e^{-\ii \bsl{p}\cdot (\bsl{R}-\bsl{R}')} V(\bsl{p}) \widetilde{n}(-\bsl{p}) U^\dagger_{c,\beta'}(\eta') U_{f,\alpha'}(-\bsl{p}+\eta') \\
&=\frac{1}{\sqrt{N}} e^{-\ii (\eta'+\bsl{q}')\cdot\bsl{R}'} \widetilde{K}'_{\eta'\beta'\alpha'}(\bsl{R}-\bsl{R}') \ ,
} 
and
\eq{
\widetilde{K}'_{\eta'\beta'\alpha'}(\bsl{R}-\bsl{R}') = \frac{1}{\N \Omega} \sum_{\bsl{p}} e^{-\ii \bsl{p}\cdot (\bsl{R}-\bsl{R}')} V(\bsl{p}) \widetilde{n}(\bsl{p}) U^\dagger_{c,\beta'}(\eta') U_{f,\alpha'}(\bsl{p}+\eta') \ .
}
Thus, we obtain
\eq{
H_{int,6} \approx \sum_{\bsl{R}\bsl{R}' } \sum_{\eta'\beta'\alpha'} \left\{  :\rho_f(\bsl{R}):  \ ,\  : c^\dagger_{\eta',\bsl{R}',\beta'} f_{\bsl{R}',\alpha'} : \right\} \sqrt{\Omega} \widetilde{K}'_{\eta'\beta'\alpha'}(\bsl{R}-\bsl{R}') + h.c. \ . 
}

$\widetilde{K}'_{\eta'\beta'\alpha'}(\bsl{R}-\bsl{R}')\approx \widetilde{K}'_{\eta'\beta'\alpha'}(0)\delta_{\bsl{R},\bsl{R}'}$, which has dominating values around $0.001$ and thus is negligible.

\subsubsection{$H_{int,7}$}

\eqa{
H_{int,7} & = \int d^2 r d^2 r' V(\bsl{r}-\bsl{r}') \left\{  \sum_{\bsl{R} }  c^\dagger_{\bsl{r}} \chi(\bsl{r}-\bsl{R})f_{\bsl{R}} \ ,\ :c^\dagger_{\bsl{r}'} y(\bsl{r}')c_{\bsl{r}'}: \right\} +h.c.\\
& = \frac{1}{\Omega\sqrt{N^3}} \sum_{\bsl{q}\bsl{q}_1'\bsl{q}_2'}^{\Lambda} \sum_{\eta \beta \alpha} \sum_{\eta_1' \beta_1' \eta_2' \beta_2'} \left\{ \sum_{\bsl{R}} c^\dagger_{\eta+\bsl{q},\beta} f_{\bsl{R},\alpha}, :c^\dagger_{\eta_1'+\bsl{q}_1',\beta_1'} c_{\eta_2'+\bsl{q}_2',\beta_2'}: \right\}\sum_{\bsl{G}'}e^{-\ii (\eta_1'+\bsl{q}_1'-\eta_2'-\bsl{q}_2'-\bsl{G}'-\eta-\bsl{q})\cdot\bsl{R}} \\ 
& \qquad \times V(\eta_1'+\bsl{q}_1'-\eta_2'-\bsl{q}_2'-\bsl{G}') U^\dagger_{c,\beta}(\eta)U_{f,\alpha}(\eta+\bsl{q}+\eta_1'+\bsl{q}_1'-\eta_2'-\bsl{q}_2'-\bsl{G}') U_{c,\beta_1'}^\dagger(\eta_1') U_{c,\beta_2'}(\eta_2'+\bsl{G}')+h.c.\\
& \approx \Omega^{3/2} \sum_{\eta\beta\alpha} \sum_{\eta_1' \beta_1' \eta_2' \beta_2'} \left\{ \sum_{\bsl{R}} c^\dagger_{\eta,\bsl{R},\beta} f_{\bsl{R},\alpha} , :c^\dagger_{\eta_1',\bsl{R},\beta_1'} c_{\eta_2',\bsl{R},\beta_2'}: \right\} \widetilde{K}_{\eta\beta\alpha,\eta_1'\beta_1'\eta_2'\beta_2'} \ ,
}
where
\eq{
\widetilde{K}_{\eta\beta\alpha,\eta_1'\beta_1'\eta_2'\beta_2'} = \frac{1}{\Omega} \sum_{\bsl{G}'}V(\eta_1'-\eta_2'-\bsl{G}') U^\dagger_{c,\beta}(\eta) U_{f,\alpha}(\eta+\eta_1'-\eta_2'-\bsl{G}') U^\dagger_{c,\beta_1'}(\eta_1') U_{c,\beta_2}(\eta_2'+\bsl{G}') \ .
}

We find that 
\eqa{
\label{eq:K_expression}
& \widetilde{K}_{\eta\beta\alpha,\eta_1'\beta_1'\eta_2'\beta_2'} = \delta_{\eta_1'\eta}\delta_{\eta_2'\bar{\eta}}\widetilde{K}_{\eta\beta\alpha,\eta\beta_1'\bar{\eta}\beta_2'}  + \delta_{\eta_1'\bar{\eta}}\delta_{\eta_2'\eta}\widetilde{K}_{\eta\beta\alpha,\bar{\eta}\beta_1'\eta\beta_2'}\\
& \widetilde{K}_{\eta\beta\alpha,\eta\beta_1'\bar{\eta}\beta_2'} = \widetilde{K}_{\bar{\eta}\beta\alpha,\bar{\eta}\beta_1'\eta\beta_2'} \\
& \widetilde{K}_{\K\beta\alpha,\K\beta_1'\K'\beta_2'} = -\widetilde{K}^*_{\K'\beta\alpha,\K'\beta_1'\K\beta_2'} \ ,
}
and we numerically find
\eqa{
\widetilde{K}_{\K\beta\alpha,\K\beta_1'\K'\beta_2'} = \delta_{\beta\alpha} \delta_{\beta_1'\alpha}\delta_{\beta_2',\alpha+2} [\Re(K)+ (-1)^{a-1} \ii \Im(K)]  + \delta_{\beta\alpha} \delta_{\beta_1',5-\alpha}\delta_{\beta_2',3-\alpha} [-\Re(K)+ (-1)^{a-1} \ii \Im(K)] \ ,
}
where
\eq{
K\approx -2.7 + 2.7 \ii \ .
}

\subsubsection{Summary of The Interaction}

In sum, the total interaction reads
\eq{
H_{int} \approx H_U  + H_V + H_{W} + H_J + H_K \ ,
}
where
\eq{
H_U =   \sum_{\bsl{R} } U :\rho_f(\bsl{R}): :\rho_f(\bsl{R}): 
}
with $U=11.27$,
\eqa{
H_V & = \frac{1}{N \Omega} \sum_{\bsl{q}\bsl{q}'}^{\Lambda} \sum_{\eta \eta', \beta \beta'} :c^\dagger_{\eta+\bsl{q},\beta} c_{\eta+\bsl{q},\beta}: :c^\dagger_{\eta'+\bsl{q}',\beta'} c_{\eta'+\bsl{q}',\beta'}:  V(\bsl{p}=0) \\
& \qquad + \frac{1}{N\Omega} \sum_{\bsl{q}\bsl{q}'}^{\Lambda} \sum_{\eta \eta'} \sum_{\beta_1 \beta-2 \beta_3 \beta_4} :c^\dagger_{\eta+\bsl{q},\beta_1} c_{\eta+\bsl{q},\beta_2}: :c^\dagger_{\eta'+\bsl{q}',\beta_3} c_{\eta'+\bsl{q}',\beta_4}: V'_{\beta_1 \beta_2 \beta_3 \beta_4}(\eta,\eta') \ ,
}
\eq{
H_{W} =  \Omega W \sum_{\bsl{R}} :\rho_f(\bsl{R}): :c^\dagger_{\bsl{R}} c_{\bsl{R}}: + h.c. 
}
with $W=32$,
\eq{
H_{J} = \Omega J  \sum_{\bsl{R}} \sum_{\eta \eta'} \sum_{\alpha} \left[ \left\{ c^\dagger_{\eta,\bsl{R},\alpha} f_{\bsl{R},\alpha}, f^\dagger_{\bsl{R},\alpha} c_{\eta',\bsl{R},\alpha} \right\}  + ( c^\dagger_{\eta,\bsl{R}, \alpha} f_{\bsl{R},\alpha} c^\dagger_{\eta',\bsl{R},\bar{\alpha}} f_{\bsl{R},\bar{\alpha}} + h.c. ) \right]
}
with $J = 4.5$ and $\bar{\alpha}=2,1$ for $\alpha=1,2$, and
\eq{
H_K= \Omega^{3/2} \sum_{\eta\beta\alpha} \sum_{\eta_1' \beta_1' \eta_2' \beta_2'} \left\{ \sum_{\bsl{R}} c^\dagger_{\eta,\bsl{R},\beta} f_{\bsl{R},\alpha} , :c^\dagger_{\eta_1',\bsl{R},\beta_1'} c_{\eta_2',\bsl{R},\beta_2'}: \right\} \widetilde{K}_{\eta\beta\alpha,\eta_1'\beta_1'\eta_2'\beta_2'} 
}
with the expression of $\widetilde{K}_{\eta\beta\alpha,\eta_1'\beta_1'\eta_2'\beta_2'}$ in \cref{eq:K_expression}.

\subsection{Alternative Convention}

There is an alternative convention where  where $m_z$ and $C_{2x}$ do not have momentum shifts as \eqnref{eq:sym_rep_moire_momentum_basis}.
It is achieved by doubling the unit cell, and we will discuss it in this subsection for completeness.
Define
\eq{
\psi^\dagger_{\pm,\bsl{r}} = \frac{1}{\sqrt{2}}\left( \psi^\dagger_{t,\bsl{r}} \pm \psi^\dagger_{b,\bsl{r}}\right)\ .
}
Then, the Hamiltonian reads
\eq{
H =\sum_{\alpha=\pm} H_{\alpha} \ ,
}
with 
\eq{
H_{\alpha} = \int d^2 r  \psi^\dagger_{\alpha,\bsl{r}} \left[ -t \partial_x \partial_y \tau_x - t' (\partial_x^2 - \partial_y^2) \tau_y + \alpha W(\bsl{r}) \right]  \psi_{\alpha,\bsl{r}}\ , 
}
and the symmetry operations read 
\eqa{
& C_{4z} \psi^\dagger_{\alpha,\bsl{r}} C_{4z}^{-1} = \psi^\dagger_{\alpha,C_{4z}\bsl{r}} (-\ii \tau_z) \\
& C_{2x} \psi^\dagger_{\alpha,\bsl{r}} C_{2x}^{-1} = \psi^\dagger_{\alpha,C_{2x}\bsl{r}} (\alpha \tau_y) \\
& \TR \psi^\dagger_{\alpha,\bsl{r}} \TR^{-1} = \psi^\dagger_{\alpha,\bsl{r}} (\ii \tau_x) \\
& T_{\bsl{R}_{M}'}\psi^\dagger_{\alpha,\bsl{r}} T_{\bsl{R}_{M}'}^{-1} =  e^{-\ii M_{-}\cdot\bsl{R}_{M}' } \psi^\dagger_{\bsl{r} + \bsl{R}_{M}',\alpha} \\
& T_{\bsl{a}_{M,1}}\psi^\dagger_{\alpha,\bsl{r}} T_{\bsl{a}_{M,1}}^{-1} =  - e^{-\ii M_{-}\cdot\bsl{a}_{M,1} } \psi^\dagger_{-\alpha,\bsl{r} + \bsl{a}_{M,1}}\\
& m_z \psi^\dagger_{\alpha,\bsl{r}} m_z^{-1} = \psi^\dagger_{\alpha,\bsl{r}} \alpha \\
}
where $\bsl{R}_{M}' = \mat { \bsl{a}_{M,1}' & \bsl{a}_{M,2}' } \dsZ^2$ with $\bsl{a}_{M,1}' = \bsl{a}_{M,1} + \bsl{a}_{M,2}$ and $\bsl{a}_{M,2}' = -\bsl{a}_{M,1} + \bsl{a}_{M,2}$.
The new $\bsl{a}_{M,1}'$ and $\bsl{a}_{M,2}'$ show that the unit cell is doubled.

By defining 
\eqa{
& \bsl{b}_{M,1}' = \bsl{q}_1 \\
& \bsl{b}_{M,2}' = \bsl{q}_2 \\
& \bsl{G}_M' = \mat { \bsl{b}_{M,1}' & \bsl{b}_{M,2}' } \dsZ^2 \ ,
}
we have
\eqa{
& C_{4z} \psi^\dagger_{\alpha,\bsl{k}- \bsl{G}_{M}'} C_{4z}^{-1} = \psi^\dagger_{\alpha,C_{4z}\bsl{k}- C_{4z}\bsl{G}_{M}'} (-\ii \tau_z) \\
& C_{2x} \psi^\dagger_{\alpha,\bsl{k}- \bsl{G}_{M}'} C_{2x}^{-1} = \psi^\dagger_{\alpha,C_{2x}\bsl{k}- C_{2x}\bsl{G}_{M}'} (\alpha \tau_y) \\
& \TR \psi^\dagger_{\alpha,\bsl{k}- \bsl{G}_{M}'} \TR^{-1} = \psi^\dagger_{\alpha,-\bsl{k}+ \bsl{G}_{M}'} (\ii \tau_x) \\
& m_z \psi^\dagger_{\alpha,\bsl{k}- \bsl{G}_{M}'} m_z^{-1} = \alpha \psi^\dagger_{\alpha,\bsl{k}- \bsl{G}_{M}'}  \ .
}

This convention shows that the twisted Checkerboard model can be treated as two copies of \refcite{KaiSun2023THF}.

\section{Lieb Lattice and Analytical Wannier States from the Instanton Action}
\label{app:liebSP}

The Lieb lattice is built on the $1b = (1/2,1/2)$ and $2c = (1/2,0), (0,1/2)$ positions. The real space model with nearest neighbor hoppings is
\bea
H &= t \sum_{\mbf{R}} \lp c^\dag_{\mbf{R},b} (c^\dag_{\mbf{R},c_1}+c^\dag_{\mbf{R}+\hat{x},c_1}+c^\dag_{\mbf{R},c_2}+c^\dag_{\mbf{R}+\hat{y},c_2})  \rp + h.c.  \ .\\
\eea
It forms an $S$-matrix model in the order basis $b,c_1,c_2$ with
\bea
h(\mbf{k}) = 2t \bpm 
 & \cos k_y/2 & \cos k_x/2 \\
  \cos k_y/2 & & \\
  \cos k_x/2 & &  \\
\epm
\eea
which has chiral symmetry $S = \text{diag }(1,-1,-1)$, $C_{4z} = 1 \oplus \sigma_1$, and $\mathcal{T} = K$. This Hamiltonian is very simple, and its spectrum can be obtained analytically:
\bea
E_n = 0, \pm t \sqrt{4 + 2(\cos k_x + \cos k_y)} 
\eea
with eigenvectors 
\bea
U_0(\mbf{k}) &= (0, - \cos \frac{k_x}{2}, \cos \frac{k_y}{2})^T/\sqrt{\cos^2 \frac{k_x}{2}+\cos^2 \frac{k_y}{2}} \\
U_{\pm 1}(\mbf{k}) &= (\pm \frac{1}{\sqrt{2}},\frac{\cos k_y/2}{\sqrt{2+\cos k_x+\cos k_y}},\frac{\cos k_x/2}{\sqrt{2 + \cos k_x+\cos k_y}})^T  \ .\\
\eea

Symmetries play a crucial role in identifying the heavy fermion wavefunction. We will use the language of topological quantum chemistry \cite{Bradlyn2017TQC} (see Ref. \cite{2021ARCMP..12..225C} for a pedagogical review with an emphasis on group theory, or \cite{Herzog2022FlatBandQuantumGeometry} for a brief introduction to the formalism in tight-binding models). The momentum space representations of the atomic basis in the space group $G = p41'$ (number 75 in the BNS notation) 
 are written
\bea
\label{eq:irrepsK}
L': \qquad  A_{1b} \uparrow G &= \Gamma_1 + X_2 + M_2 \\ 
L: \qquad  A_{2c} \uparrow G &= \Gamma_1 \oplus \Gamma_2 + X_1 \oplus X_2 + M_3 M_4 \\
\eea
where $L$ is the larger sublattice consisting of $s$ orbitals at the 2c = $(1/2,0), (0,1/2)$ position, and $L'$ is the smaller sublattice consisting of an $s$ orbital at the 1b = $(1/2,1/2)$ position. These real space orbitals induce the momentum space irreps \EqJHA{eq:irrepsK} where $\Gamma, X, M$ denote the high-symmetry momenta with little group $41', 2,$ and $41'$ respectively. Their character tables are
\bea
\begin{tabular}{c | c c c c c}
 & $1$ & $C_{4}$ & $C_2$ & $C_{4}^3$ \\
 \hline
$\Gamma_1, M_1$ & $1$ & $1$ & $1$ & $1$ \\
$\Gamma_2, M_2$ & $1$ & $-1$ & $1$ & $-1$ \\
$\Gamma_3\Gamma_4, M_3M_4$ & $2$ & $0$ & $-2$ & 0 \\
\end{tabular}, \qquad \begin{tabular}{c | c c c }
 & $1$ & $C_{2}$ & \\
 \hline
$X_1$ & $1$ & $1$  \\
$X_2$ & $1 $ & $-1$ \\
\end{tabular} \ .
\eea

As discussed in the Main Text, the $\Gamma$ and $X$ irreps uniquely restrict any possible Wannier state to transform in a single irrep. For completeness, we enumerate all atomic limits in $p41'$. They are
\bea
A_{1a} &= \Gamma_1 + X_1 + M_1, \quad  B_{1a} = \Gamma_2 + X_1 + M_2, \quad  {}^1E{}^2E_{1a} = \Gamma_3 \Gamma_4 + X_2 + M_3 M_4 \\
A_{1b} &= \Gamma_1 + X_2 + M_2, \quad  B_{1b} = \Gamma_2 + X_2 + M_1, \quad  {}^1E{}^2E_{1b} = \Gamma_3 \Gamma_4 + X_1 + M_3 M_4 \\
A_{2c} &= \Gamma_1 \oplus \Gamma_2 +X_1\oplus X_2 + M_3 M_4, \quad B_{2c} = \Gamma_3 \Gamma_4 +X_1\oplus X_2 + M_1 M_2 \ .
\eea

Since the Wannier state must carry the $\Gamma_2, X_1$ irreps of the flat band, only the $B_{1a}$ irrep is possible as one can see above. Thus our ansatz for the HF wannier state is the following
\bea
\label{eq:Omegaf}
U_f(\mbf{k}) &= U_0(\mbf{k}) \cos \th(\mbf{k}) + \bpm 1 \\ 0 \\ 0 \epm \sin \th(\mbf{k}) = U(\mbf{k}) \bpm -\frac{1}{\sqrt{2}} \sin \th \\ \cos \th \\ \frac{1}{\sqrt{2}}  \sin \th \epm 
\eea
where $U(\mbf{k})$ is the $3 \times 3$ eigenvector matrix whose columns are $U_{-1}(\mbf{k}), U_0(\mbf{k}), U_{+1}(\mbf{k})$. Since the $f$-mode wavefunction $U_f(\mbf{k})$ of the Wannier state must occupy the $M_2$ irrep at $M$ and then transition into the flat band, we expect $\th \to \pi/2$ at $\mbf{k}=M$ with decay to $\th = 0$ away from $M$. The functional form of $\th(\mbf{k})$ can now be determined. 

We now compute the Wannier spread of $U_f(\mbf{k})$. Since $U_f(\mbf{k})$ is $C_2\mathcal{T}$-symmetric and $D[C_2\mathcal{T}] = K$, we can choose $U_f(\mbf{k})$ to be real so its Berry connection vanishes, and thus the Wannier spread is the single integral
\bea
\frac{1}{2} \braket{r^2} = \int \frac{d^2k}{(2\pi)^2} \frac{1}{2} \del_i U_f^\dag \del_i U_f = \int \frac{d^2k}{(2\pi)^2} \lp \frac{1}{2}(\nabla \theta)^2 + \frac{1}{2} g(\mbf{k}) \cos^2 \th \rp, \qquad g(\mbf{k}) = (\del U_{0})^2 = \frac{1-\cos k_x \cos k_y}{2(2+ \cos k_x +\cos k_y )^2} \geq 0 \ .
\eea
Here $g(\mbf{k})$ is the Fubini-Study quantum metric. In general, $g(\mbf{k})$ is given by the equivalent expressions
\bea
g(\mbf{k}) &= \frac{1}{2} \Tr \, \del_i P(\mbf{k})\del_i P(\mbf{k}) = \del_i U^\dag \del_i U + U^\dag \del_i U U^\dag \del_i U
\eea
where $P = UU^\dag$, and here we have used $U^\dag \del_i U = 0$ since $U$ is real. 

The key observation is that minimization of $\braket{r^2}$ (the maximal localization condition of the Wannier function) can be phrased as an Euler-Lagrange problem. To make the problem analytically tractable, we focus near $\mbf{k} = M$ where $\th \sim \pi/2$ and $g(M+\mbf{k}) \sim \frac{1}{|k|^2}$. We fix a cutoff $\Lambda$ (the Wannier disentanglement window) and impose Dirichlet boundary conditions $\th(\mbf{k}+M) = 0$ for $|\mbf{k}| = \Lambda$. The rotational symmetry emerges naturally in the low energy limit around $M$. Thus we have
\bea
\frac{1}{2} \braket{r^2} &= S +  \int_{\mbf{k}: |\mbf{k}-M| \geq \Lambda } \frac{d^2k}{(2\pi)^2} \frac{1}{2} g(\mbf{k})
\eea
where $S$ is the action
\bea
S &= \int_\Lambda \frac{d^2k}{(2\pi)^2} \lp \frac{1}{2}(\nabla \th)^2 + \frac{1}{2 |k|^2} \cos^2 \th \rp \ .
\eea
Note that $(\nabla \th)^2 = (\del_{|k|} \th)^2 + \frac{1}{|k|^2}(\del_\phi \th)^2$ in polar coordinates, so the minimum is achieved when $\del_\phi \th = 0$ (the $s$-wave circular harmonic). Thus we can assume $\th(\mbf{k}) = \th(|k|)$. Since 
the Lagrangian is scale-invariant, we are motivated to change variables to $|k| = e^t$ so that $|k| = 0$ maps to $t = - \infty$. 
We find
\bea
S &= \int \frac{e^{2t} \, dt}{(2\pi)^2}  \lp \frac{1}{2}(e^{-t} \dot \th)^2 + \frac{1}{2 e^{2t}} \cos^2 \th \rp \\
&= \int \frac{dt}{(2\pi)^2}  \lp \frac{1}{2}\dot \th^2 + \frac{1}{2} \cos^2 \th \rp \\
\eea
which is just the Lagrangian of a particle in the periodic potential $- \frac{1}{2} \cos^2 \th$. At time $t= -\infty$, we must have $\cos \th = 0$, which is at the top of the potential. We can now use known techniques from the theory of instantons to reduce the second order Euler-Lagrange equations to a first order differential equation which is immediately integrable. First we observe
\bea
S &= \frac{1}{2} \int \frac{dt}{(2\pi)^2}  \lp \dot \th^2 +\cos^2 \th \rp  = \frac{1}{2} \int \frac{dt}{(2\pi)^2}  \lp \dot \th \pm \cos \th \rp^2 \mp \int \frac{dt}{(2\pi)^2}  \dot \th \cos \th \\
&= \frac{1}{2} \int \frac{dt}{(2\pi)^2}  \lp \dot \th \pm \cos \th \rp^2 \mp \int \frac{d\th}{(2\pi)^2} \cos \th \ . \\
\eea
The $\theta$ integral we can perform since we know $\th$ interpolates from $\pi/2$ at $M$ to $0$ on the boundary. Thus we have
\bea
S &= \frac{1}{2} \int \frac{dt}{(2\pi)^2}  \lp \dot \th + \cos \th \rp^2 - \int_{\pi/2}^{0} \frac{d\th}{(2\pi)^2} \cos \th \\
&= \frac{1}{2} \int \frac{dt}{(2\pi)^2}  \lp \dot \th + \cos \th \rp^2 + \frac{1}{(2\pi)^2} \\
\eea
which shows the action is lower-bounded. This lower bound can be reached by taking
\bea
\dot \th = - \cos \th
\eea
which can be directly integrated to find
\bea
\th(t) = -2 \arctan (\tanh \frac{t - t_0}{2}) \\
\eea
where $t_0$ is the only undetermined constant. Note that this instanton solution obeys
\bea
\lim_{t\to -\infty} \th(t) &= -2 \arctan (-1) = \frac{\pi}{2} \\ 
\lim_{t\to t_0} \th(t) &= -2 \arctan 0 = 0 \\
\eea
so for any value of the cutoff $\Lambda  = e^{t_0}$, we can produce a function that strictly minimizes the action. In terms of the original variables, we find
\bea
\th &= -2 \arctan \frac{|k|/\Lambda-1}{|k|/\Lambda+1}, \qquad \cos \th &= \frac{2|k|/\Lambda}{(|k|/\Lambda)^2+1}  \\
\eea
where $M + \mbf{k} = |k| (\cos \phi, \sin \phi)$. 

Note that $\cos \th \sim |k|$ is not smooth in $\mbf{k}$. However, the flat band wavefunction is also not smooth, and obeys $U_0(M+\mbf{k}) \to (0, \cos \th, - \sin \phi)$. Thus we see that
\bea
U_f(\mbf{k}) &= U_0(\mbf{k}) \cos \th(\mbf{k}) + \bpm 1 \\ 0 \\ 0 \epm \sin \th(\mbf{k}) 
\eea
which approaches $(1, |k| \cos \phi ,|k| \sin \phi)$ with $M + \mbf{k} = |k| (\cos \phi, \sin \phi)$ as $|k| \to 0$. We see that the wavefunction is smoothed at $M$ through the instanton profile, and hence can give a localized Wannier function. 

\subsection{Conduction electrons and Hamiltonian}

We will now determine the conduction electron wavefunctions by requiring smoothness and orthogonality with the HF fermion wavefunction $U_f(\mbf{k})$. Recall from \EqJHA{eq:Omegaf} that $U_f(\mbf{k}) = U(\mbf{k}) \Omega_f(\th(\mbf{k}))$ where $\Omega_f(\th) = (- \frac{1}{\sqrt{2}} \sin \th, \cos \th,  \frac{1}{\sqrt{2}} \sin \th)$ is the transformation from the band basis (ordered $1, 0, -1$) to the $f$-mode wavefunction. $\Omega_f(\th)$ is one column of the full unitary matrix transforming the band basis to the $c$-$f$ basis. The two vectors $\Omega_1 = (1,0,1)^T/\sqrt{2},\Omega_2 = (\cos \th ,-\sqrt{2} \sin \th,-\cos \th)^T/\sqrt{2}$ form a complete orthonormal basis with $\Omega_f$. We then choose the time-reversal symmetric gauge $\Omega_\pm = e^{\pm i \phi}(\pm i \Omega_1 + \Omega_2)$ to form the following unitary matrix $\Omega(\th(\mbf{k})) = (\Omega_+ , \Omega_-, \Omega_f)$
\bea
\Omega(\mbf{k}) = \bpm
 \frac{e^{i \phi}}{2} (1- i \cos \th) &   \frac{e^{-i \phi}}{2} (1+ i \cos \th) & - \sin \th / \sqrt{2} \\
- \frac{i e^{i \phi}}{\sqrt{2}} \sin \th & \frac{i e^{-i \phi}}{\sqrt{2}} \sin \th & \cos \th \\
 \frac{e^{i \phi}}{2} (1+ i \cos \th) &  \frac{e^{-i \phi}}{2} (1- i \cos \th) & \sin \th / \sqrt{2}\\
 \epm, \quad \mbf{k} = M + |k|(\cos \phi, \sin \phi), \quad \th = \th(|k|) = -2 \arctan \frac{|k|/\Lambda -1}{|k|/\Lambda +1} \\
\eea
defines a smooth, symmetry-preserving transformation into $\gamma^\dag_{\mbf{k},c+},\gamma^\dag_{\mbf{k},c-},\gamma^\dag_{\mbf{k},f}$, the two conduction electrons and the Wannier state respectively. One can verify that the conduction electron wavefunctions are smooth using the explicit form of the band eigenvectors, 
\bea
U(\mbf{k}) &= \left(
\begin{array}{ccc}
-1/\sqrt{2} & 0 & 1/\sqrt{2} \\
 \frac{\cos k_y/2}{\sqrt{2 + \cos k_x + \cos k_y}} & -\frac{\sqrt{2} \cos k_x/2}{\sqrt{2 + \cos k_x + \cos k_y}} & \frac{\cos k_y/2}{\sqrt{2 + \cos k_x + \cos k_y}} \\
 \frac{\cos k_x/2}{\sqrt{2 + \cos k_x + \cos k_y}} & \frac{\sqrt{2} \cos k_y/2}{\sqrt{2 + \cos k_x + \cos k_y}} & \frac{\cos k_x/2}{\sqrt{2 + \cos k_x + \cos k_y}} \\
\end{array}
\right) \to \left(
\begin{array}{ccc}
 -\frac{1}{\sqrt{2}} & 0 & \frac{1}{\sqrt{2}} \\
 -\frac{\sin \phi}{\sqrt{2}} & \cos \phi & -\frac{\sin \phi}{\sqrt{2}} \\
 -\frac{\cos \phi}{\sqrt{2}} & -\sin \phi & -\frac{\cos \phi}{\sqrt{2}} \\
\end{array}
\right)
\eea
approaching the singular point at $M$. Note that individually, $\Omega$ and $U$ are discontinuous at $M$. However, their product is smooth:
\bea
\label{eq:smooth}
U(\mbf{k}) \Omega(\mbf{k}) &=\left(
\begin{array}{ccc}
 \frac{i \sqrt{2} k}{\Lambda } & -\frac{i \sqrt{2} \bar{k}}{\Lambda } & 1 \\
 -\frac{i}{\sqrt{2}} & \frac{i}{\sqrt{2}} & \frac{2 k_x }{\Lambda } \\
 -\frac{1}{\sqrt{2}} & -\frac{1}{\sqrt{2}} & -\frac{2 k_y }{\Lambda } \\
\end{array}
\right) + O(k^2), \qquad k_x + i k_y = k, \, k_x - i k_y = \bar{k} \ .
\eea
 Thus we see explicitly that the HF states
\bea
\gamma^\dag_{\mbf{k},\mu} = \sum_\al c^\dag_{\mbf{k},\al} [U(\mbf{k}) \Omega(\mbf{k})]_{\al \mu}
\eea
have smooth wavefunctions everywhere on the BZ for any value of the cutoff $\Lambda$. It is now a simple matter to write down the single-particle Hamiltonian in the HF basis. Defining $E(\mbf{k}) = \text{diag}(E_+(\mbf{k}),0, E_-(\mbf{k}))$, we have 
\bea
H_0 &= \sum_{\al \be} c^\dag_{\mbf{k},\al} h_{\al \beta}(\mbf{k}) c_{\mbf{k},\be} = \sum_{\mu \nu} \gamma^\dag_{\mbf{k},\mu} [\Omega^\dag(\mbf{k}) E(\mbf{k}) \Omega(\mbf{k})]_{\mu \nu} \gamma_{\mbf{k},\nu} = \sum_{\mu \nu} \gamma^\dag_{\mbf{k},\mu} h^{HF}_{\mu \nu}(\mbf{k}) \gamma_{\mbf{k},\nu}  \\
h^{HF}(\mbf{k}) &= \Omega^\dag(\mbf{k}) E(\mbf{k}) \Omega(\mbf{k})= t 
\bpm
 & i 2 \bar{k}^2/\Lambda & - \bar{k}/\sqrt{2}  \\
-i 2 k^2/\Lambda & & - k/\sqrt{2}  \\
- k/\sqrt{2} & - \bar{k}/\sqrt{2} & \\
\epm + \dots 
\eea
so we identify the conduction electron Hamiltonian as a $2\pi$ Berry phase topological semi-metal with a quadratic band touching, the heavy electron block being completely flat, and the linear coupling between them giving the Dirac band structure $0,\pm t|k|$.  Note that \EqJHA{eq:smooth} shows there is a spinless representation of time-reversal,
\bea
\mathcal{T} \gamma^\dag_{\mbf{k},c\pm} \mathcal{T}^{-1} = \gamma^\dag_{-\mbf{k},c\mp},
\eea
as is appropriate for spinor wavefunctions with $2\pi$ Berry phase.

\subsection{Anderson Model}

Choosing the Hubbard interaction, we have
\bea
H_{int} &= U \sum_{\mbf{R}\al} n_{\mbf{R},\al ,\uparrow} n_{\mbf{R},\al ,\downarrow} = U \sum_{\mbf{q} \al} \rho_{\mbf{q},\al,\uparrow} \rho_{-\mbf{q},\al,\downarrow}
\eea
where the density operator is 
\bea
\rho_{\mbf{q},\al,\sigma} = \frac{1}{\sqrt{N}}\sum_{\mbf{k} \mu \nu} M^\al_{\mu \nu}(\mbf{k},\mbf{q})\gamma^\dag_{\mbf{k}+\mbf{q},\mu,\sigma} \gamma_{\mbf{k},\nu,\sigma}, \qquad M_{\mu \nu}^\al(\mbf{k},\mbf{q}) = U^*_{\al \mu}(\mbf{k}+\mbf{q})U_{\al \nu}(\mbf{k})
\eea
and $N$ is the number of unit cells. Here $\mu,\nu$ denotes the $f,c$ basis. 

We compute a series of terms by seperating out the various flavour interactions. First we have the $ffff$ term:
\bea
H_{U} &= \frac{U}{N} \sum_{\mbf{q} \mbf{k} \mbf{k}' \mu \nu \mu' \nu' \al} M^\al_{ff}(\mbf{k},\mbf{q})f^\dag_{\mbf{k}+\mbf{q},\uparrow} f_{\mbf{k},\uparrow} M^\al_{ff}(\mbf{k}',-\mbf{q})f^\dag_{\mbf{k}'-\mbf{q},\uparrow} f_{\mbf{k}',\downarrow} \\
&=  \sum_{\mbf{R}_1 \mbf{R}_1' \mbf{R}_2 \mbf{R}_2'} U_{\mbf{R}_1,\mbf{R}_1',\mbf{R}_2,\mbf{R}_2'} f^\dag_{\mbf{R}_1,\uparrow} f_{\mbf{R}_1',\uparrow} f^\dag_{\mbf{R}_2,\downarrow} f_{\mbf{R}_2',\downarrow} \\
U_{\mbf{R}_1,\mbf{R}_1',\mbf{R}_2,\mbf{R}_2'} &= \frac{U}{N^3} \sum_{\mbf{q} \mbf{k} \mbf{k}',\al} M^\al_{ff}(\mbf{k},\mbf{q})M^\al_{ff}(\mbf{k}',-\mbf{q}) e^{- i (\mbf{k}+\mbf{q}) \cdot \mbf{R}_1 + i \mbf{k} \cdot \mbf{R}_1' - i (\mbf{k}'-\mbf{q}) \cdot \mbf{R}_2 + i \mbf{k}' \cdot \mbf{R}_2'}
\eea
where the real-space interaction $U_{\mbf{R}_1,\mbf{R}_2,\mbf{R}_3,\mbf{R}_4}$ is large when $|\mbf{R}_i-\mbf{R}_j| \leq 1$ due to the lobe-like structure of the Wannier state with large onsite and next-nearest neighbor overlap as shown in the Main Text.  

To gain a more intuitive understanding of the $f$-mode density-density terms, we analyze them directly in real space in the approximation
\bea
\label{eq:flocalapprox}
f^\dag_{\mbf{R},\sigma} &\approx \frac{1}{2} \sum_{\pmb{\delta}} (-1)^{2\pmb{\delta}\cdot \hat{y}} c^\dag_{\mbf{R}+\pmb{\delta},\sigma}, \qquad \pmb{\delta} \in \{\pm \frac{1}{2}\hat{x},\pm \frac{1}{2}\hat{y} \} \ .
\eea
This approximation is motivated by the large amplitudes of the $f$-mode on the bonds of its plaquette, and much smaller amplitudes (by a factor of 10) otherwise. Fourier transforming to momentum space, \EqJHA{eq:flocalapprox} is equivalent to the approximation
\bea
U_{f,\al}(\mbf{k}) &\approx (0, - \cos \frac{k_x}{2}, \cos \frac{k_y}{2})_\al  \ .
\eea
We then obtain an expression for the density operator (throughout, $\bar{\mathcal{O}}^f$ denotes the projection of $\mathcal{O}$ onto the $f$ mode basis)
\bea
\label{eq:nfintegral}
\bar{n}^f_{\mbf{R},\al,\sigma} &\approx \frac{1}{N}\sum_{\mbf{k}\mbf{k}'} e^{- i (\mbf{k}-\mbf{k}')\cdot(\mbf{R}+\mbf{r}_\al)} \bar{c}^{\dag f}_{\mbf{k},\al,\sigma}\bar{c}^f_{\mbf{k}',\al,\sigma} \\
&= \frac{1}{N}\sum_{\mbf{k}\mbf{k}'} e^{- i (\mbf{k}-\mbf{k}')\cdot(\mbf{R}+\mbf{r}_\al)} f^\dag_{\mbf{k},\sigma}U^\dag_{f,\al}(\mbf{k})U_{f,\al}(\mbf{k}')f_{\mbf{k}',\sigma} \\
&= \frac{1}{N^2} \sum_{\mbf{L},\mbf{L}'}\sum_{\mbf{k}\mbf{k}'} e^{- i (\mbf{k}-\mbf{k}')\cdot(\mbf{R}+\mbf{r}_\al) + i \mbf{k} \cdot \mbf{L}- i \mbf{k}' \cdot \mbf{L}'} f^\dag_{\mbf{L},\sigma}U^\dag_{f,\al}(\mbf{k})U_{f,\al}(\mbf{k}')f_{\mbf{L}',\sigma} \\
&= \sum_{\mbf{L},\mbf{L}'}f^\dag_{\mbf{L},\sigma} f_{\mbf{L}',\sigma} \lp \frac{1}{N^2} 
 \sum_{\mbf{k}\mbf{k}'} e^{- i (\mbf{k}-\mbf{k}')\cdot(\mbf{R}+\mbf{r}_\al) + i \mbf{k} \cdot \mbf{L}- i \mbf{k}' \cdot \mbf{L}'} U^\dag_{f,\al}(\mbf{k})U_{f,\al}(\mbf{k}') \rp \\
 &= \sum_{\mbf{L},\mbf{L}'}f^\dag_{\mbf{L},\sigma} f_{\mbf{L}',\sigma} \lp \frac{1}{N} 
 \sum_{\mbf{k}} e^{- i \mbf{k}\cdot(\mbf{R}-\mbf{L}+\mbf{r}_\al)} U^\dag_{f,\al}(\mbf{k})\rp \lp \frac{1}{N} 
 \sum_{\mbf{k}'} e^{i \mbf{k}'\cdot(\mbf{R}-\mbf{L}'+\mbf{r}_\al)} U_{f,\al}(\mbf{k}') \rp \\
 &= \sum_{\mbf{L},\mbf{L}'}f^\dag_{\mbf{L},\sigma} f_{\mbf{L}',\sigma} \frac{1}{4} (\delta_{\mbf{R}-\mbf{L},\mbf{0}} + \delta_{\mbf{R}-\mbf{L},-2\mbf{r}_\al})(\delta_{\mbf{R}-\mbf{L}',\mbf{0}} + \delta_{\mbf{R}-\mbf{L}',-2\mbf{r}_\al}), \qquad \al = 2,3 \\
 \bar{n}^f_{\mbf{R},\al,\sigma} &\approx \frac{1}{4} (f^\dag_{\mbf{R},\sigma}+f^\dag_{\mbf{R}+2\mbf{r}_\al,\sigma})(f_{\mbf{R},\sigma}+f_{\mbf{R}+2\mbf{r}_\al,\sigma}) 
\eea
and $\bar{n}^f_{\mbf{R},1,\sigma} \approx 0$, since the $f$-modes have small weight on the $L$ ($\al =1$) sublattice. To write this in a simple way, we define the bond-centered operators
\bea
f^\dag_{\braket{\mbf{R} \mbf{R}'},\sigma} &= \frac{f^\dag_{\mbf{R},\sigma}+f^\dag_{\mbf{R}',\sigma}}{\sqrt{2}}, \qquad n_{\braket{\mbf{R} \mbf{R}'},\sigma} = f^\dag_{\mbf{R} \mbf{R}',\sigma}f_{\mbf{R} \mbf{R}',\sigma} \ .
\eea
We now obtain
\bea
H_U &= U \sum_{\mbf{R} \al} \bar{n}^f_{\mbf{R},\al,\uparrow}\bar{n}^f_{\mbf{R},\al,\downarrow} \approx \frac{U}{4} \sum_{\braket{\mbf{R} \mbf{R}'}} n_{\braket{\mbf{R} \mbf{R}'},\uparrow}n_{\braket{\mbf{R} \mbf{R}'},\downarrow}
\eea

We now consider the other interactions terms. Since each band operator is broken into $f$ and $c$ terms and the interaction contains 4 band operators, there are in total 16 terms. We will give explicit expressions for the dominant terms in the TBG Anderson-`+'' model, which are the $W$ and $J$ terms to be explained below. 

The $W$ term is a density-density interaction between the $c$ modes and $f$ modes. It can be written
\bea
H_W &= U \sum_{\mbf{R} \al} (\bar{n}^f_{\mbf{R},\al,\uparrow} \bar{n}^c_{\mbf{R},\al,\downarrow} + \bar{n}^f_{\mbf{R},\al,\downarrow} \bar{n}^c_{\mbf{R},\al,\uparrow} )
\eea
where $\bar{n}^c_{\mbf{R},\al,\sigma}$ is the projection of the density onto the $c$-electrons:
\bea
\bar{n}^c_{\mbf{R},\al,\sigma} &= \frac{1}{N}\sum_{\mbf{k}\mbf{k}',\mu \nu = c_{\pm}} e^{- i (\mbf{k}-\mbf{k}')\cdot(\mbf{R}+\mbf{r}_\al)} \gamma^\dag_{\mbf{k},\mu,\sigma}U^\dag_{\mu,\al}(\mbf{k})U_{\nu ,\al}(\mbf{k}')\gamma_{\mbf{k}', \nu,\sigma} \ . \\ 
\eea
A convenient approximation is to now drop the $\mbf{k}$ dependence in the $c$-electron wavefunction, replacing $U_{\mu,\al}(\mbf{k}) \to U_{\mu ,\al}(M)$ since away from the $M$ point, the $c$-electrons are high-energy and errors in their wavefunctions will be irrelevant. We find that
\bea
e^{- i (\mbf{k}-\mbf{k}') \cdot \mbf{r}_\al} U^\dag_{\mu,\al}(\mbf{k})U_{\nu ,\al}(\mbf{k}') \to \frac{1}{2} (-1)^{\delta_{\al,2} (1-\delta_{\mu \nu})}, \qquad (\al = 2,3)
\eea
and thus the $W$ term can be written
\bea
H_W &= \frac{U}{4N} \sum_{\mbf{R} \al, \mbf{k}\mbf{k}, \mu \nu} n_{\braket{\mbf{R}, \mbf{R}+2\mbf{r}_\al},\uparrow} e^{- i (\mbf{k}-\mbf{k}')\cdot \mbf{R}} \gamma^\dag_{\mbf{k},\mu,\downarrow}(-1)^{\delta_{\al,2} (1-\delta_{\mu \nu})}\gamma_{\mbf{k}', \nu,\downarrow}  + (\uparrow \, \leftrightarrow \, \downarrow) \ .
\eea
The last term we will consider is the $J$ term, which is an $fc$-exchange term taking the form
\bea
H_J &= \frac{U}{N} \sum_{\mbf{q} \al} \sum_{\mbf{k} \mu \nu,\mbf{k}'} M^\al_{f \nu}(\mbf{k},\mbf{q})M^\al_{\mu f}(\mbf{k}',-\mbf{q}) f^\dag_{\mbf{k}+\mbf{q},\uparrow} \gamma_{\mbf{k},\nu,\uparrow} \gamma^\dag_{\mbf{k}'-\mbf{q},\mu',\downarrow} f_{\mbf{k}',\downarrow} + h. c. \\
&= \frac{U}{N} \sum_{\mbf{q} \al} \sum_{\mbf{k} \mu \nu,\mbf{k}'} M^\al_{f \nu}(\mbf{k}-\mbf{q},\mbf{q})M^\al_{\mu f}(\mbf{k}',-\mbf{q}) f^\dag_{\mbf{k},\uparrow} \gamma_{\mbf{k}-\mbf{q},\nu,\uparrow} \gamma^\dag_{\mbf{k}'-\mbf{q},\mu',\downarrow} f_{\mbf{k}',\downarrow} + h. c. \\
&\sim \frac{U}{N} \sum_{\mbf{q}} \sum_{\mbf{k} \mu \nu,\mbf{k}'} \lp \sum_\al U^*_{\al f}(\mbf{k})U_{\al \nu}(M) U^*_{ \al \mu}(M) U_{\al f}(\mbf{k}') 
\rp f^\dag_{\mbf{k},\uparrow} \gamma_{\mbf{k}-\mbf{q},\nu,\uparrow} \gamma^\dag_{\mbf{k}'-\mbf{q},\mu',\downarrow} f_{\mbf{k}',\downarrow} + h. c. \\
\eea
where we have again replaced the conduction mode wavefunctions with their value at the band crossing point.

\bibliographystyle{apsrev4-2}
\bibliography{bibfile_references,tbg_ref_DH}

\end{document}